\documentclass[floats,twocolumn,pra]{revtex4}%
\usepackage{graphics}
\usepackage{amsmath}
\usepackage{graphicx}
\usepackage{amsfonts}
\usepackage{amssymb}
\usepackage{eurosym}
\usepackage{graphics}
\usepackage{float}
\usepackage{longtable}
\usepackage{epsfig}
\usepackage{latexsym}
\usepackage{theorem}
\usepackage{bbm}
\usepackage{bm}
\usepackage{psfrag}
\usepackage{color}%
\newtheorem{theorem}{\emph{Theorem}}

\newtheorem{remark}[theorem]{\emph{Remark}}

\begin{document}
\title{Limits and Security of Free-Space Quantum Communications}
\author{Stefano Pirandola}
\affiliation{Department of Computer Science, University of York, York YO10 5GH, United Kingdom}

\begin{abstract}
The study of free-space quantum communications requires tools from quantum
information theory, optics and turbulence theory. Here we combine these tools
to bound the ultimate rates for key and entanglement distribution through a
free-space link, where the propagation of quantum systems is generally
affected by diffraction, atmospheric extinction, turbulence, pointing errors,
and background noise. Besides establishing ultimate limits, we also show that
the composable secret-key rate achievable by a suitable (pilot-guided and
post-selected) coherent-state protocol is sufficiently close to these limits,
therefore showing the suitability of free-space channels for high-rate quantum
key distribution. Our work provides analytical tools for assessing the
composable finite-size security of coherent-state protocols in general
conditions, from the standard assumption of a stable communication channel (as is
typical in fiber-based connections) to the more challenging scenario of a
fading channel (as is typical in free-space links).

\end{abstract}
\maketitle

\section{Introduction}

In a future vision where quantum technologies are expected to be developed on
a large scale, hybrid and flexible architectures represent a key strategy for
their success~\cite{NatureComment}. Quantum communications will need to
involve mixed scenarios where fiber connections, good for fixed ground
stations, are merged and interfaced with free-space links, clearly more
suitable for mobile devices. Currently, fiber-based implementations are well
studied, but free-space quantum channels are clearly under-developed from the
point of view of theoretical analysis, both in terms of ultimate limits and
rigorous security assessment. Indeed they require a more demanding study due
to the presence of many effects, such as diffraction, atmospheric extinction,
turbulence effects, pointing errors etc.

In this work we consider all these aspects by combining tools from quantum
information theory~\cite{NCbook,RMP},
optics~\cite{Goodman,Siegman,svelto,Huffman} and turbulence
theory~\cite{Tatarskii,Majumdar,AndrewsBook,Hemani}. In this way, we
investigate the ultimate limits of free-space quantum communications,
establishing upper and lower bounds on the maximum number of secret key bits
(and entanglement bits) that can be shared by two remote parties. Such
analysis explicitly accounts for the fading nature of the free-space channels
together with their typical background noise. Our treatment is mainly
developed for the relevant regime of weak turbulence, but we also discuss how
to extend the results to stronger fluctuations.

Besides investigating the ultimate limits achievable in free-space quantum
communications, we also analyze the practical secret-key rates that are
achievable in such conditions by continuous-variable (CV) protocols of quantum
key distribution (QKD)~\cite{QKDreview}. To this aim we develop a general
theory for assessing the composable finite-size security of coherent-state
protocols~\cite{Noswitch,GG02}, starting from the standard assumption of a
stable communication channel (e.g., as typical in fiber-based connections) to
considering the more challenging scenario of a free-space fading channel,
whose transmissivity rapidly fluctuates.

In particular, we have designed a coherent-state protocol, aided by pilot
pulses and a suitable post-selection procedure, which is able to achieve high
secret-key rates in conditions of weak turbulence, within one order of
magnitude of the ultimate bounds. In this way, we show that
generally-turbulent free-space channels are indeed able to support high-rate
QKD, with immediate consequences for wireless quantum communications.

The manuscript is structured as follows. In Sec.~\ref{Sec_PLOBfree} we provide
the general bounds and capacities for free-space quantum communications. In
Sec.~\ref{Sec_stableCVQKD} we provide a general formulation of composable
finite-size security for CV-QKD. In Sec.~\ref{Sec_freeCVQKD} we extend this
formulation to free-space, showing that suitably-high key rates can indeed be
achieved. Finally, Sec.~\ref{SEC_conclusios} is for conclusions.

\section{Bounds for free-space quantum communications\label{Sec_PLOBfree}}

\subsection{Diffraction-limited bounds\label{diff_bounds_SEC}}

Consider two remote parties separated by distance $z$, one acting as a
transmitter (Alice) and the other as a receiver (Bob). They are located
approximately at the same altitude $h$ on Earth's surface. We consider
free-space quantum communication mediated by a quasi-monochromatic bosonic
mode ($\Delta\lambda$-nm large and $\Delta t$-sec long) represented by a
Gaussian beam, with carrier wavelength $\lambda$, curvature $R_{0}$, and field
spot size $w_{0}$~\cite{svelto,Siegman,Andrews93,Andrews94}. The beam is
prepared by the transmitter (whose aperture is sufficiently larger than
$w_{0}$) and directed towards the receiver, whose aperture is circular with
radius $a_{R}$. Due to free-space diffraction, the receiver gets a beam whose
spot size is increased to
\begin{equation}
w_{z}^{2}=w_{0}^{2}\left[  \left(  1-z/R_{0}\right)  ^{2}+\left(
z/z_{R}\right)  ^{2}\right]  ,
\end{equation}
where $z_{R}:=\pi w_{0}^{2}\lambda^{-1}$ defines the Rayleigh range. Because
the receiver only collects a portion $a_{R}$ of the spread beam, there is a
diffraction-induced transmissivity associated with the channel, given by
\begin{equation}
\eta_{\text{d}}=1-e^{-2a_{R}^{2}/w_{z}^{2}}. \label{etadMAIN}%
\end{equation}
See Appendix~\ref{APP_free_basics} for a brief review on the basic theory of
free-space propagation with Gaussian beams.

Let us apply the point-to-point repeaterless
Pirandola-Laurenza-Ottaviani-Banchi (PLOB)\ bound~\cite{QKDpaper}
$\Phi(x):=-\log_{2}(1-x)$ to $\eta_{\text{d}}$, which provides the secret key
capacity and the two-way quantum/entanglement distribution capacity of the
pure loss channel with transmissivity $\eta_{\text{d}}$. Then, we find that
the maximum rate $K$ of secret key bits that can be distributed per
transmitted mode through the free-space channel must satisfy
\begin{equation}
K\leq\mathcal{U}(z):=\frac{2}{\ln2}\frac{a_{R}^{2}}{w_{z}^{2}}%
.\label{diffractivePLOBonly}%
\end{equation}
(See Appendix~\ref{APP_diffraction_limited} for an explicit proof). Let us
stress that, because entanglement bits (or ebits) are a specific type of
private bits, this inequality also provides an upper bound for the maximum
rate of ebits per mode $E\leq K$ that is achievable by protocols of
entanglement distribution. The diffraction-limited bound $\mathcal{U}(z)$ is
simple, depending only on the ratio between the receiver's aperture $a_{R}$
and the spot size of the beam at the receiver $w_{z}$. Furthermore, it is not
restricted to the far field ($z\gg z_{R}$).

We can check that $\mathcal{U}(z)$ is maximized by a focused beam ($R_{0}=z$),
so that $\mathcal{U}_{\text{foc}}(z)=2f_{0R}/\ln2$, where $f_{0R}:=[\pi
w_{0}a_{R}/(\lambda z)]^{2}$ is the Fresnel number product of the beam and the
receiver. However, this solution is typically restricted to short distances. A
more robust solution, suitable for any distance, is to employ a collimated
beam ($R_{0}=\infty$). In such a case, we write the bound
\begin{equation}
\mathcal{U}_{\text{coll}}(z)=\frac{2}{\ln2}\frac{a_{R}^{2}}{w_{0}^{2}%
[1+z^{2}/z_{R}^{2}]}.
\end{equation}
This formula is simple but may be too optimistic, not including other
important physical aspects of free-space communication. We progressively
include them below.

\subsection{Atmospheric extinction and setup efficiency}

Besides free-space geometric loss $\eta_{\text{d}}$ due to diffraction, there
are other inevitable effects to consider which include atmospheric extinction.
In fact, while a Gaussian beam is propagating through the atmosphere, it is
subject to both absorption and scattering. For a \textit{fixed} altitude $h$
above the ground/sea-level, the overall atmospheric transmissivity is modelled
by the Beer-Lambert extinction equation
\begin{equation}
\eta_{\text{atm}}(h,z)=\exp[-\alpha(h)z], \label{BLmain}%
\end{equation}
where $z$ is the path length in the atmosphere, and $\alpha(h)=N(h)\sigma$ is
the extinction factor~\cite[Ch.~11]{Huffman}. Here $N(h)$ is the mean number
of particles per unit volume at altitude $h$, and $\sigma=\sigma_{\text{abs}%
}+\sigma_{\text{sca}}$ is the total cross section associated with molecular
and aerosol absorption ($\sigma_{\text{abs}}$) and scattering ($\sigma
_{\text{sca}}$)~\cite[Ch. 2]{Hemani}. In general, both Rayleigh and Mie
scattering give contributions to $\sigma_{\text{sca}}$.

Assuming a standard model of atmosphere, one can write its mean density at
altitude $h$ as~\cite{Duntley}
\begin{equation}
N(h)=N_{0}\exp(-h/\tilde{h}),
\end{equation}
where $\tilde{h}=6600~$m and $N_{0}=2.55\times10^{25}~$m$^{-3}$ is the density
at sea level. As a result, we may similarly write
\begin{equation}
\alpha(h)=\alpha_{0}\exp(-h/\tilde{h}),
\end{equation}
where $\alpha_{0}\simeq5\times10^{-6}$ m$^{-1}$ is a good estimate of the
extinction factor at the sea-level for the optical wavelength $\lambda=800$~nm
(see also Ref.~\cite[Sec.~III.C]{Vasy19}).

Besides extinction, there is also a fixed constant contribution associated
with the local transmissivities of the setups. At the receiver, we may have
non-unit transmissivity $\eta_{\text{eff}}$, as a result of fiber couplings
and limited quantum efficiency of the detector. In a realistic implementation,
one may reach values of $\eta_{\text{eff}}\simeq0.5$~\cite{Jovanovic,BrussSAT}%
. At the transmitter, there may be an additional loss $\eta_{T}$ due to the
diffraction caused by the finite radius $a_{T}$ of its aperture. For the sake
of simplicity, in our treatment we assume that $a_{T}\geq2w_{0}$, so that we
can safely set $\eta_{T}\simeq1$ (see
Appendix~\ref{APP_diffraction_subsection}). Small deviations from this
assumption can be considered by explicitly re-inserting parameter $\eta_{T}$
into the model. In our study, we generally assume the worst-case scenario
where $\eta_{\text{eff}}$ may cause leaks to a potential eavesdropper
(suitable relaxations of this assumption into scenarios of trusted loss/noise
for the receiver are discussed afterwards).

Atmospheric extinction and setup efficiency cause several modifications to the
general diffraction-limited bounds discussed in Sec.~\ref{diff_bounds_SEC}
above. In fact, we need to consider the combined transmissivity $\eta
_{\text{d}}\eta_{\text{atm}}\eta_{\text{eff}}$, which leads to the revised
upper bound
\begin{align}
K  &  \leq-\log_{2}(1-\eta_{\text{d}}\eta_{\text{atm}}\eta_{\text{eff}%
})\label{farfieldBB}\\
&  =-\log_{2}\left[  1-\eta_{\text{eff}}\left(  1-e^{-2a_{R}^{2}/w_{z}^{2}%
}\right)  e^{-\alpha(h)z}\right] \\
&  \simeq\frac{2\eta_{\text{eff}}}{\ln2}\frac{a_{R}^{2}}{w_{z}^{2}}%
e^{-\alpha(h)z},
\end{align}
where the latter expansion is obtained in the far field, so that we can use
$\eta_{\text{d}}\simeq2a_{R}^{2}/w_{z}^{2}\ll1$ and the linear approximation
of the PLOB\ bound $\Phi(x)\simeq x/\ln2$.

It is important to remark that the combined transmissivity $\eta_{\text{d}%
}\eta_{\text{atm}}\eta_{\text{eff}}$ still misses an important aspect: the
process of channel fading induced by atmospheric turbulence and pointing
errors, a process that was pioneered in seminal works from the late 60s and
early 70s~\cite{Esposito,Fried73,Titterton}.

\subsection{Turbulence and pointing errors}

\subsubsection{Broadening and wandering of the beam}

Assuming weak turbulence, we can identify physical processes with different
time-scales~\cite{Fante75}. On a fast time-scale, we have the broadening of
the beam waist due to the interaction with smaller turbulent eddies; for this
reason, $w_{z}$ becomes a larger \textquotedblleft
short-term\textquotedblright\ spot size $w_{\text{st}}$. On a slow time-scale,
we have the deflection of the beam due to the interaction with the larger
eddies. This causes the\ random Gaussian wandering of the beam centroid with
variance $\sigma_{\text{TB}}^{2}$. Its dynamics is of the order of
$10-100~$ms~\cite{Burgoin}, which means that it can be resolved by a
sufficiently fast detector (e.g., with a realistic bandwidth of $100~$MHz).
Pointing error from jitter and imprecise tracking also causes centroid
wandering with a slow time-scale. For a typical $1~\mu$rad error at the
transmitter, it contributes with a variance $\sigma_{\text{P}}^{2}%
\simeq\left(  10^{-6}z\right)  ^{2}$, so that the centroid wanders with total
variance $\sigma^{2}=\sigma_{\text{TB}}^{2}+\sigma_{\text{P}}^{2}$. The
characterization of $w_{\text{st}}$ and $\sigma_{\text{TB}}^{2}$ needs
specific tools from turbulence theory that we introduce below.

For a beam with wave-number $k=2\pi/\lambda$ and propagation distance $z$, one
defines the spherical-wave coherence length~\cite[Eq.~(38)]{Fante75}
\begin{equation}
\rho_{0}=(0.548k^{2}C_{n}^{2}z)^{-3/5},
\end{equation}
where $C_{n}^{2}$ is the refraction index structure constant (measuring the
strength of the fluctuations in the refraction index caused by spatial
variations of temperature and pressure). Parameter $C_{n}^{2}$ is typically
described by the Hufnagel-Valley model of atmospheric
turbulence~\cite{Stanley,Valley} (see Appendix~\ref{TurbSECTION} for details).
For an horizontal path, the structure constant takes a fixed value which
depends on the specific altitude, besides the time of day and weather
conditions. In particular, its value is typically larger during the day,
meaning that the effects of turbulence are more pronounced for day-time
operation. For slightly-slant paths, it is a good approximation to average
$C_{n}^{2}$ over the various altitudes or, alternatively, to take its highest
value along the path, typically at the lowest altitude. (In our following
numerical investigations, we assume a horizontal path with $h=30$~m.)

Then, the regime of weak turbulence can be expressed by the condition
\begin{equation}
z\lesssim k\left[  \min\{2a_{R},\rho_{0}\}\right]  ^{2}. \label{YuraWRmain}%
\end{equation}
or, alternatively, it can be more stringently expressed in terms of the Rytov
parameter as
\begin{equation}
\sigma_{\text{Rytov}}^{2}=1.23C_{n}^{2} k^{7/6}z^{11/6}<1.
\end{equation}
For weak turbulence and setting $\phi:=0.33(\rho_{0}/w_{0})^{1/3}$, we may
write the analytical approximations~\cite{Yura73}
\begin{equation}
w_{\text{st}}^{2}\simeq w_{z}^{2}+2\left(  \frac{\lambda z}{\pi\rho_{0}%
}\right)  ^{2}(1-\phi)^{2},~\sigma_{\text{TB}}^{2}\simeq\frac{0.1337\lambda
^{2}z^{2}}{w_{0}^{1/3}\rho_{0}^{5/3}}. \label{Yurass}%
\end{equation}

These analytical expressions are rigorous for $\phi\ll1$ and represent very
good approximations for $\rho_{0}/w_{0}<1$. For $\rho_{0}/w_{0} \gtrsim1$,
they need to be replaced by numerical estimates (see
Appendix~\ref{TurbSECTION} for details). For $\rho_{0}/w_{0}\gg1$,
$\sigma_{\text{TB}}^{2}$ is negligible and $w_{\text{st}}^{2}$ is equal to the
long-term spot size $w_{\text{lt}}^{2}=w_{z}^{2}+2\left[  \lambda z/(\pi
\rho_{0})\right]  ^{2}$~\cite{Fante75}. Let us also note that, in the limit of
negligible turbulence $C_{n}^{2} \rightarrow0$, we have $\rho_{0}
\rightarrow\infty$. In such a case, Yura's analytical expansions  are just
replaced by $\sigma_{\text{TB}}\simeq0$ and $w_{\text{lt}} \simeq
w_{\text{st}} \simeq w_{z}$ (which all come  from the collapse of the
long-term spot-size $w_{\text{lt}}^{2}=w_{\text{st}}^{2}+ \sigma_{\text{TB}%
}^{2}$ into its diffraction component $w_{z}^{2}$).

\subsubsection{Incorporating short-term effects and deflection}

The first mathematical modification induced by turbulence is that the
diffraction-limited transmissivity $\eta_{\text{d}}$ needs to be replaced by a
more general expression $\eta_{\text{st}}$ in terms of the short-term waist
$w_{\text{st}}$, i.e.,
\begin{equation}
\eta_{\text{st}}=1-e^{-2a_{R}^{2}/w_{\text{st}}^{2}}\simeq\frac{2a_{R}^{2}%
}{w_{\text{st}}^{2}}:=\eta_{\text{st}}^{\text{far}}, \label{etaSTmain}%
\end{equation}
where the expansion is valid in the far field ($z\gg z_{R}$). The new loss
parameter
\begin{equation}
\eta:=\eta_{\text{st}}\eta_{\text{atm}}\eta_{\text{eff}}%
\end{equation}
represents the maximum value of the link-transmissivity when the beam centroid
$\vec{x}_{C}$\ is perfectly aligned with the center $\vec{x}_{R}$ of the
receiver's aperture.

Because the beam centroid wanders following a Gaussian probability with
variance $\sigma^{2}$, the actual instantaneous value of the transmissivity
varies over time and can only be $\leq\eta$. This leads to the second
modification associated with the fading process: the maximum transmissivity
$\eta$ needs to be replaced by a distribution $P_{0}(\tau)$ of instantaneous
transmissivities $\tau\leq\eta$. Here we first connect the instantaneous
transmissivity $\tau$ to the deflection value $r:=\left\Vert \vec{x}_{C}%
-\vec{x}_{R}\right\Vert \geq0$; we will then super-impose the random walk in
$r$ to describe the fading process affecting $\tau$ (discussed in the next subsection).

\begin{figure}[t]
\vspace{-1.4cm}
\par
\begin{center}
\includegraphics[width=1\columnwidth] {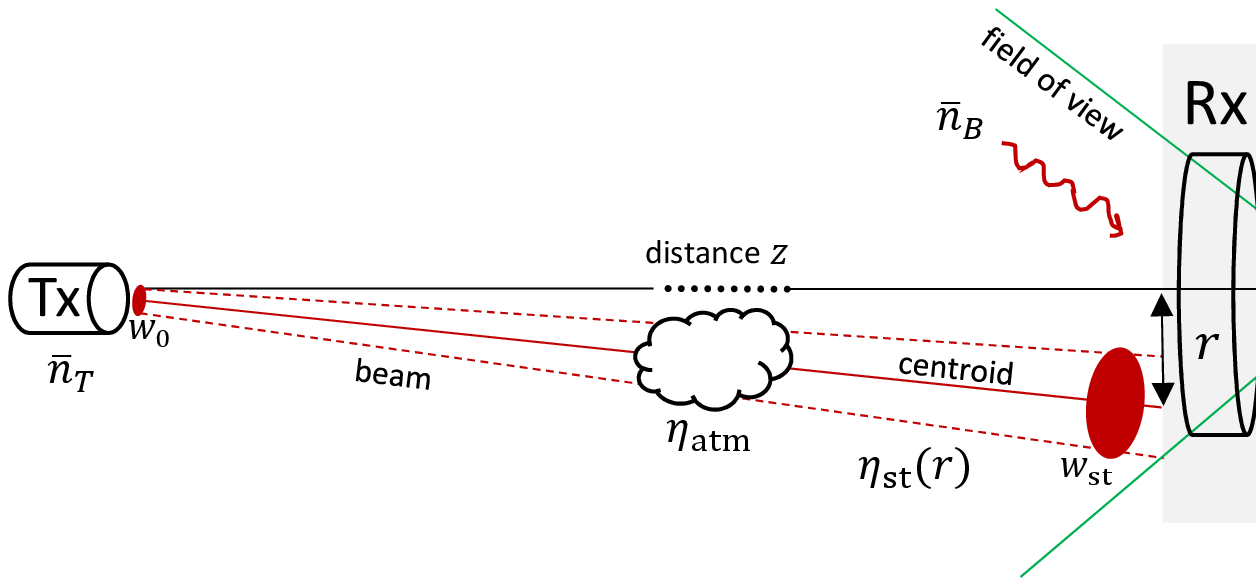}
\end{center}
\par
\vspace{-2.1cm}\caption{Free-space communication from a transmitter (Tx) to a
receiver (Rx) separated by distance $z$. The transmitter generates a Gaussian
beam with spot-size $w_{0}$ and mean number of photons $\bar{n}_{T}$. The
propagation of the beam is affected by diffraction, atmospheric extinction
$\eta_{\text{atm}}$, and turbulence/pointing errors, so that its short-term
spot-size $w_{\text{st}}$ is randomly deflected by $r$ from the aperture
center of the receiver, with an associated transmissivity $\eta_{\text{st}%
}(r)$. The beam is also affected by an additional attenuation, given by the
efficiency $\eta_{\text{eff}}$ of the receiver. In total, transmitter and
receiver are connected by an instantaneous lossy channel with transmissivity
$\tau(r)=\eta_{\text{st}}(r)\eta_{\text{atm}}\eta_{\text{eff}}$ as in
Eq.~(\ref{etaFIRST}). Besides loss, we also consider noise. In particular,
thermal noise $\bar{n}_{B}$ is collected by the field of view of the Rx and
further noise $\bar{n}_{\text{ex}}$ may be locally generated by setup
imperfections. As a result, the detector is hit by $\bar{n}_{R}=\tau(r)\bar
{n}_{T}+\bar{n}$ mean photons whose $\bar{n}=\eta_{\text{eff}}\bar{n}_{B}%
+\bar{n}_{\text{ex}}$ are due to thermal noise.}%
\label{detectorPIC}%
\end{figure}

As also depicted in Fig.~\ref{detectorPIC}, for each value of the deflection
$r$, there is an associated transmissivity
\begin{equation}
\tau(r)=\eta_{\text{st}}(r)\eta_{\text{atm}}\eta_{\text{eff}},
\label{etaFIRST}%
\end{equation}
where $\eta_{\text{st}}(r)$ accounts for the misalignment and reads
\begin{equation}
\eta_{\text{st}}(r)=e^{-\frac{4r^{2}}{w_{\text{st}}^{2}}}Q_{0}\left(
\frac{2r^{2}}{w_{\text{st}}^{2}},\frac{4ra_{R}}{w_{\text{st}}^{2}}\right)  .
\label{etadiff}%
\end{equation}
In the expression above, the factor $Q_{0}(x,y)$ is an incomplete Weber
integral~\cite{Agrest}%
\begin{equation}
Q_{0}(x,y):=(2x)^{-1}e^{x}\int_{0}^{y}dt~te^{-t^{2}/4x}I_{0}(t),
\end{equation}
where the notation $I_{n}$ denotes a modified Bessel function of the first
kind with order $n$. Note that Eq.~(\ref{etadiff}) is obtained by adapting a
previous result~\cite[Eq.~(D2)]{Vasy12}.

Following Ref.~\cite{Vasy12}, we have that $\eta_{\text{st}}(r)$ can be
well-approximated by the analytical expression
\begin{equation}
\eta_{\text{st}}(r) \simeq \eta_{\text{st}}\exp\left[  -\left(  \frac{r}{r_{0}%
}\right)  ^{\gamma}\right]  , \label{etaSTanalytical}%
\end{equation}
where $\gamma$ and $r_{0}$\ are shape and scale (positive) parameters, given
by the following functionals
\begin{align}
\gamma &  =\frac{4\eta_{\text{st}}^{\text{far}}\Lambda_{1}(\eta_{\text{st}%
}^{\text{far}})}{1-\Lambda_{0}(\eta_{\text{st}}^{\text{far}})}\left[  \ln
\frac{2\eta_{\text{st}}}{1-\Lambda_{0}(\eta_{\text{st}}^{\text{far}})}\right]
^{-1},\label{expGG1}\\
r_{0}  &  =a_{R}\left[  \ln\frac{2\eta_{\text{st}}}{1-\Lambda_{0}%
(\eta_{\text{st}}^{\text{far}})}\right]  ^{-\frac{1}{\gamma}}, \label{expGG2}%
\end{align}
with $\Lambda_{n}(x):=\exp\left(  -2x\right)  I_{n}\left(  2x\right)  $. As a
result, combining Eqs.~(\ref{etaFIRST}) and~(\ref{etaSTanalytical}), we may
write%
\begin{equation}
\tau(r)=\eta\exp\left[  -\left(  \frac{r}{r_{0}}\right)  ^{\gamma}\right]  .
\label{Eqnst}%
\end{equation}

\subsubsection{Incorporating beam wandering}

Beam wandering is modelled by treating the position of the centroid as a
stochastic variable, which can be taken to be Gaussian~\cite{Dowling} with
variance $\sigma^{2}$ around the center of the receiver's aperture, where
$\sigma^{2}$ is the sum of two independent contributions: the variance
$\sigma_{\text{TB}}^{2}$ due to large-scale turbulence, and the variance
$\sigma_{\text{P}}^{2}$ due to pointing error. In general, one may also assume
that the wandering is around an average deflection point at a non-zero
distance $d$ from the center of the receiver's aperture. For the sake of
simplicity, here we consider the optimal working condition of $d=0$, which can
always be realized by means of sufficiently-fast adaptive optics.

The Gaussian random walk around the receiver's center induces a Weibull
distribution for the deflection $r$, expressed by the zero-mean density
function
\begin{equation}
P_{\text{WB}}(r)=\frac{r}{\sigma^{2}}\exp\left(  -\frac{r^{2}}{2\sigma^{2}%
}\right)  . \label{WeibullDISTRIBUTION}%
\end{equation}
In turn, the Weibull distribution over $r$ induces a corresponding probability
density for $\tau=\tau(r)$, given by%
\begin{equation}
P_{0}(\tau)=\frac{r_{0}^{2}}{\gamma\sigma^{2}\tau}\left(  \ln\frac{\eta}{\tau
}\right)  ^{\frac{2}{\gamma}-1}\exp\left[  -\frac{r_{0}^{2}}{2\sigma^{2}%
}\left(  \ln\frac{\eta}{\tau}\right)  ^{\frac{2}{\gamma}}\right]  ,
\label{P0tau}%
\end{equation}
as also discussed in Appendix~\ref{APP_centroid}.

The random fluctuation of the effective transmissivity $\tau$ creates a fading
channel from transmitter to receiver that can be described by the ensemble
$\mathcal{E}:=\{P_{0}(\tau),\mathcal{E}_{\tau}\}$, where the lossy channel
$\mathcal{E}_{\tau}$ with transmissivity $\tau$\ is randomly selected with
probability density $P_{0}(\tau)$. Using the convexity properties of the
relative entropy of entanglement (REE)~\cite{REE1,REE2,REE3} over an ensemble of
channels as in Ref.~\cite[Eq.~(17)]{QKDpaper}, we can bound the secret key
capacity of the fading channel $\mathcal{E}$ by means of the following average%
\begin{equation}
K\leq\int_{0}^{\eta}d\tau~P_{0}(\tau)\Phi(\tau):=\mathcal{B}(\eta,\sigma),
\label{convexityP2old}%
\end{equation}
where $\Phi(\tau)=-\log_{2}(1-\tau)$ is the PLOB bound associated with the
instantaneous channel $\mathcal{E}_{\tau}$.

The integral in Eq.~(\ref{convexityP2old}) can be simplified by working with
the variable $\ln(\eta/\tau)$ and then solving by parts. In this way, we find
that the maximum secret key rate achievable through the free-space channel is
bounded by%
\begin{equation}
K\leq\mathcal{B}(\eta,\sigma)=-\Delta(\eta,\sigma)\log_{2}(1-\eta),
\label{PLOBdelta}%
\end{equation}
where the correction factor $\Delta$\ is given by%
\begin{equation}
\Delta(\eta,\sigma)=1+\frac{\eta}{\ln(1-\eta)}\int_{0}^{+\infty}dx\frac
{\exp\left(  -\frac{r_{0}^{2}}{2\sigma^{2}}x^{2/\gamma}\right)  }{e^{x}-\eta}.
\label{deltaEXPRESSION}%
\end{equation}
The formula in Eq.~(\ref{PLOBdelta}) is our main result: It bounds the secret
key capacity $K$ and the entanglement-distribution capacity $E$ of a
free-space lossy channel $\mathcal{E}$ affected by diffraction, extinction,
setup-loss, and fading, the latter being induced by turbulence and pointing errors.

We can further simplify the upper bound $\mathcal{B}(\eta,\sigma)$ for high
loss $\eta\ll1$. In fact, in such a case, we can reduce the $\Delta
$-correction and write the approximate bound
\begin{align}
\mathcal{B}(\eta,\sigma)  &  \simeq\frac{\eta\Lambda(\eta,\sigma)}{\ln
2},\label{plobWAN}\\
\Lambda(\eta,\sigma)  &  :=1-\int_{0}^{+\infty}dx~\exp\left(  -\frac{r_{0}%
^{2}}{2\sigma^{2}}x^{2/\gamma}-x\right)
\end{align}
Note that the condition $\eta\ll1$ is not necessarily achieved in the far
field, because $\eta=\eta_{\text{st}}\eta_{\text{atm}}\eta_{\text{eff}}$ and
the factors $\eta_{\text{atm}}\eta_{\text{eff}}$ may decrease the overall
value of the transmissivity already in the near field. In the far field ($z\gg
z_{R}$), we may use both $\eta\ll1$ and the expansion $\eta_{\text{st}}%
\simeq2a_{R}^{2}w_{\text{st}}^{-2}$, so that we can write%
\begin{equation}
\mathcal{B}(\eta,\sigma)\simeq\frac{\eta_{\text{atm}}\eta_{\text{eff}}}{\ln
2}\frac{2a_{R}^{2}}{w_{\text{st}}^{2}}\Lambda(\eta,\sigma).
\end{equation}

In our model above, the free-space channel $\mathcal{E}$ is an ensemble
$\{P_{0}(\tau),\mathcal{E}_{\tau}\}$ of instantaneous pure-loss channels
$\mathcal{E}_{\tau}$ with probability $P_{0}(\tau)$. For all these channels
the upper bound $\Phi(\tau)$ is achievable by their (bosonic) reverse coherent
information~\cite{RCI,CInfo}, which corresponds to the optimal rate of
entanglement distribution protocols assisted by one-way classical
communication (see Appendix~\ref{App_achievability} for details). Averaging
over $P_{0}(\tau)$ implies that the upper bound in Eq.~(\ref{PLOBdelta}) is
achievable by these entanglement distribution protocols and, therefore, we may
write $E=K=-\Delta\log_{2}(1-\eta)$, where $E\leq K$ is the entanglement
distribution capacity of the link.

In conclusion, as long as we can neglect thermal noise and consider a
pure-loss fading process, the bound in Eq.~(\ref{PLOBdelta}) represents both
the secret-key and entanglement distribution capacity of the free-space link.
In particular, note that the formulas in Eqs.~(\ref{PLOBdelta})
and~(\ref{plobWAN}) have a clear structure. They are given by the capacity
$-\log_{2}(1-\eta)\simeq\eta/\ln2$ achievable with a perfectly-aligned link
with no wandering, multiplied by a free-space correction factor which accounts
for the wandering effects ($\Delta\simeq\Lambda$).

One can check that, with the assumptions of negligible turbulence and pointing
error (so that $\sigma\simeq0$ and $\eta_{\text{st}} \simeq\eta_{\text{d}}$),
we have $\Delta\simeq1$ in Eq.~(\ref{deltaEXPRESSION}), and
Eq.~(\ref{PLOBdelta}) reduces to Eq.~(\ref{farfieldBB}). If we further assume
no atmospheric extinction and unit setup efficiency, Eq.~(\ref{PLOBdelta})
reduces to Eq.~(\ref{diffractivePLOBonly}) which only accounts for free-space diffraction.

\subsection{Thermal noise\label{SEC_thermal_noise}}

The quantity $-\Delta\log_{2}(1-\eta)$ in Eq.~(\ref{PLOBdelta}) provides an
upper bound even in the presence of thermal noise. The reason is because any
instantaneous thermal-loss channel $\mathcal{E}_{\tau,\bar{n}}$ adding a mean
number of photons $\bar{n}$ can be written as a decomposition of a pure-loss
channel $\mathcal{E}_{\tau}$ followed by a suitable additive-Gaussian noise
channel~\cite{RMP}. Because the PLOB bound $\Phi$ is based on the REE, it is monotonic over such decompositions, so that its
value $\Phi(\tau,\bar{n})$ computed over $\mathcal{E}_{\tau,\bar{n}}$\ cannot
exceed its value $\Phi(\tau)$ over $\mathcal{E}_{\tau}$. Thus, the loss-based
upper bound in Eq.~(\ref{PLOBdelta}) is still valid in the presence of thermal
noise (no matter if this noise is trusted or untrusted). However, it is no
longer guaranteed to be achievable. For this reason, we derive a tighter upper
bound and a corresponding lower bound (technical details about the following
derivations are in Appendix~\ref{APP_ThermalBounds}).

Assume that the receiver collects a non-trivial amount of thermal noise which
couples into the output mode. The natural source is the brightness of the sky
$B_{\lambda}^{\text{sky}}$ which varies between $\simeq1.5\times10^{-6}$ and
$\simeq1.5\times10^{-1}$~W m$^{-2}$ nm$^{-1}$ sr$^{-1}$, from clear night to
cloudy day-time~\cite{Miao} (and assuming that the field of view does not
include the Moon or the Sun). For a receiver with aperture $a_{R}$, angular
field of view $\Omega_{\text{fov}}$, and using a detector with time window
$\Delta t$ and spectral filter $\Delta\lambda$ around $\lambda$, the number of
background thermal photons per mode is given by~\cite{Miao,BrussSAT}%
\begin{equation}
\bar{n}_{B}=\frac{\pi\lambda\Gamma_{R}}{hc}B_{\lambda}^{\text{sky}}%
,~\Gamma_{R}:=\Delta\lambda\Delta t\Omega_{\text{fov}}a_{R}^{2}, \label{downT}%
\end{equation}
where $h$ is Planck's constant and $c$ is the speed of light.

As an example, for a $100~$MHz\ detector ($\Delta t=10~$ns) with a filter
$\Delta\lambda=1~$nm around $\lambda=800~$nm, and a telescope with $a_{R}%
=5~$cm and $\Omega_{\text{fov}}=10^{-10}~$sr, the value of $\bar{n}_{B}$
ranges between $\simeq4.75\times10^{-8}$ photons/mode (at night)\ and
$\simeq4.75\times10^{-3}$ photons/mode (during a cloudy day). A fraction
$\eta_{\text{eff}}\bar{n}_{B}$ of these photons is detected by a receiver with
limited efficiency $\eta_{\text{eff}}$. See Fig.~\ref{detectorPIC}.

It is important to note that the number of photons in the natural background
$\bar{n}_{B}$\ may be higher than that expected from Eq.~(\ref{downT}), as a
consequence of the presence of bright sources of light within the field of
view of the receiving telescope. Our formalism accounts for such deviations,
even though we consider Eq.~(\ref{downT}) in our numerical simulations. In
general, all the (detected) photons coming from the outside channel must be
ascribed to Eve in the worst-case scenario, even though this is a
over-pessimistic assumption due to the line-of-sight configuration in
free-space communication. However, such an assumption must be made because Eve
might inject and hide her photons in the background.

Besides the natural background, excess photons $\bar{n}_{\text{ex}}$\ may be
created by imperfections in the receiver setup (e.g., due to electronic noise
and other errors), so that the receiver sees a total of $\bar{n}%
=\eta_{\text{eff}}\bar{n}_{B}+\bar{n}_{\text{ex}}$ thermal photons. Thus,
assuming that $\bar{n}_{T}$ mean photons are generated at the transmitter and
$\tau$ is the overall instantaneous transmissivity of the channel, the
receiver's detector gets $\bar{n}_{R}=\tau\bar{n}_{T}+\bar{n}$ mean photons
(per mode). See Fig.~\ref{detectorPIC}.

The free-space process in Fig.~\ref{detectorPIC} can be described by an
overall thermal-loss channel $\mathcal{E}_{\tau,\bar{n}}$ with instantaneous
transmissivity $\tau$ and output thermal noise $\bar{n}$. This channel is
equivalent to a beam-splitter mixing the signal mode with an input thermal
mode with $\bar{n}_{e}:=\bar{n}(1-\tau)^{-1}$ mean photons. In the worst-case
scenario, Eve controls all the input noise and collects all the photons that
are leaked from the other output of the beam-splitter (which means that she
collects photons leaking from both the channel and the receiver setup).

In order to account for the centroid wandering, we adopt the distribution
$P_{0}(\tau)$ for the transmissivity $\tau$\ while keeping the output thermal
noise $\bar{n}$ as a constant. The latter is in fact composed of a fraction
$\bar{n}_{B}$ which is independent from the fading process, while the other
contribution $\bar{n}_{\text{ex}}$ can always be assumed to be optimized over
such a process (see discussion in Appendix~\ref{APP_subsection_Thermal} for
more details). For this reason, the free-space fading channel can be
represented by the ensemble $\mathcal{E}=\{P_{0}(\tau),\mathcal{E}_{\tau
,\bar{n}}\}$.

For a free-space fading channel $\mathcal{E}$ with maximum transmissivity
$\eta$\ and thermal noise $\bar{n}\leq\eta$, we compute the following tighter
upper bound for the secret key capacity%
\begin{equation}
K\leq-\Delta(\eta,\sigma)\log_{2}(1-\eta)-\mathcal{T}(\bar{n},\eta,\sigma),
\label{UBTTTT}%
\end{equation}
where the thermal correction $\mathcal{T}$ is given by\
\begin{align}
\mathcal{T}(\bar{n},\eta,\sigma)  &  =\left\{  1-e^{-\frac{r_{0}^{2}}%
{2\sigma^{2}}\left[  \ln\left(  \eta/\bar{n}\right)  \right]  ^{2/\gamma}%
}\right\}  \left[  \frac{\bar{n}\log_{2}\bar{n}}{1-\bar{n}}+h\left(  \bar
{n}\right)  \right] \nonumber\\
&  -\Delta(\bar{n},\sigma)\log_{2}(1-\bar{n}), \label{ThermalCORRECTION}%
\end{align}
and we have used the entropic function
\begin{equation}
h\left(  x\right)  :=(x+1)\log_{2}(x+1)-x\log_{2}x. \label{hFUNCTIONmain}%
\end{equation}
We also compute the following achievable rate (lower bound) for entanglement
distribution and, therefore, secret key generation%
\begin{equation}
E\geq-\Delta(\eta,\sigma)\log_{2}(1-\eta)-h\left(  \frac{\bar{n}}{1-\eta
}\right)  . \label{LBTTTT}%
\end{equation}

For negligible noise $\bar{n}$, the bounds in Eqs.~(\ref{UBTTTT})
and~(\ref{LBTTTT}) collapse to the bound in Eq.~(\ref{PLOBdelta}). By
contrast, for strong noise $\bar{n}=\eta$, the thermal correction in
Eq.~(\ref{ThermalCORRECTION}) becomes predominant and we get $K\leq0$\ from
Eq.~(\ref{UBTTTT}). The threshold condition $\bar{n}=\eta$ implies the
existence of a maximum security distance $z_{\text{max}}$ for free-space QKD
in the presence of thermal noise. A simple bound on this maximum distance is
achieved imposing $\bar{n}=\eta_{\text{d}}$. In fact, for a collimated beam,
this leads to
\begin{equation}
2f_{0R}(z_{\max})\geq-\ln(1-\bar{n}),
\end{equation}
where $f_{0R}$ is the Fresnel number product of the beam and the receiver (see
Sec.~\ref{diff_bounds_SEC}).

\begin{figure*}[t]
\vspace{-0cm}
\par
\begin{center}
\includegraphics[width=1\textwidth] {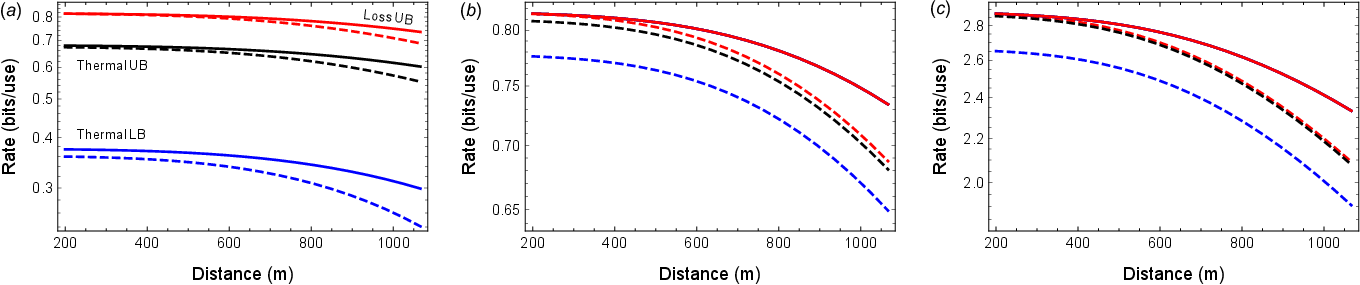}
\end{center}
\par
\vspace{-0.5cm}\caption{Performance of free-space quantum communications in
terms of bits per channel use versus distance. (a)~We consider the general
worst-case scenario with untrusted loss and noise at the receiver
($\eta_{\text{eff}}=0.5$, $\bar{n}_{\text{ex}}=0.05$). We plot the ultimate
loss-based upper bound of Eq.~(\ref{PLOBdelta}) for night time (top red line)
and day time (red dashed line). This is compared with the bounds explicitly
accounting for thermal noise $\bar{n}=\eta_{\mathrm{eff}} \bar{n}_{B}+\bar
{n}_{\mathrm{ex}}$. In particular, we plot the thermal upper bound of
Eq.~(\ref{UBTTTT}) for night time (black solid line) and day time (black
dashed line), as well as the thermal lower bound of Eq.~(\ref{LBTTTT}) for
night time (blue solid line) and day time (blue dashed line). (b)~Same
comparison as in (a) but considering a noise-less receiver ($\eta_{\text{eff}%
}=0.5$, $\bar{n}_{\text{ex}}=0$). For night time, the upper- and lower-
thermal bounds coincide with loss-based upper bound (solid red line). For day
time, the performances are instead separate (dashed lines). (c)~Same
comparison as in (a) but considering an ideal loss-less and noise-less
receiver ($\eta_{\text{eff}}=1$, $\bar{n}_{\text{ex}}=0$). As in (b), the two
thermal bounds collapse in the loss-based upper bound during night time (solid
red line). Performances are different during day time (dashed lines). Other
parameters are: $R_{0}=\infty$ (collimated Gaussian beam), $\lambda=800~$nm,
$w_{0}=a_{R}=5~$cm, $\Omega_{\text{fov}}=10^{-10}~$sr, $\Delta t=10~$ns and
$\Delta\lambda=1~$nm. We consider $h=30~$m, so that $C_{n}^{2}\simeq
1.28(2.06)\times10^{-14}~$m$^{-2/3}$ for night (day),\ and we have $\bar
{n}_{B}\simeq4.75\times10^{-8}$ ($\times10^{-3}$) at night (cloudy day).}%
\label{FSpics}
\end{figure*}

\subsection{Analysis of the ultimate bounds\label{AnalysisSEC}}

In order to study our bounds, we consider different possibilities which depend
on the treatment of loss and noise present in the setup of the receiver. In
the worst-case scenario assumed so far, we explicitly account for the
non-ideal values of the receiver parameters $\eta_{\text{eff}}$ and $\bar
{n}_{\text{ex}}$, assuming that Eve may access that leakage and control that
noise. This setting can be used to bound the performance of all protocols
where both leakage and local noise in the receiving setup are considered to be
untrusted. We may then consider the case where the local noise $\bar
{n}_{\text{ex}}$ is set to zero, i.e., a noiseless-receiver. This setting can
be used to bound all protocols where such local noise is considered to be
trusted (trusted-noise scenario). Finally, we may also consider the optimal
case of $\bar{n}_{\text{ex}}=0$ and $\eta_{\text{eff}}=1$, i.e., an ideal
loss-less and noise-less receiver. This can be used to bound all those
protocols where local noise and limited efficiency of the receiver are both
considered to be trusted (trusted-loss-and-noise scenario).

Numerical behavior of the bounds is shown in Fig.~\ref{FSpics}. For the chosen
parameters, the condition of weak turbulence $\sigma_{\text{Rytov}}^{2}<1$
limits day-time distance to a range of $z\lesssim1~$km. As we can see from
Fig.~\ref{FSpics}(a), there is a clear gap between the ultimate loss-based
upper bound of Eq.~(\ref{PLOBdelta}) and the two thermal bounds in
Eqs.~(\ref{UBTTTT}) and~(\ref{LBTTTT}). This is created by the presence of
thermal noise $\bar{n}$. During the night, when the background contribution
$\bar{n}_{B}$ is negligible, it is the presence of untrusted setup noise
$\bar{n}_{\text{ex}}$ to create the gap in the performances [see solid lines
in Fig.~\ref{FSpics}(a)]. During the day, there is a higher turbulence on the
ground as quantified by the higher value of the structure constant $C_{n}^{2}%
$; mainly for this reason, we have a degradation of all the day-time rates
with respect to their night-time counterparts [compare dashed with solid lines
in Fig.~\ref{FSpics}(a)]. For the thermal bounds this degradation is slightly
increased due to the additional contribution of the thermal background
$\bar{n}_{B}$, which is non-negligible during the day.

In the case of a noise-less receiver as in Fig.~\ref{FSpics}(b), thermal noise
is only coming from the external background $\bar{n}_{B}$. For night-time
operation, this background is negligible and the two thermal bounds in
Eqs.~(\ref{UBTTTT}) and~(\ref{LBTTTT}) collapse into the loss-bound of
Eq.~(\ref{PLOBdelta}), which therefore represents the secret key capacity (and
entanglement distribution capacity) of the night-time link [see red solid line
in Fig.~\ref{FSpics}(b)]. However, during the day, the external background
$\bar{n}_{B}$ is not negligible and this creates a small gap in the
performance, so that there is no collapse of the thermal bounds [black and
blue dashed lines in Fig.~\ref{FSpics}(b)] into the upper loss-based bound
[red dashed line in Fig.~\ref{FSpics}(b)]. In the case of an ideal (loss- and
noise-less) receiver, we have basically the same situation but with higher
rates, as shown in Fig.~\ref{FSpics}(c).

\begin{figure}[h]
\vspace{0.15cm}
\par
\begin{center}
\includegraphics[width=0.35\textwidth] {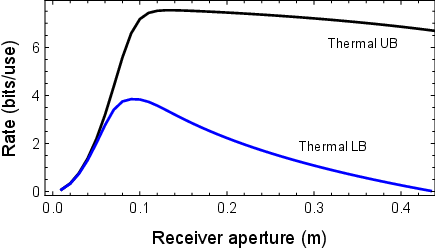}
\end{center}
\par
\vspace{-0.5cm}\caption{For day time and fixed distance $z=1~$km, we plot the
thermal bounds in Eqs.~(\ref{UBTTTT}) and~(\ref{LBTTTT}) as a function of the
receiver's aperture $a_{R}$. We assume an ideal receiver ($\bar{n}_{\text{ex}%
}=0$ and $\eta_{\text{eff}}=1$). Other parameters are as in Fig.~\ref{FSpics}%
.}%
\label{AperturePIC}%
\end{figure}

An interesting observation for day-time operation is the trade-off between
Eq.~(\ref{etaSTmain}), where $a_{R}$ increases the transmissivity, and
Eq.~(\ref{downT}), where $a_{R}$ increases thermal noise. For this reason, the
optimal performance is achieved when the receiver's aperture $a_{R}$ takes an
intermediate value. For instance consider the case of an ideal receiver, and
let us study the behavior of the two thermal bounds in Eqs.~(\ref{UBTTTT})
and~(\ref{LBTTTT}) as a function of $a_{R}$ at some fixed distance, say
$z=1$~km. As we can see from Fig.~\ref{AperturePIC} we find an optimal working
point at around $a_{R}\simeq10~$cm for the specific regime considered. This is
true as long as the other parameters of the receiver are fixed, such as its
field of view $\Omega_{\text{fov}}$ which intervenes in Eq.~(\ref{downT}).
Note that the field of view does not directly depend on $a_{R}$, but decreases
with the focal length of the receiver's telescope $f$ and increases with the
area of the detector $a$. For instance, for a rectilinear optical system
focused at $\infty$, it is easy to check that the angle of view satisfies
$\Omega_{\text{fov}}^{1/2}$ $\simeq2\arctan[\sqrt{a}/(2f)]$, which is also a
good approximation for a spherical optical system.

\subsubsection{Noise filtering\label{NoiseFilterSubsection}}

It is important to note that the behavior of the thermal bounds is strongly
dependent on the filter $\Delta\lambda$. So far, numerical investigations have
assumed a value of $\Delta\lambda=1~$nm, which is the value of the narrow-band
filter typically considered in studies with discrete variables. At $800~$nm,
the value $\Delta\lambda=1~$nm corresponds to a relatively-large bandwidth of
$\Delta\nu=c\lambda^{-2}\Delta\lambda\simeq470~$GHz. However, in the setting
of continuous variables, much narrower filters are possible by exploiting
suitable interferometric procedures at the receiver, so that the effective
value of $\Delta\nu$ becomes equivalent to the bandwidth of the transmitted pulses.

An important ingredients in experiments with CV systems is the local
oscillator (LO). They are typically performed with a transmitted LO (TLO),
where each quantum signal is multiplexed in polarization with an associated LO
and both are sent to the receiver. At the receiver, signal and LO are
demultiplexed via a polarizing beam splitter and made interfered on a beam
splitter before detection (in a homodyne or heterodyne setup). Alternatively,
CV experiments may be performed with a local local oscillator (LLO), where
quantum signals are interleaved with strong reference pulses, the latter being
used by the receiver to reconstruct the local oscillator \textquotedblleft
locally\textquotedblright\ (with some imperfection~\cite{LLO,LLO2}).

It is important to note that, in a homodyne measurement, the output of the
detector is proportional to $\sqrt{\bar{n}_{\text{LO}}}\hat{x}$, where
$\hat{x}$ is the generic quadrature of the signal and $\bar{n}_{\text{LO}}$ is
the number of photons from the LO. The value of $\bar{n}_{\text{LO}}$ can be
very high. In fact, considering $10~$ns-long pulses from a $100$~mW laser at
$\lambda=800~$nm, we have that each pulse contains $\bar{n}_{\text{LO}}%
\simeq4\times10^{9}$ photons. Even if we pessimistically assume $20$dB of loss
($\tau\simeq10^{-2}$), we see that about $\mathcal{O}(10^{7})$ photons reach
the receiver.

Thanks to the large pre-factor $\sqrt{\bar{n}_{\text{LO}}}$, only the
contribution of thermal noise mode-matching with the LO will survive in the
output. This means that the interferometric process introduces an effective
filter which is given by the bandwidth $\Delta\nu$ of the LO. Compatibly with
the time-bandwidth product $\Delta t\Delta\nu\geq0.44$ (for Gaussian pulses),
one can make $\Delta\nu$ very small. As an example, for a $10~$ns pulse, we
may consider $\ \Delta\nu=50~$MHz corresponding to just $\Delta\lambda=0.1~$pm
around $800$~nm; this filter is $4$ orders of magnitude narrower than the one
considered above. With respect to $\Delta\lambda=1~$nm, such a narrow filter
realizes a corresponding $10^{-4}$ suppression of the background noise
$\bar{n}_{B}$, which therefore becomes negligible (day-time noise becomes
$\bar{n}_{B}\simeq10^{-7}$). As a result, the detector would only experience
locally-generated noise, i.e., $\bar{n}\simeq\bar{n}_{\text{ex}}$.

From the point of view of the rates, with a narrow filter $\Delta\lambda
=0.1~$pm, we have an increase of the day-time thermal bounds in
Fig.~\ref{FSpics}. In particular, for a noise-less setup ($\bar{n}_{\text{ex}%
}=0$) we have $\bar{n}\simeq0$. In this case, the day-time thermal bounds
computed from Eqs.~(\ref{UBTTTT}) and~(\ref{LBTTTT}) collapse into the
day-time loss-bound given by Eq.~(\ref{PLOBdelta}), which therefore becomes
the secret-key capacity (and entanglement distribution capacity) of the
day-time link. This means that the black and blue dashed lines in
Fig.~\ref{FSpics}(b) collapse into the upper red dashed line. The same happens
in Fig.~\ref{FSpics}(c) which refers to a loss-less and noise-less setup, but
with higher rates.

It is worth stressing that, if we optimize over the receiver so to make the
total thermal noise $\bar{n}$ negligible (as a result of a noise-less setup
$\bar{n}_{\text{ex}}\simeq0$ and noise-filtering $\bar{n}_{B}\simeq0$), then
the loss-bound of Eq.~(\ref{PLOBdelta}) is achievable no matter what the
external conditions are (night- or day-time). It is also clear that this bound
can be further optimized by assuming no pointing error at the transmitter and
unit quantum efficiency at the receiver. The result of these optimizations
(implicit in our formula) provides a bound/capacity which uniquely depends on
the external free-space channel between the two remote parties (affected by
diffraction, extinction and turbulence).


\subsection{Extension of the bounds}

\subsubsection{Slow detection}

So far, we have considered the situation where the detector of the receiver is
fast enough to resolve the wandering of the centroid. In general, this
dynamics has two components: on the one hand, there are the fluctuations
induced by atmospheric turbulence, with a time scale of the order of
10-100~ms; on the other hand, there is pointing error (from jitter and
imprecise tracking) that fluctuates over a slightly slower time scale, of the
order of 0.1-1~s. For detection, we can therefore identify three different
regimes: (i) fast detectors able to resolve all the dynamics above; (ii)
intermediate detectors, able to solve part of the dynamics, i.e.,
pointing-error wandering but not turbulence-induced fluctuations; and (iii)
slow detectors, not able to resolve any of the wandering dynamics. For
instance, the latter situation may occur when the measurement time is
intentionally increased with the aim of increasing the detection efficiency.
In all cases, we assume that the pulses have a temporal length perfectly
matching the bandwidth of the detector.

In the case of an intermediate detector (ii), we integrate over the fast
fading process induced by turbulence. As a result, we have an overall fading
channel which is only generated by the pointing error, and whose instantaneous
transmissivity is now determined by the long-term spot size $w_{\text{lt}}%
^{2}=w_{\text{st}}^{2}+\sigma_{\text{TB}}^{2}$. Let us set
\begin{align}
\eta_{\text{int}}  &  =\eta_{\text{lt}}\eta_{\text{atm}}\eta_{\text{eff}},\\
\eta_{\text{lt}}  &  :=1-\exp\left(  -2a_{R}^{2}/w_{\text{lt}}^{2}\right)
\simeq\frac{2a_{R}^{2}}{w_{\text{lt}}^{2}}:=\eta_{\text{lt}}^{\text{far}}.
\end{align}
Then we may write the upper bound%
\begin{equation}
K_{\text{int}}\leq\mathcal{B}_{\text{int}}:=-\Delta(\eta_{\text{int}}%
,\sigma_{\text{P}})\log_{2}(1-\eta_{\text{int}}), \label{kappaINTER}%
\end{equation}
where $\Delta$ of Eq.~(\ref{deltaEXPRESSION}) has to be computed over
$\eta_{\text{int}}$ and $\sigma_{\text{P}}$ (with parameters $r_{0}$ and
$\gamma$ to be computed over $\eta_{\text{lt}}$ and $\eta_{\text{lt}%
}^{\text{far}}$). Similarly, the thermal upper bound takes the form%
\begin{equation}
K_{\text{int}}\leq\mathcal{B}_{\text{int}}-\mathcal{T}(\bar{n},\eta
_{\text{int}},\sigma_{\text{P}}). \label{kappaINTER2}%
\end{equation}
Basically, we obtain the modified formulas by setting $\sigma_{\text{TB}}%
^{2}\simeq0$ and replacing $w_{\text{st}}$ with the long-term spot size
$w_{\text{lt}}$ in the bounds of Eqs.~(\ref{PLOBdelta}), (\ref{UBTTTT})
and~(\ref{LBTTTT}).

Assuming a slower detector~(iii), we need to integrate over the entire fading
process induced by turbulence and pointing error. Instead of a fading channel,
we now have an average lossy channel with transmissivity $\eta_{\text{tot}}$
which is determined by the long-term spot size $w_{\text{lt}}^{2}%
=w_{\text{st}}^{2}+\sigma_{\text{TB}}^{2}$ together with the variance of the
pointing error $\sigma_{\text{P}}^{2}$, besides $\eta_{\text{atm}}$ and
$\eta_{\text{eff}}$. In other words, we have~\cite{NoteIntegral}%
\begin{align}
\eta_{\text{tot}}  &  =\left[  1-\exp\left(  -2a_{R}^{2}/w_{\text{tot}}%
^{2}\right)  \right]  \eta_{\text{atm}}\eta_{\text{eff}},\\
w_{\text{tot}}^{2}  &  :=w_{\text{lt}}^{2}+\sigma_{\text{P}}^{2}=w_{\text{st}%
}^{2}+\sigma_{\text{TB}}^{2}+\sigma_{\text{P}}^{2}.
\end{align}
As a result, the upper bound in Eq.~(\ref{PLOBdelta}) simplifies to%
\begin{equation}
K_{\text{slow}}\leq-\log_{2}(1-\eta_{\text{tot}})\leq\frac{2}{\ln2}\frac
{a_{R}^{2}}{w_{\text{lt}}^{2}+\sigma_{\text{P}}^{2}}. \label{ineq2}%
\end{equation}
Similarly, the thermal upper bound of Eq.~(\ref{UBTTTT}) becomes
\begin{align}
K_{\text{slow}}  &  \leq-\log_{2}\left[  (1-\eta_{\text{tot}})\eta
_{\text{tot}}^{\overline{n}^{\ast}}\right]  -h\left(  \overline{n}^{\ast
}\right)  ,\label{ineq22}\\
\overline{n}^{\ast}  &  :=\bar{n}/(1-\eta_{\text{tot}}),
\end{align}
for $\bar{n}\leq\eta_{\text{tot}}$, and is equal to zero otherwise. Note that
this formula is a direct modification of Ref.~\cite[Eq.~(23)]{QKDpaper}.

It is important to note that, in order to fairly compare
Eqs.~(\ref{kappaINTER}), (\ref{kappaINTER2}), (\ref{ineq2}) and~(\ref{ineq22})
with the previous fast-detection bounds, we need to account for the clock of
the system. In fact, in such a comparison, one should explicitly account for
the integration time which smooths the fluctuations but also reduces the final
rate (or throughput) in terms of bits per second. In fact, given a rate $K$ in
terms of bits/use, we need to plug a clock $C$ (uses/second) which depends on
the bandwidth of the detector and the repetition rate of the source. The
effective rate (bits/second) would then be $CK$. For instance, using a
detector with bandwidth $W=100$~MHz, we may work with $10~$ns pulses and use a
clock of $C=W/3\simeq3.3\times10^{7}$ uses (pulses) per second. If we assume a
slow detector (and corresponding longer pulses) with a detection time of
$100$~ms, we then have a clock of about $3.3$ uses per second, leading to
orders-of-magnitude lower rate in terms of bits per second. Furthermore, long
detection times also lead to higher background noise, which may become a major
problem for day time.

\subsubsection{Intermediate and strong turbulence}

The previous bounds for slow detection can be stated for increasing levels of
turbulence. From a physical point of view, stronger values of turbulence can
be associated with an increasingly-faster averaging process so that the
receiver loses the ability to resolve the fading dynamics. The effect is similar to having an increasingly-slower detector. However,
besides this averaging process, there is also the appearance of scintillation
effects and other effects of beam deformation, so that the transition from
weak to stronger regimes of turbulence cannot be described in simple
mathematical terms. That being said, the concept of long-term spot size is
robust and applies to the various regimes of turbulence, from weak to
strong~\cite[Sec.~IIIA]{Fante75}. In fact, even when the beam is broken up in
multiple patches (e.g., see case~4 of~\cite[Sec.~IIIA]{Fante75}), the
long-term spot size provides the mean square radius of the region containing
the patches.

In virtue of these considerations, we may rely on the robustness of the notion
of long-term spot size to extend our upper bounds beyond the weak
($\sigma_{\text{Rytov}}^{2}<1$) and the weak-intermediate ($\sigma
_{\text{Rytov}}^{2}\simeq1$) regimes of turbulence (see also
Appendix~\ref{TurbSECTION} for a discussion of these regimes in terms of the
ratio $\rho_{0}/w_{0}$). At intermediate-strong turbulence ($\sigma
_{\text{Rytov}}^{2}>1$), the variance $\sigma_{\text{TB}}^{2}$ becomes
relatively small, while the short-term spot size $w_{\text{st}}$ tends to
approximate the long-term value $w_{\text{lt}}$. If the pointing error is
non-negligible, then we may write the upper bounds in Eqs.~(\ref{kappaINTER})
and (\ref{kappaINTER2}). However, if pointing error $\sigma_{\text{P}}^{2}$ is
also negligible (with respect to $w_{\text{lt}}^{2}$), then we directly
consider the upper bounds in Eqs.~(\ref{ineq2}) and (\ref{ineq22}). For high
values of turbulence ($\sigma_{\text{Rytov}}^{2}\gg1$), we may certainly
assume $\sigma_{\text{P}}^{2}\simeq\sigma_{\text{TB}}^{2}\ll w_{\text{lt}}%
^{2}$, so that we write the upper bounds in Eqs.~(\ref{ineq2}) and
(\ref{ineq22}) for the strong-turbulence secret-key capacity $K_{\text{strong}%
}$. Because these bounds do not come from an operational reduction of the
detection time, the value $C$ of the system of clock can be high here.

\section{Composable security and key rates for CV-QKD\label{Sec_stableCVQKD}}

In this second part of the manuscript we study practical rates for free-space
CV-QKD, therefore providing state-of-the-art lower bounds for the free-space
secret key capacities discussed in the first part of the manuscript
(Sec.~\ref{Sec_PLOBfree}). In this specific section, we first develop a
general and simplified theory of composable security that applies to CV-QKD
protocols with a stable channel (fixed transmissivity), as is the typical case
in fiber-based implementations or even certain free-space links where
turbulence and other fading effects are negligible. This theory is the basis
for the next Sec.~\ref{Sec_freeCVQKD}, where we extend it to the case of
CV-QKD protocols over a fading channel (variable transmissivity) as is the
general case of free-space links affected by pointing errors and turbulence.
The latter is a more difficult scenario but with interesting implications for
both ground- and satellite-based
communications~\cite{PirandolaSAT,UsenkoSTRATEGY1,UsenkoSTRATEGY2,PanosFading,LeverrierSAT,RalphFREE,Malaney,Masoud}%
.

\subsection{Description of the protocol\label{descriptionPROTsec}}

\begin{figure}[t]
\vspace{-0.5cm}
\par
\begin{center}
\includegraphics[width=1\columnwidth] {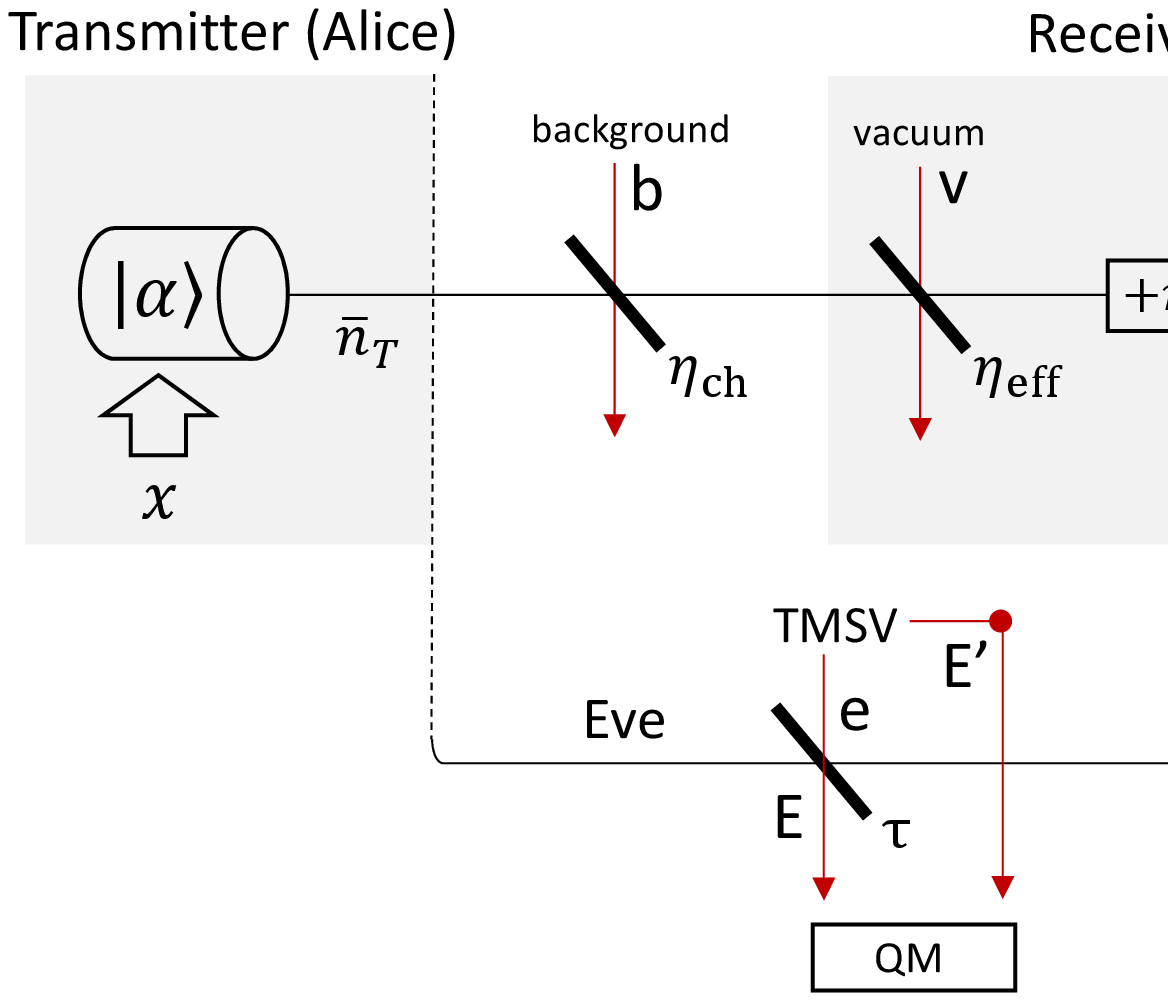}
\end{center}
\par
\vspace{-0.8cm}\caption{General description of the protocol and worst-case
eavesdropping scenario. Alice's modulated coherent state $\left\vert
\alpha\right\rangle $, with $\bar{n}_{T}$\ mean photons, is subject to channel
loss $\eta_{\text{ch}}$ and background noise $\bar{n}_{B}$, before entering
the receiver with quantum efficiency $\eta_{\text{eff}}$ and setup noise
$\bar{n}_{\text{ex}}$. Bob's detects $\bar{n}_{R}=\tau\bar{n}_{T}+\bar{n}$ mean
photons, where $\tau=\eta_{\text{ch}}\eta_{\text{eff}}$ is the total
transmissivity of the link and $\bar{n}=\eta_{\text{eff}}\bar{n}_{B}+\bar
{n}_{\text{ex}}$ is the total number of thermal photons. The input-output
relation for the quadratures is given by Eq.~(\ref{IOnew}). In the worst-case
scenario, Eve~collects all the leakage and controls all thermal noise. This is
equivalent to assume that she replaces the channel with a beam-splitter with
transmissivity $\tau$\ and thermal input $\bar{n}_{e}=\bar{n}/(1-\tau)$. The
latter is part of a TMSV\ state in her hands, whose output is stored in a
quantum memory.}%
\label{setupandWC}%
\end{figure}

Let us study a Gaussian-modulated coherent-state protocol with a fixed
transmissivity between Alice (the transmitter) and Bob (the
receiver)~\cite{QKDreview}. The general scenario is the one depicted
Fig.~\ref{setupandWC}. Alice encodes classical information in a bosonic mode
by preparing a coherent state $\left\vert \alpha\right\rangle $ whose
amplitude $\alpha$ is modulated according to a complex Gaussian distribution
with zero mean and variance $\mu-1$. Note that we may write $\alpha=(q+ip)/2$,
where $x=q$ or $p$ is the mean value of the generic quadrature operator
$\hat{x}=\hat{q}$ or $\hat{p}$ with $[\hat{q},\hat{p}]=2i$~\cite{RMP}.
Therefore, the generic quadrature of the mode can be decomposed as $\hat
{x}=\hat{x}_{0}+x$, where $\hat{x}_{0}$\ corresponds to vacuum noise and the
displacement $x$ is a real Gaussian variable with zero mean and variance
$\sigma_{x}^{2}=\mu-1$.

The coherent state contains $\bar{n}_{T}=\left\vert \alpha\right\vert ^{2}$
mean number of photons and it is transmitted through a channel with
transmissivity $\eta_{\text{ch}}$ and environmental noise $\bar{n}_{b}=\bar
{n}_{B}(1-\eta_{\text{ch}})^{-1}$, so that $\bar{n}_{B}$ thermal photons are
injected in the channel. (In terms of the free-space configuration of
Fig.~\ref{detectorPIC}, parameter $\eta_{\text{ch}}$ corresponds to the
instantaneous value $\eta_{\text{atm}}\eta_{\text{st}}(r)$, and $\bar{n}_{B}$
is the thermal background.) The output state is then measured by a receiver
with limited efficiency $\eta_{\text{eff}}$ and affected by thermal noise,
such to add $\bar{n}_{\text{ex}}$ extra mean photons. As a result, the final
(ideal) detection is reached by $\bar{n}_{R}=\tau\bar{n}_{T}+\bar{n}$ mean
photons, where $\tau=\eta_{\text{ch}}\eta_{\text{eff}}$ is the total
transmissivity and
\begin{equation}
\bar{n}=\eta_{\text{eff}}\bar{n}_{B}+\bar{n}_{\text{ex}} \label{nBARvalue}%
\end{equation}
is the total number of thermal photons. See Fig.~\ref{setupandWC}.

The final detection is either a randomly-switched measurement of $\hat{q}$ or
$\hat{p}$\ (homodyne) or a joint measurement of $\hat{q}$ and $\hat{p}$
(heterodyne). In both cases, there is an outcome $y$ corresponding to Alice's
classical input $x$. A single pair $(x,y)$ per mode is generated by the
homodyne protocol~\cite{GG02}, while two pairs per mode are generated by the
heterodyne protocol~\cite{Noswitch}. For both protocols, we may compactly
write the input-output relation
\begin{equation}
y=\sqrt{\tau}x+z, \label{IOnew}%
\end{equation}
where the noise variable is given by
\begin{align}
z  &  =\sqrt{\eta_{\text{eff}}(1-\eta_{\text{ch}})}\hat{x}_{b}+\sqrt{\tau}%
\hat{x}_{0}\nonumber\\
&  +\sqrt{1-\eta_{\text{eff}}}\hat{x}_{v}+\xi_{\text{ex}}+\xi_{\text{det}}.
\end{align}
Here $\hat{x}_{b}$ is the quadrature of the background thermal mode, $\hat
{x}_{v}$ is the quadrature of a setup vacuum mode, $\xi_{\text{ex}}$ is a
Gaussian variable with variance $2\bar{n}_{\text{ex}}$, and $\xi_{\text{det}}$
is an additional variable whose variance depends on the specific type of final
detection, i.e., we have $\mathrm{var}(\xi_{\text{det}})=0$ for homodyne, and
$\mathrm{var}(\xi_{\text{det}})=1$ for heterodyne. It is useful to introduce
the \textquotedblleft quantum duty\textquotedblright\ or \textquotedblleft
qu-duty\textquotedblright\ $\nu_{\text{det}}$ to pay by the detector, which is
$\nu_{\text{det}}=1$ for homodyne (due to the vacuum noise in the state) and
$\nu_{\text{det}}=2$ for heterodyne (which is increased due to the
simultaneous measurements of the two conjugate quadratures). Thus, in total,
the noise variable $z$ has variance%
\begin{equation}
\sigma_{z}^{2}=2\bar{n}+\nu_{\text{det}}.
\end{equation}

Alice and Bob's mutual information $I(x:y)$ is the same in direct
reconciliation (Bob inferring $x$ from $y$) and reverse reconciliation (Alice
inferring $y$ from $x$). This is easy to compute under ideal post-processing
techniques, able to reach the Shannon capacity of the additive-noise Gaussian
channel. In fact, from $\mathrm{var}(y)=\tau\sigma_{x}^{2}+\sigma_{z}^{2}$ and
$\mathrm{var}(y|x)=\sigma_{z}^{2}$, one derives%
\begin{equation}
I(x:y)=\frac{\nu_{\text{det}}}{2}\log_{2}\left(  1+\frac{\sigma_{x}^{2}}{\chi
}\right)  , \label{MutualINFOeq}%
\end{equation}
where $\chi:=\sigma_{z}^{2}/\tau$ is the equivalent noise, given by
\begin{equation}
\chi=\frac{2\bar{n}_{B}}{\eta_{\text{ch}}}+\frac{\nu_{\text{det}}+2\bar
{n}_{\text{ex}}}{\tau}. \label{chiEQ}%
\end{equation}
In particular, note that the first term in Eq.~(\ref{chiEQ}) is the specific
contribution of the channel to the excess noise
\begin{equation}
\varepsilon_{\text{ch}}:=\frac{2\bar{n}_{B}}{\eta_{\text{ch}}}=\frac{2(\bar
{n}-\bar{n}_{\text{ex}})}{\tau}. \label{excessCHnoise}%
\end{equation}
For the homodyne and heterodyne protocols, we may explicitly write%
\begin{align}
I^{\text{hom}}(x  &  :y)=\frac{1}{2}\log_{2}\left(  1+\frac{\tau\sigma_{x}%
^{2}}{2\bar{n}+1}\right)  ,\label{homoIm}\\
I^{\text{het}}(x  &  :y)=\log_{2}\left(  1+\frac{\tau\sigma_{x}^{2}}{2\bar
{n}+2}\right)  . \label{heteroIm}%
\end{align}

Before proceeding with the security analysis and the derivation of the
asymptotic key rate, it is important to clarify the most relevant noise
contributions that are present in the setup noise $\bar{n}_{\text{ex}}$. In
our study, we assume the worst-case scenario where this noise is considered to
be untrusted, even though it may be estimated or calibrated by the parties.
This robust approach allows us to lower-bound the performances that are
achievable by CV-QKD in general, including those situations where some of the
setup noise is considered to be trusted (as it might be the case for some
tolerable level of electronic noise).

\subsection{Practical observations on the receiver setup}

Here we discuss the contributions to the setup noise, that may be broken up as
$\bar{n}_{\text{ex}}=\bar{n}_{\text{LO}}+\bar{n}_{\text{el}}+\bar
{n}_{\text{other}}$, where $\bar{n}_{\text{LO}}$ are thermal photons generated
by imperfection in the LO (phase errors), $\bar{n}_{\text{el}}$ is electronic
noise, and $\bar{n}_{\text{other}}$ is any other uncharacterized and
independent noise source that might appear in the setup (that we numerically
neglect here). In general, the setup noise $\bar{n}_{\text{ex}}$ will depend
on the channel transmissivity. Below we start by describing $\bar
{n}_{\text{LO}}$\ which has a different behavior depending on the type of LO.
Afterwards, we discuss the expression of $\bar{n}_{\text{el}}$.

\subsubsection{Local oscillator (TLO and LLO)\label{LLOTLO}}

In order to encode and decode information with the quadratures of a bosonic
mode, the reference frames of the transmitter and receiver need to be
phase-locked. There are two possible ways to achieve this: either via a
TLO\ or an LLO. In the experimental practice, the use of a TLO is the simplest
solution. One the one hand, it introduces negligible phase error $\bar
{n}_{\text{TLO}}\simeq0$ and guarantees that the spatial modes of the signal
and LO pulses are the same, so that the mode matching is ideal at the
receiver. On the other hand, the fact that the LO transmitted together with
the signal means that it may also be the subject of attacks. This problem can
be mitigated by real-time monitoring of the LO intensity and properties, so as
to match the values expected by the parties~\cite{QKDreview}.

The other solution of a LLO excludes channel attacks against the LO, but
inevitably introduces non-trivial phase errors in the receiver setup.
These phase errors provide a contribution to the excess noise equal to
\begin{equation}
\varepsilon_{\text{LLO}}\simeq2\pi\sigma_{x}^{2}C^{-1}l_{\text{W}%
},\label{LLOeq}%
\end{equation}
where $C$ is the clock and $l_{\text{W}}$ is the laser linewidth. This formula
is derived from Ref.~\cite{LLO} assuming that signal pulses and LO-reference
pulses are generated with the same coherence time $\tau_{\text{coh}}\simeq(\pi
l_{\text{W}})^{-1}$. More generally, in Eq.~(\ref{LLOeq}) one needs to
consider the average linewidth $(l_{\text{W}}^{\text{signal}}+l_{\text{W}%
}^{\text{LO}})/2$, but we omit this technicality here.

From the formula, it is clear that the noise decreases for higher clocks and
narrower linewidths. In general, this approach requires better hardware than
the TLO. In our analysis, we have $\sigma_{x}^{2}\lesssim10$, so that a
reasonably low value $\varepsilon_{\text{LLO}}\lesssim0.02$ can be reached by
$C=5~$MHz and $l_{\text{W}}\simeq1.6$~KHz or, alternatively, by $C=100~$MHz
and $l_{\text{W}}\simeq32$~KHz (e.g., together with a $1$~GHz homodyne
receiver for detecting $0.1C^{-1}\simeq1$~ns pulses~\cite{NoteCLOCK}). In
other words, very good cw-lasers and detectors are needed. Refined analyses
suggest that highly-performant amplitude modulators are also required in order
to avoid the introduction of other noise contributions~\cite{LLO2,LLO3}.

To account for the LLO in our theoretical treatment, we recall that Alice and
Bob's mutual information takes the form in Eq.~(\ref{MutualINFOeq}) where the
equivalent noise $\chi$ is broken down as in Eq.~(\ref{chiEQ}), i.e., we write%
\begin{equation}
\chi=\varepsilon_{\text{ch}}+\frac{\nu_{\text{det}}+2\bar{n}_{\text{ex}}}%
{\tau}, \label{chiEQ2}%
\end{equation}
where $\varepsilon_{\text{ch}}:=2\bar{n}_{B}/\eta_{\text{ch}}$ is channel's
excess noise. The introduction of the LLO contribution consists of making the
replacement $\chi\rightarrow\chi+\varepsilon_{\text{LLO}}$ in the formula
above. Because this type of noise is within the local setup of the receiver,
we make it a contribution to $\bar{n}_{\text{ex}}$ by writing%
\begin{equation}
\bar{n}_{\text{LLO}}=\frac{\tau\varepsilon_{\text{LLO}}}{2}=\pi\tau\sigma
_{x}^{2}C^{-1}l_{\text{W}}. \label{UBnex}%
\end{equation}

Some observations are in order. The basic implementation of LLO considers the
regular alternation between signal and LO-reference pulses. In such a setting,
one may argue that the actual rate per second (throughput) is halved with
respect to the TLO. However, it is worth noticing that this factor $1/2$ may
be compensated if the signals are encoded in both polarizations for each
channel use (not possible for a TLO due to its multiplexing in polarization).
Another observation is about the use of homodyne or heterodyne at the
receiver. Because of the regular signal-reference alternation, the receiver
may use a dedicated heterodyne detector for the LO references and another
detector for the signals (heterodyne or randomly-switched homodyne). However,
if the receiver is limited to a single homodyne detector, then the transmitter
can send two LO-reference pulses with orthogonal polarizations and rotated by
$\pi/2$ in phase space. At the receiver, these pulses can be demultiplexed,
delayed and sequentially homodyned to give the complete phase information.

\subsubsection{Electronic noise}

One of the typical and unavoidable sources of noise within the setup of the
receiver is electronic noise, with associated variance $\nu_{\text{el}}$ or
equivalent number of photons $\bar{n}_{\text{el}}=\nu_{\text{el}}/2$. This
depends on the noise equivalent power (NEP) of the amplifiers and photodiodes
to be used in the homodyne detectors, besides the detection bandwidth $W$, the
duration of the LO\ pulses $\Delta t_{\text{LO}}$, the LO\ power at the
detector $P_{\text{LO}}^{\text{det}}$, and the frequency of the light $\nu$.
In fact, one can write the formula~\cite{elenoise1,elenoise2}%
\begin{equation}
\nu_{\text{el}}=\frac{\nu_{\text{det}}\mathrm{NEP}^{2}W\Delta t_{\text{LO}}%
}{h\nu P_{\text{LO}}^{\text{det}}}. \label{elEQ}%
\end{equation}

At $W=100~$MHz, we may consider $\mathrm{NEP}=6~$pW/$\sqrt{\text{Hz}}$. Then,
assuming $\nu\simeq3.75\times10^{14}$~Hz ($\lambda=800~$nm) and $\Delta
t_{\text{LO}}=10~$ns, we may write $\nu_{\text{el}}=1.45\times10^{-4}%
\nu_{\text{det}}/P_{\text{LO}}^{\text{det}}$. In a TLO setup, we have
$P_{\text{LO}}^{\text{det}}=\tau P_{\text{LO}}$, where $P_{\text{LO}}$ is the
initial LO power at the transmitter. Setting $P_{\text{LO}}=100~$mW, we
derive
\begin{equation}
\nu_{\text{el}}(\tau)\leq\frac{2.9\times10^{-3}}{\tau}, \label{reduceEL}%
\end{equation}
where the bound is taken by assuming the worst-case scenario of heterodyne
detection ($\nu_{\text{det}}=2$). As we can see from Eq.~(\ref{reduceEL}), the
noise is small at short ranges but may become non-trivial at long distances,
e.g., $\nu_{\text{el}}\leq0.29$ at $20$~dB, i.e., for $\tau=10^{-2}$.

In the case of an LLO setup, where the LO\ pulse is locally generated, we have
$P_{\text{LO}}^{\text{det}}=P_{\text{LO}}$ in Eq.~(\ref{elEQ}). This means
that $\nu_{\text{el}}$ becomes independent from the transmissivity and its
value can be very low. In our numerical example, Eq.~(\ref{reduceEL}) is
replaced by $\nu_{\text{el}}\leq2.9\times10^{-3}$. Thus, the LLO setup
provides an advantage with respect to the TLO in terms of reduced electronic
noise (to be balanced with the negative effect of introducing phase errors).

\subsubsection{Setup noise versus channel transmissivity\label{subSEC_options}%
}

As we see from the discussion above, the setup noise $\bar{n}_{\text{ex}}$
also depends on the transmissivity of the channel $\tau$, \ due to the fact
that the value of $\tau$ is relevant for both the LO\ power and the
(attenuated) modulation of the signals at the receiver. Let us make the
notation more compact by introducing the term
\begin{equation}
\Theta_{\text{el}}:=\frac{\nu_{\text{det}}\mathrm{NEP}^{2}W\Delta
t_{\text{LO}}}{2h\nu P_{\text{LO}}}.\label{eleCONTRIBUT}%
\end{equation}
Then, the setup noise $\bar{n}_{\text{ex}}$ has different monotonicity in
$\tau$ depending on the use of a TLO\ or an LLO. In fact, we can write the
following%
\begin{equation}
\bar{n}_{\text{ex}}^{\text{TLO}}(\tau)=\frac{\Theta_{\text{el}}}{\tau}%
,~\bar{n}_{\text{ex}}^{\text{LLO}}(\tau)=\Theta_{\text{el}}+\pi\tau\sigma
_{x}^{2}C^{-1}l_{\text{W}},
\end{equation}
so that $\bar{n}_{\text{ex}}^{\text{TLO}}$ is decreasing in $\tau$, while
$\bar{n}_{\text{ex}}^{\text{LLO}}$ is increasing.

\subsection{Asymptotic key rate\label{subsectionAKR}}

Once we have clarified the various contributions to thermal noise, we proceed
with the security analysis assuming that the various imperfections of the
receiver are untrusted, both in terms of setup noise $\bar{n}_{\text{ex}}$ and
quantum efficiency $\eta_{\text{eff}}$. Thus, our approach assumes the
worst-case scenario where Eve not only perturbs the outside channel (with
transmissivity $\eta_{\text{ch}}$ and background noise $\bar{n}_{B}$), but
also collects the fraction $1-\eta_{\text{eff}}$ of photons leaked by the
receiver, and potentially tampers with its setup noise $\bar{n}_{\text{ex}}$
(which might be exploited to insert Trojan-horse photons). As already said
before, this is a conservative approach which allows us to lower-bound the
performance of CV-QKD and to remove the exploitation of potential loopholes in
the practical devices.

In the worst-case scenario, Alice and Bob ascribe the entirety of loss
$\tau=\eta_{\text{ch}}\eta_{\text{eff}}$ and thermal noise $\bar{n}%
=\eta_{\text{eff}}\bar{n}_{B}+\bar{n}_{\text{ex}}$ to Eve. See
Fig.~\ref{setupandWC}. In other words, Eve is assumed to have the total
control of the environmental dilation of the thermal-loss channel
$\mathcal{E}_{\tau,\bar{n}}$ that is observed by the parties and leading to
the input-output relation of Eq.~(\ref{IOnew}). Such a dilation corresponds to
a beam-splitter of transmissivity $\tau$ that mixes each signal mode with an
environmental mode carrying $\bar{n}_{e}=\bar{n}/(1-\tau)$ thermal photons,
which is in turn part of a two-mode squeezed vacuum (TMSV) state prepared by
Eve. For each incoming signal, a fresh TMSV state is prepared and used in the
interaction. After interaction, the signal output of the beam splitter is
released to Bob, while the environmental output is stored in a quantum memory,
to be jointly measured by Eve at the end of the protocol. This is a collective
entangling-cloner attack which is the most practical and relevant collective
Gaussian attack~\cite{collectiveG}.

In this scenario, let us compute Eve's Holevo information, i.e., the maximum
amount of information that she can steal per use of the channel. It is
convenient to work in the entanglement-based representation, where Alice's
Gaussian-modulated coherent states with variance $\sigma_{x}^{2}=\mu-1$ are
realized by heterodyning the idler mode $A$ of a TMSV state~\cite{RMP} with
covariance matrix (CM)
\begin{equation}
\mathbf{V}_{AA^{\prime}}=\left(
\begin{array}
[c]{cc}%
\mu\mathbf{I} & \sqrt{\mu^{2}-1}\mathbf{Z}\\
\sqrt{\mu^{2}-1}\mathbf{Z} & \mu\mathbf{I}%
\end{array}
\right)  ,
\end{equation}
where $\mathbf{I}:=\mathrm{diag}(1,1)$ and $\mathbf{Z}:=\mathrm{diag}(1,-1)$.
After the action of the thermal-loss channel on the transmitted mode
$A^{\prime}$, we have that Alice and Bob share a zero-mean Gaussian state with
CM%
\begin{equation}
\mathbf{V}_{AB}=\left(
\begin{array}
[c]{cc}%
\mu\mathbf{I} & \mathbf{C}\\
\mathbf{C}^{T} & b\mathbf{I}%
\end{array}
\right)  ,~%
\begin{array}
[c]{l}%
\mathbf{C:=}\sqrt{\tau(\mu^{2}-1)}\mathbf{Z},\\
b:=\tau(\mu-1)+2\bar{n}+1.
\end{array}
\end{equation}

Because the total output state $\rho_{AB\mathbf{E}}$ of Alice $A$, Bob $B$ and
Eve $\mathbf{E}=EE^{\prime}$ is a pure state, we can compute Eve's Holevo
bound from Alice's and Bob's von Neumann entropies $S(\cdots)$. In reverse
reconciliation, Eve's Holevo bound with respect to Bob's variable $y$ is given
by
\begin{equation}
\chi(\mathbf{E}:y):=S(\mathbf{E})-S(\mathbf{E}|y)=S(AB)-S(A|y),
\label{ChiJOINT}%
\end{equation}
where $S(\mathbf{E})=S(AB)$ comes from the total purity, and $S(\mathbf{E}%
|y)=S(A|y)$\ comes from the fact that Bob's measurement is a rank-1 projection
(homodyne/heterodyne), so that Alice and Eve's conditional state
$\rho_{A\mathbf{E}|y}$ is pure.

It is easy to compute the entropies above starting from Alice and Bob's output
CM $\mathbf{V}_{AB}$. Let us call $\nu_{\pm}$ the two symplectic eigenvalues
of\ $\mathbf{V}_{AB}$. Then, we may write
\begin{equation}
S(AB)=H(\nu_{+})+H(\nu_{-}),~H(x):=h[\left(  x-1\right)  /2],
\end{equation}
where $H(x)$ is defined using Eq.~(\ref{hFUNCTIONmain}). The value of
$S(A|y)$\ is given by computing $H(x)$\ over the symplectic eigenvalue of the
conditional CM $\mathbf{V}_{A|y}$, whose explicit expression depends on the
type of detection.

Let us set $\boldsymbol{\Pi}:=\mathrm{diag}(1,0)$. For the homodyne protocol,
Alice's CM conditioned\ on Bob's outcome $y$ is~\cite{RMP,GaeCM,GaeCM2}%
\begin{equation}
\mathbf{V}_{A|y}^{\text{hom}}=\mu\mathbf{I}-b^{-1}\mathbf{C}\boldsymbol{\Pi
}\mathbf{C}^{T},
\end{equation}
and its symplectic eigenvalue is given by%
\begin{equation}
\nu^{\text{hom}}=\sqrt{\det\mathbf{V}_{A|y}^{\text{hom}}}=\sqrt{\mu^{2}%
-\frac{\mu\tau(\mu^{2}-1)}{b}}.
\end{equation}

For the heterodyne protocol, we have instead~\cite{RMP,GaeCM,GaeCM2}%
\begin{equation}
\mathbf{V}_{A|y}^{\text{het}}=\mu\mathbf{I}-(b+1)^{-1}\mathbf{CC}^{T}%
=\nu^{\text{het}}\mathbf{I},
\end{equation}
with symplectic eigenvalue%
\begin{equation}
\nu^{\text{het}}=\mu-\frac{\tau(\mu^{2}-1)}{b+1}.
\end{equation}
As a result, we have
\begin{align}
\chi^{\text{hom}}(\mathbf{E}  &  :y)=S(AB)-H\left(  \nu^{\text{hom}}\right)
,\label{homoCHI}\\
\chi^{\text{het}}(\mathbf{E}  &  :y)=S(AB)-H\left(  \nu^{\text{het}}\right)  .
\label{heteroCHI}%
\end{align}

For a realistic reconciliation efficiency $\beta\in\lbrack0,1]$, accounting
for the fact that data-processing may not reach the Shannon limit, we write
the asymptotic key rate
\begin{equation}
R_{\text{asy}}(\tau,\bar{n})=\beta I(x:y)_{\tau,\bar{n}}-\chi(\mathbf{E}%
:y)_{\tau,\bar{n}}~, \label{rateASYm}%
\end{equation}
where the explicit expressions for the homodyne protocol~\cite{GG02}
($R_{\text{asy}}^{\text{hom}}$) and the heterodyne protocol~\cite{Noswitch}
($R_{\text{asy}}^{\text{het}}$) derive from the corresponding expressions for
the mutual information $I^{\text{hom}}$ and $I^{\text{het}}$ [cf.
Eqs.~(\ref{homoIm}) and~(\ref{heteroIm})] and the Holevo bound $\chi
^{\text{hom}}$ and $\chi^{\text{het}}$ [cf. Eqs.~(\ref{homoCHI})
and~(\ref{heteroCHI})]. In an experimental implementation, the term $\beta I$
in Eq.~(\ref{rateASYm}) is determined by the empirical entropy associated with
the key and the specific code used for error correction.

It is important to observe that the rate in Eq.~(\ref{rateASYm}) can be
computed by Alice and Bob once they know the values of the total
transmissivity $\tau$ and the total thermal noise $\bar{n}$. In a practical
setting, the values of $\tau$ and $\bar{n}$\ are not known but must be
evaluated during the protocol via a dedicated procedure of parameter
estimation. Because a realistic protocol runs for a finite number of times,
this estimation is not perfect and decreases the rate.

Up to an error probability $\varepsilon_{\text{pe}}$, Alice and Bob derive
worst-case estimators $\tau^{\prime}\simeq\tau-f(\tau,\bar{n})$ and $\bar
{n}^{\prime}\simeq\bar{n}+g(\bar{n})$, for suitable monotonic functions $f$
and $g$ (both increasing in $\bar{n}$). Thus, they use $\tau^{\prime}$ and
$\bar{n}^{\prime}$ to compute the parameter-estimation-based version of the
rate
\begin{equation}
R_{\text{pe}}(\tau^{\prime},\bar{n}^{\prime})=\beta I(x:y)_{\tau^{\prime}%
,\bar{n}^{\prime}}-\chi(\mathbf{E}:y)_{\tau^{\prime},\bar{n}^{\prime}}.
\label{Rpe1untrusted}%
\end{equation}
Below we clarify the explicit expressions for $f$ and $g$.

\subsection{Details of parameter estimation\label{PEmainSEC}}

Here we go into the fine details of parameter estimation, also clarifying the
explicit forms of the functions $f$ and $g$ that are used above. For
implementing this step of the protocol, Alice and Bob jointly choose a random
subset of $m$ channel uses. By publicly comparing the corresponding
input-output values, they estimate the relevant channel parameters ($\tau$ and
$\bar{n}$) whose knowledge is crucial for applying the most appropriate
procedures of error correction and privacy amplification.

\subsubsection{Estimators}

Alice and Bob randomly choose $m$ signals whose encoding $x$ and decoding $y$
are publicly disclosed. This means that the parties compare $m_{p}%
:=\nu_{\text{det}}m$ pairs of values $\{x_{i},y_{i}\}_{i=1}^{m_{p}}$ related
by Eq.~(\ref{IOnew}). These pairs are $m$ for the homodyne protocol, and $2m$
for the heterodyne protocol. Under the assumption of a collective Gaussian
attack, they are Gaussian as well as independent and identically distributed (iid).

From the $m_{p}$ disclosed pairs, the parties construct an estimator $\hat{T}$
of $T:=\sqrt{\tau}$ as follows~\cite{LevDELTA,UsenkoFinite}%
\begin{equation}
\hat{T}:=\frac{\sum_{i=1}^{m_{p}}x_{i}y_{i}}{\sum_{i=1}^{m_{p}}x_{i}^{2}%
}\simeq\frac{\sum_{i=1}^{m_{p}}x_{i}y_{i}}{m_{p}\sigma_{x}^{2}},
\end{equation}
which is Gaussianly distributed for sufficiently large $m_{p}$. Equivalently,
one may write
\begin{equation}
\hat{T}=\frac{\widehat{C_{xy}}}{\sigma_{x}^{2}},~~\widehat{C_{xy}}=m_{p}%
^{-1}\sum_{i=1}^{m_{p}}x_{i}y_{i}, \label{alternativeT}%
\end{equation}
where $\widehat{C_{xy}}$ estimates the covariance $C_{xy}:=\langle
xy\rangle=\sqrt{\tau}\sigma_{x}^{2}$.

It is easy to check that $\hat{T}$ is unbiased since we have%
\begin{equation}
\langle\hat{T}\rangle\simeq\frac{\sum_{i=1}^{m_{p}}\langle x_{i}y_{i}\rangle
}{m_{p}\sigma_{x}^{2}}\simeq\frac{\langle xy\rangle}{\sigma_{x}^{2}}=T.
\end{equation}
For the variance, we may compute
\begin{equation}
\sigma_{T}^{2}:=\mathrm{var}(\hat{T})=\frac{\sum_{i=1}^{m_{p}}\mathrm{var}%
(x_{i}y_{i})}{m_{p}^{2}\sigma_{x}^{4}}\simeq\frac{\sigma_{z}^{2}}{m_{p}%
\sigma_{x}^{2}}+\frac{2\tau}{m_{p}}, \label{UsenkoVersion}%
\end{equation}
where we use that $x_{i}y_{i}$ are iid (so that $\mathrm{var}\sum
=\sum\mathrm{var}$), the fact that the noise has zero mean $\langle
z\rangle=0$, and finally that $\langle x^{4}\rangle=3\sigma_{x}^{4}$ for a
zero-mean Gaussian variable.

From the square-root transmissivity, Alice and Bob can derive the estimator of
the transmissivity as $\hat{\tau}=(\hat{T})^{2}$, which is unbiased with
variance%
\begin{equation}
\sigma_{\tau}^{2}:=\mathrm{var}(\widehat{\tau})\simeq\frac{4\tau^{2}}{m_{p}%
}\left(  2+\frac{\sigma_{z}^{2}}{\tau\sigma_{x}^{2}}\right)  +\mathcal{O}%
(m_{p}^{-2}). \label{sigmatVAL}%
\end{equation}
This is shown by noting that, for a Gaussian variable $X\sim\mathcal{N}%
(\bar{x},\sigma)$, one has $\mathrm{var}(X^{2})=2\sigma^{2}(2\bar{x}%
+\sigma^{2})$. Alternatively, one uses Eq.~(\ref{alternativeT}) and notes that
$\widehat{\gamma_{xy}}:=$ $(\widehat{C_{xy}})^{2}/\sigma_{\text{cov}}^{2}$
with
\begin{equation}
\sigma_{\text{cov}}^{2}:=\mathrm{var}(\widehat{C}_{xy})\simeq m_{p}^{-1}%
\tau\sigma_{x}^{4}\left[  2+\sigma_{z}^{2}/(\tau\sigma_{x}^{2})\right]
\label{sigmaCOVV}%
\end{equation}
is a non-central chi-square distribution $\chi^{2}(1,\lambda_{\text{nc}})$,
having $1$ degree of freedom and non-centrality parameter $\lambda_{\text{nc}%
}=C_{xy}^{2}/\sigma_{\text{cov}}^{2}$ (so that its mean is $1+\lambda
_{\text{nc}}$ and its variance is $2+4\lambda_{\text{nc}}$). Computing the
variance of $\widehat{\tau}=\widehat{\gamma_{xy}}(\sigma_{\text{cov}}%
^{2}/\sigma_{x}^{4})$ up to $\mathcal{O}(m_{p}^{-2})$, one gets
Eq.~(\ref{sigmatVAL}).

Note that Eq.~(\ref{UsenkoVersion}) is in line with the derivation of
Ref.~\cite{UsenkoFinite}, while Ref.~\cite{LevDELTA} resorts to a further
approximation that would lead to the removal of the term $2\tau/m_{p}$ in the
expression above. Here we follow the most conservative choice (approach of
Ref.~\cite{UsenkoFinite}) which implies a larger uncertainty for the value of
the transmissivity.

For the variance of the thermal noise $\sigma_{z}^{2}$, Alice and Bob build an
estimator%
\begin{equation}
\widehat{\sigma_{z}^{2}}:=\frac{1}{m_{p}}\sum_{i=1}^{m_{p}}(y_{i}-\hat{T}%
x_{i})^{2}=\frac{1}{m_{p}}\sum_{i=1}^{m_{p}}z_{i}^{2}. \label{sigmaZesimates}%
\end{equation}
For large $m_{p}$, the variable $Y_{z}:=m_{p}\widehat{\sigma_{z}^{2}}%
/\sigma_{z}^{2}$ follows a chi-square distribution $\chi^{2}(m_{p})$ with
$m_{p}$ degrees of freedom (mean value $m_{p}$ and variance $2m_{p}$), so that
we have
\begin{equation}
\langle\widehat{\sigma_{z}^{2}}\rangle\simeq\sigma_{z}^{2},~\mathrm{var}%
(\widehat{\sigma_{z}^{2}})\simeq\frac{2\sigma_{z}^{4}}{m_{p}}.
\label{varsigmazz}%
\end{equation}
Equivalently, they can build the estimator for the thermal number $\bar{n}$
defined by%
\begin{equation}
\widehat{\bar{n}}:=\left(  \widehat{\sigma_{z}^{2}}-\nu_{\text{det}}\right)
/2, \label{photonEST2}%
\end{equation}
with mean value $\langle\widehat{\bar{n}}\rangle\simeq\bar{n}$ and variance
\begin{equation}
\sigma_{\bar{n}}^{2}=\frac{\mathrm{var}(\widehat{\sigma_{z}^{2}})}{4}%
\simeq\frac{\sigma_{z}^{4}}{2m_{p}}. \label{varnbar}%
\end{equation}
Because the number of degrees of freedom is typically very large, the
chi-square distribution $\chi^{2}(m_{p})$ can also be approximated by a
Gaussian distribution with the same mean value and variance. As a result, the
estimators $\widehat{\sigma_{z}^{2}}$ and $\widehat{\bar{n}}$ can be
considered to be asymptotically Gaussian.

It is important to observe that, from an experimental point of view, the
variances in Eqs.~(\ref{UsenkoVersion}), (\ref{sigmatVAL}), (\ref{varsigmazz})
and~(\ref{varnbar}) can be computed by replacing/using the estimators $\hat
{T}$ and $\widehat{\sigma_{z}^{2}}$ in the right-hand sides of the equations.

\subsubsection{Worst-case estimators}

From the estimators, Alice and Bob construct suitable worst-case estimators by
assuming a certain number $w$ of confidence intervals, for some acceptable
error probability $\varepsilon_{\text{pe}}$. For the square-root
transmissivity they build
\begin{equation}
T^{\prime}:=\hat{T}-w\sigma_{T}\simeq T-w\sqrt{\frac{2\tau+\sigma_{z}%
^{2}/\sigma_{x}^{2}}{m_{p}}}. \label{wcparT}%
\end{equation}
The probability $\varepsilon_{\text{pe}}$ that the actual value $T$ is less
than $T^{\prime}$ is given by
\begin{align}
\varepsilon_{\text{pe}}  &  =\mathrm{prob}(T<\hat{T}-w\sigma_{T})\\
&  =\mathrm{prob}\left[  \frac{\hat{T}-T}{\sigma_{T}}>w\right]  =1-\Phi
_{\text{CND}}(w),\nonumber
\end{align}
where $\Phi_{\text{CND}}(x)=[1+\operatorname{erf}(x/\sqrt{2})]/2$ is the
cumulative of the standard normal distribution. Equivalently, for a given
value of $\varepsilon_{\text{pe}}$, one derives%
\begin{equation}
w=\sqrt{2}\operatorname{erf}^{-1}(1-2\varepsilon_{\text{pe}}).
\label{wSTVALUE}%
\end{equation}

From Eq.~(\ref{wcparT}), one can immediately construct the worst-case
estimator for the transmissivity $\tau$ by taking the square $\tau^{\prime
}=(T^{\prime})^{2}$ so that we obtain%
\begin{equation}
\tau^{\prime}\simeq\tau-2w\sqrt{\frac{2\tau^{2}+\tau\sigma_{z}^{2}/\sigma
_{x}^{2}}{m_{p}}}+\mathcal{O}(m_{p}^{-1}). \label{wscase1}%
\end{equation}
Equivalently, this is derived by writing $\tau^{\prime}:=\hat{\tau}%
-w\sigma_{\tau}$, and then using $\hat{\tau}\simeq\tau$ together with
$\sigma_{\tau}$ from Eq.~(\ref{sigmatVAL}).

Because $\widehat{\sigma_{z}^{2}}$ and $\widehat{\bar{n}}$ are asymptotically
Gaussian, Alice and Bob can build corresponding worst-case estimators for
which they connect the number $w$ of confidence intervals with the error
probability $\varepsilon_{\text{pe}}$ according to Eq.~(\ref{wSTVALUE}). In
particular, they build the worst-case estimator for the thermal number
$\bar{n}^{\prime}:=\widehat{\bar{n}}+w\sigma_{\bar{n}}$, where $w$\ is such
that $\varepsilon_{\text{pe}}=\mathrm{prob}(\bar{n}>\bar{n}^{\prime})$. We
easily compute%
\begin{equation}
\bar{n}^{\prime}\simeq\bar{n}+\frac{w\sigma_{z}^{2}}{\sqrt{2m_{p}}}.
\label{wscase2}%
\end{equation}

As a result, up to an error probability $\varepsilon_{\text{pe}}%
=\varepsilon_{\text{pe}}(w)$, Alice and Bob are able to bound the actual
values of $\tau$ and $\bar{n}$ with the worst-case estimators in
Eqs.~(\ref{wscase1}) and~(\ref{wscase2}). In the notation of
Sec.~\ref{subsectionAKR}, this means that we have $\tau^{\prime}\simeq
\tau-f(\tau,\bar{n})$ and $\bar{n}^{\prime}\simeq\bar{n}+g(\bar{n})$, where
\begin{align}
f(\tau,\bar{n})  &  =2w\sqrt{\frac{2\tau^{2}+\tau\sigma_{x}^{-2}(2\bar{n}%
+\nu_{\text{det}})}{m_{p}}},\label{wcest1}\\
g(\bar{n})  &  =\frac{w}{\sqrt{2m_{p}}}(2\bar{n}+\nu_{\text{det}}).
\label{wcest2}%
\end{align}
Note that $\varepsilon_{\text{pe}}$ is here defined for each basic parameter
to be estimated, so that the total error associated with the two parameters
$\tau$ and $\bar{n}$ is given by $\varepsilon_{\text{pe}}(1-\varepsilon
_{\text{pe}})+(1-\varepsilon_{\text{pe}})\varepsilon_{\text{pe}}%
+\varepsilon_{\text{pe}}^{2}\simeq2\varepsilon_{\text{pe}}$. Also note that,
for the typical choice $\varepsilon_{\text{pe}}=2^{-33}\simeq10^{-10}$, we
have $w\simeq6.34$.

\subsubsection{Tail bounds}

When the value of $\varepsilon_{\text{pe}}$ is chosen to be very low
($\leq10^{-17}$), the approach above creates divergences ($w\rightarrow\infty
$). In this case, we must resort to suitable tail bounds. Let us start by
analyzing the estimation of the thermal noise. For the central chi-square
variable $Y_{z}\sim\chi^{2}(m_{p})$, we may write the following tail
bound~\cite[Lemma~1]{TailBound}%
\begin{equation}
\mathrm{prob}\left[  Y_{z}\leq m_{p}-2\sqrt{m_{p}x}\right]  \leq e^{-x},
\end{equation}
for any $x$. Let us combine the latter with Eq.~(\ref{photonEST2}). With
probability $\leq e^{-x}$, the estimator $\widehat{\bar{n}}$ satisfies%
\begin{equation}
\widehat{\bar{n}}\leq\bar{n}-\sigma_{z}^{2}\sqrt{\frac{x}{m_{p}}},
\end{equation}
or, equivalently, the actual value $\bar{n}$ satisfies
\begin{equation}
\bar{n}\geq\widehat{\bar{n}}+\sigma_{z}^{2}\sqrt{\frac{x}{m_{p}}}%
\simeq\widehat{\bar{n}}+\sigma_{\bar{n}}\sqrt{2x}.
\end{equation}

Let us set $x=\ln(1/\varepsilon_{\text{pe}})$. Then, with probability
$\leq\varepsilon_{\text{pe}}$, we have%
\begin{equation}
\bar{n}\gtrsim\widehat{\bar{n}}+\sigma_{\bar{n}}\sqrt{2\ln(1/\varepsilon
_{\text{pe}})}.
\end{equation}
Thus, the worst-case value takes the form $\bar{n}^{\prime}:=\widehat{\bar{n}%
}+w\sigma_{\bar{n}}$ as before but now with%
\begin{equation}
w=\sqrt{2\ln(1/\varepsilon_{\text{pe}})}. \label{wTAIL}%
\end{equation}
Note that, in this case, $\varepsilon_{\text{pe}}=2^{-33}$ corresponds to
$w\simeq6.76$, slightly larger than before. However, now we can also deal with
smaller values of the error probability; e.g., $\varepsilon_{\text{pe}%
}=10^{-43}$ corresponds to $w\simeq14$.

Similar extensions can be derived with other tail bounds~\cite[App.~6.1]%
{Kolar}. In particular, the derivation can immediately be adapted to the
transmissivity. For a variable $X\sim\chi^{2}(d,\lambda_{\text{nc}})$ with $d$
degrees of freedom and non-centrality parameter $\lambda_{\text{nc}}$, we may
write~\cite{Birge} (see also Ref.~\cite[Lemma~8]{Kolar})%
\begin{equation}
\mathrm{prob}\left[  X\leq(d+\lambda_{\text{nc}})-2\sqrt{(d+2\lambda
_{\text{nc}})x}\right]  \leq e^{-x}.
\end{equation}
Setting $x=\ln(1/\varepsilon_{\text{pe}})$, we then write%
\begin{equation}
\mathrm{prob}\left[  X\leq(d+\lambda_{\text{nc}})-2\sqrt{(d+2\lambda
_{\text{nc}})\ln\frac{1}{\varepsilon_{\text{pe}}}}\right]  \leq\varepsilon
_{\text{pe}}.
\end{equation}

Take $X=\widehat{\gamma_{xy}}\sim\chi^{2}(1,C_{xy}^{2}/\sigma_{\text{cov}}%
^{2})$. With probability $\leq\varepsilon_{\text{pe}}$, this estimator
satisfies%
\begin{equation}
\widehat{\gamma_{xy}}\leq1+\frac{C_{xy}^{2}}{\sigma_{\text{cov}}^{2}}%
-2\sqrt{\left(  1+2\frac{C_{xy}^{2}}{\sigma_{\text{cov}}^{2}}\right)  \ln
\frac{1}{\varepsilon_{\text{pe}}}}.
\end{equation}
With the same probability, $\widehat{\tau}=\widehat{\gamma_{xy}}%
(\sigma_{\text{cov}}^{2}/\sigma_{x}^{4})$ satisfies%
\begin{align}
\widehat{\tau}  &  \leq\frac{\sigma_{\text{cov}}^{2}+C_{xy}^{2}}{\sigma
_{x}^{4}}-\frac{2}{\sigma_{x}^{4}}\sqrt{\left(  \sigma_{\text{cov}}%
^{4}+2\sigma_{\text{cov}}^{2}C_{xy}^{2}\right)  \ln\frac{1}{\varepsilon
_{\text{pe}}}}\\
&  \overset{(\ast)}{\simeq}\tau-2\tau\sqrt{2m_{p}^{-1}\left(  2+\frac
{\sigma_{z}^{2}}{\tau\sigma_{x}^{2}}\right)  \ln\frac{1}{\varepsilon
_{\text{pe}}}}+\mathcal{O}(m_{p}^{-1})\\
&  =\tau-\sigma_{\tau}\sqrt{2\ln\frac{1}{\varepsilon_{\text{pe}}}}%
+\mathcal{O}(m_{p}^{-1}), \label{eqtulast}%
\end{align}
where in $(\ast)$ we have used $C_{xy}^{2}\simeq\tau\sigma_{x}^{4}$, the
scaling $\sigma_{\text{cov}}^{2}\simeq\mathcal{O}(m_{p}^{-1})$ and
Eq.~(\ref{sigmaCOVV}). More precisely, the approximation in $(\ast)$ is
certainly valid for $2(m_{p}-1)\gg\sigma_{z}^{2}/(\tau\sigma_{x}^{2})$ which
is the typical regime of parameters. From Eq.~(\ref{eqtulast}) we see that,
for the transmissivity, we have again $\tau^{\prime}:=\hat{\tau}-w\sigma
_{\tau}$ but where $w$ is now given in Eq.~(\ref{wTAIL}).

\subsection{Finite-size composable key rate}

So far we have considered the effect of parameter estimation on the key rate,
so that its expression takes the form $R_{\text{pe}}$ in
Eq.~(\ref{Rpe1untrusted}), where the worst-case estimators $\tau^{\prime}$ and
$\bar{n}^{\prime}$\ are computed according to Eqs.~(\ref{wscase1})
and~(\ref{wscase2}) with a confidence parameter $w$ as in Eq.~(\ref{wSTVALUE})
[or Eq.~(\ref{wTAIL}) for smaller values of $\varepsilon_{\text{pe}}$]. Now we
further develop the security analysis and derive a formula for the composable
key rate of a coherent-state protocol that is valid under conditions of
stability for the quantum channel (no fading). From this point of view, the
results of this section provides the basic tool for the composable security
analysis of a CV-QKD protocol that is implemented over a stable channel, as
typical in fiber-based implementations.

Assume that the parties exchange $N$ signals over the quantum channel. Because
$m$ are publicly sacrificed for parameter estimation, there are remaining
$n=N-m$ signals to be used for key generation. Besides parameter estimation,
any realistic QKD implementation needs to consider error correction and
privacy amplification, which also come with their own imperfections. First of
all, there is a probability of successful error correction $p_{\text{ec}}$
which is less than $1$, so that only an average of $np_{\text{ec}}$ signals
are processed into a key. This means that final secret-key rate will be
rescaled by the pre-factor
\begin{equation}
r:=\frac{np_{\text{ec}}}{N}=\left(  1-\frac{m}{N}\right)  p_{\text{ec}}.
\label{prefactorRR}%
\end{equation}

Various imperfections arise in the finite-size scenario, which are summarized
in the overall $\varepsilon$-security of the protocol with additive
contributions from parameter estimation, error correction\ and privacy
amplification. Besides $\varepsilon_{\text{pe}}$, the protocol has an
associated $\varepsilon$-correctness $\varepsilon_{\text{cor}}$ (which bounds
the residual probability that the strings are different after passing error
correction) and an associated $\varepsilon$-secrecy $\varepsilon_{\text{sec}}$
(which bounds the distance between the final key and an ideal output
classical-quantum state that is completely decoupled from the eavesdropper).
More technically, one writes $\varepsilon_{\text{sec}}=\varepsilon_{\text{s}%
}+\varepsilon_{\text{h}}$, where $\varepsilon_{\text{s}}$ is a smoothing
parameter and $\varepsilon_{\text{h}}$ is a hashing parameter. All these
parameters are set to be small (e.g., $2^{-33}\simeq10^{-10}$) and provide the
overall security parameter
\begin{equation}
\varepsilon=2p_{\text{ec}}\varepsilon_{\text{pe}}+\varepsilon_{\text{cor}%
}+\varepsilon_{\text{sec}}.
\end{equation}
Note that $p_{\text{ec}}$ explicitly multiplies $\varepsilon_{\text{pe}}$ due
to the fact that error correction occurs after parameter estimation. Also note
the factor $2$ before $\varepsilon_{\text{pe}}$ which accounts for the
estimation of two basic channel parameters.

For a Gaussian-modulated coherent-state protocol~\cite{GG02,Noswitch} with
success probability $p_{\text{ec}}$ and $\varepsilon$-security against
collective (Gaussian) attacks~\cite{collectiveG}, we write the following
composable key rate in terms of secret bits per use of the channel (see
Appendix~\ref{MoreDetailsAPP} for its proof)%
\begin{equation}
R\geq r\left(  R_{\text{pe}}-\frac{\Delta_{\text{aep}}}{\sqrt{n}}+\frac
{\Theta}{n}\right)  , \label{sckeee}%
\end{equation}
where $R_{\text{pe}}$ is given in Eq.~(\ref{Rpe1untrusted}) and%
\begin{align}
&  \Delta_{\text{aep}}:=4\log_{2}\left(  2\sqrt{d}+1\right)  \sqrt{\log
_{2}\left(  \frac{18}{p_{\text{ec}}^{2}\varepsilon_{\text{s}}^{4}}\right)
},\label{deltaAEPPP}\\
&  \Theta:=\log_{2}[p_{\text{ec}}(1-\varepsilon_{\text{s}}^{2}/3)]+2\log
_{2}\sqrt{2}\varepsilon_{\text{h}}, \label{bigOMEGA}%
\end{align}
with $d$ representing the size of the effective alphabet after
analog-to-digital conversion of sender's and receiver's continuous variables
(quadrature encodings and outcomes). Note that one typically chooses a $5$-bit
digitalization ($d=2^{5}=32$), so that there is a negligible discrepancy
between the information quantities computed over discretized and continuous variables.

In ground-based QKD\ experiments, the total number of data points
(signals/uses of the channel) can be of the order of $10^{12}$~\cite{LeoEXP}.
Thus, data points are split in blocks of suitable size for data processing,
typically of the order of $10^{6}-10^{7}$ points. The success probability
$p_{\text{ec}}$ represents the frequency with which a block is successfully
processed into key generation, and this can also be written as $p_{\text{ec}%
}=1-\mathrm{FER}$, where $\mathrm{FER}$ is known as `frame error rate'.

\subsection{Key rate under general coherent attacks\label{SecKEYgeneral}}

The rate in Eq.~(\ref{sckeee}) is derived for collective attacks and, in
particular, collective Gaussian attacks, since the Gaussian assumption is
adopted for parameter estimation. This level of security can be extended to
general coherent attacks under certain symmetries for the\ protocol, which are
satisfied by the no-switching protocol based on the heterodyne
detection~\cite{Noswitch}. In particular, by combining our rate in
Eq.~(\ref{sckeee}) with some of the\ tools from Ref.~\cite{Lev2017}, we derive
a simple formula for the composable finite-size key rate under general attacks.

Suppose that the coherent-state protocol $\mathcal{P}$ is $\varepsilon$-secure
with finite-size rate $R$ under collective Gaussian attacks, and $\mathcal{P}$
can be symmetrized with respect to a Fock-space representation of the group of
unitary matrices. This symmetrization is equivalent to apply an identical
random orthogonal matrix to the classical continuous variables of the two
parties (encodings and outcomes)~\cite{Lev2017}, which is certainly possible
for the heterodyne-based protocol~\cite{Noswitch}. Let us denote by
$\mathcal{\tilde{P}}$ the symmetrized protocol.

Then, let us assume that the remote parties perform an energy test
$\mathcal{T}$ on $m_{\mathrm{et}}$ randomly-chosen pairs of modes. This test
is based on two thresholds, $d_{T}$ for the transmitter, and $d_{R}$ for the
receiver. For each pair, they measure the number of photons in their local
modes and they average these quantities over their $m_{\mathrm{et}}$
measurements, so as to compute the local mean number of photons. If these
energies are below the thresholds, the test is passed (with probability
$p_{\mathrm{et}}$); otherwise the protocol aborts. Now assume that $d_{T}$ is
larger than the mean number of thermal photons $\bar{n}_{T}=(\mu-1)/2$
associated with the average thermal state generated by the transmitter.
Working with $d_{T}\gtrsim\bar{n}_{T}+\mathcal{O}(m_{\mathrm{et}}^{-1/2})$
implies that the test is almost-certainly successful ($p_{\mathrm{et}}\simeq
1$) for sufficiently large values of $m_{\mathrm{et}}$. Also note that, for a
lossy channel with reasonably-small excess noise, the receiver will get an
average number of photons which is clearly less than that of the transmitter,
which means that a successful value for $d_{R}$ can be chosen to be equal to
$d_{T}$. (In our numerical investigations we set $d_{R}=d_{T}\simeq\bar{n}%
_{T}$).


By taking the local dimensions large enough so that $p_{\mathrm{et}}\simeq1$,
the overall success of the protocol remains unchanged, i.e., we have
$p_{\mathrm{ec}}p_{\mathrm{et}}\simeq p_{\mathrm{ec}}$. Then, the parties go
ahead with the symmetrized protocol $\mathcal{\tilde{P}}$ which will now use
$n=N-\tilde{m}$ modes for key generation, where $\tilde{m}:=m+m_{\mathrm{et}}%
$. This already introduces a modification in Eq.~(\ref{sckeee}), where the
effective number $n$ of modes for key generation will be reduced in the rate,
so that the prefactor of Eq.~(\ref{prefactorRR}) becomes%
\begin{equation}
r=\left(  1-\frac{\tilde{m}}{N}\right)  p_{\text{ec}}.
\end{equation}
By setting $m_{\mathrm{et}}=f_{\mathrm{et}}n$ for some factor $f_{\mathrm{et}%
}<1$, the total number of key generation signals takes the form
\begin{equation}
n=\frac{N-m}{1+f_{\mathrm{et}}}. \label{ncomputeTOT}%
\end{equation}

The second modification consists of an additional step of privacy
amplification which reduces the final number of secret key bits by the
following amount~\cite{Lev2017}
\begin{equation}
\Phi_{n}:=2\left\lceil \log_{2}\binom{K_{n}+4}{4}\right\rceil , \label{FInGEN}%
\end{equation}
where%
\begin{align}
K_{n}  &  =\max\left\{  1,n(d_{T}+d_{R})\Sigma_{n}\right\}  ,\label{KnGEN}\\
\Sigma_{n}  &  :=\frac{1+2\sqrt{\frac{\ln(8/\varepsilon)}{2n}}+\frac
{\ln(8/\varepsilon)}{n}}{1-2\sqrt{\frac{\ln(8/\varepsilon)}{2f_{\mathrm{et}}%
n}}}.
\end{align}

Accounting for the two modifications above, we have that the key rate
$R^{\text{het}}$ of Eq.~(\ref{sckeee}), specified for the heterodyne
protocol~\cite{Noswitch}, becomes the following%
\begin{equation}
R_{\text{gen}}^{\text{het}}\geq r\left[  R_{\text{pe}}^{\text{het}}%
-\frac{\Delta_{\text{aep}}}{\sqrt{n}}+\frac{\Theta-\Phi_{n}}{n}\right]  ,
\label{eqgenn}%
\end{equation}
where $R_{\text{pe}}^{\text{het}}$ is $R_{\text{pe}}$ of
Eq.~(\ref{Rpe1untrusted}) for the heterodyne protocol.

The rate established in Eq.~(\ref{eqgenn}) is valid for a symmetrized
coherent-state protocol $\mathcal{\tilde{P}}$ with heterodyne
detection~\cite{Noswitch} which is now secure against general coherent
attacks, with modified epsilon security equal to~\cite{Lev2017}
\begin{equation}
\varepsilon^{\prime}=K_{n}^{4}\varepsilon/50, \label{epsPRIMO}%
\end{equation}
and\ probability of success $p_{\mathrm{ec}}\simeq p_{\mathrm{ec}%
}p_{\mathrm{et}}$. Note that, because $K_{n}\simeq\mathcal{O}(n)$, we need to
start with a very small value for $\varepsilon$, so that the final
epsilon-security $\varepsilon^{\prime}$ remains well below $1$ and the term
$\Phi_{n}$ in Eq.~(\ref{eqgenn}) does not explode. In particular, this means
that $\varepsilon_{\text{pe}}$ needs to be very small (e.g., $\simeq10^{-43}$)
and the corresponding confidence parameter $w$ must be computed from
Eq.~(\ref{wTAIL}).

\section{Composable security and key rates for free-space
CV-QKD\label{Sec_freeCVQKD}}

\subsection{Preliminary considerations}

Here we extend the previous theory (Sec.~\ref{Sec_stableCVQKD}) to account for
the channel fluctuations that generally affect free-space quantum
communications. We consider free-space fading where the transmissivity $\tau$
is not stable but varies over a time-scale of the order of $100$ms or similar.
Because of this issue, the first important physical condition is that the
setups need to have system clocks and detectors that are suitably fast to
collect enough statistics while the value of $\tau$ fluctuates.

In a general fading process the instantaneous transmissivity $\tau$ between
transmitter and receiver follows a probability distribution $P_{0}(\tau)$,
which takes the specific expression in Eq.~(\ref{P0tau}) when the physical
aspects of the free-space communication are taken into account. The
probability that $\tau$ falls in a small interval $[\tau,\tau+\delta\tau]$ is
given by $p_{\delta}=p(\tau,\tau+\delta\tau)$, where we define
\begin{equation}
p(\tau_{1},\tau_{2}):=\int_{\tau_{1}}^{\tau_{2}}d\tau~P_{0}(\tau).
\label{pt1t2}%
\end{equation}
This means that only a small fraction $p_{\delta}m$ of the signals can be used
for estimating this value of $\tau$. (From now on, when we write a
post-selected quantity like $p_{\delta}m$, we implicitly mean an integer
approximation of it).

As we can see from Eq.~(\ref{sigmatVAL}),\ the error-variance $\sigma_{\tau
}^{2}$ in the estimation of $\tau$\ scales as $\mathcal{O}(m^{-1})$. Here this
becomes $\mathcal{O}[(p_{\delta}m)^{-1}]$, with the problem of leading to
insufficient statistics. We can overcome this issue by introducing energetic
pilot pulses, specifically dedicated to track the instantaneous transmissivity
of the channel, so that we can create suitable bins for collecting signals
with almost equal transmissivity. These bins are then subject to a suitable
post-processing that we call \textquotedblleft de-fading\textquotedblright.

Another preliminary consideration is about noise filtering. As already
mentioned in Sec.~\ref{NoiseFilterSubsection}, one can effectively narrow the
frequency filter of the receiver to match the bandwidth of the LO, thanks to
the interferometric process occurring in the homodyne/heterodyne setup. Thus,
instead of being limited to a physical filter of $1~$nm around $800$~nm at the
receiver's aperture, the detector imposes a much narrower filter of $0.1~$pm,
by interfering the signal with the $10$~ns-long and $50~$GHz-wide pulse of the
LO, close to the time-bandwidth product. Such a process is secure as long as
the projection of the homodyne detectors does not create correlations with the
frequencies outside the bandwidth of the LO, since these extra frequencies
could be used as Trojan-horse modes. In realistic implementations, such a
cross-talk is/can be made negligible. As a result, thanks to the use of the LO
(as TLO or LLO) the parties are able to suppress the external background noise
(down to $\bar{n}_{B}\simeq10^{-7}$ in day-light conditions with typical
parameters). For this reason, one can make the numerical approximation
\begin{equation}
\bar{n}_{B}\ll1,~\bar{n}\simeq\bar{n}_{\text{ex}}. \label{nBnex}%
\end{equation}

\subsection{Loss tracking via random pilots}


For free-space parameter estimation, the parties sacrifice not only $m$ signal
pulses (as before in Sec.~\ref{Sec_stableCVQKD}), but also additional
$m_{\text{P}}$ energetic pilot pulses. The $m_{\text{P}}$ pilots are
specifically used for the quasi-perfect estimation of the (generally-variable)
transmissivity $\tau$, so as to track its instantaneous value. In this way,
the parties can create a lattice of suitably-narrow bins of transmissivity for
signal classification (discussed in the next subsection).

The pilots are prepared in exactly the same coherent state $|\bar{n}%
_{\text{P}}^{1/2}e^{i\pi/4}\rangle$ and randomly transmitted during the
quantum communication. In a TLO setup, both signals and pilots are multiplexed
with their LOs. As previously discussed, the LO\ can be very bright, with mean
number of photons $\bar{n}_{\text{LO}}$\ of the order $10^{7}$ at the receiver
even after $20$dB of loss (this is for $10~$ns-long pulses from a $100$~mW
laser at $\lambda=800~$nm). This means that relatively-energetic pilots can be
generated with just a $10^{-4}$ fraction of the LO energy (so that $\bar
{n}_{\text{P}}\simeq10^{3}$ photons are collected by the receiver). In this
way, the pilots are bright enough to provide an excellent estimate of $\tau$,
while the LO remains so much brighter that the measurements of the pilots will
still be shot-noise limited. In an LLO setup, the reference pulses for the
local LO reconstruction are transmitted at the odd uses of the channel, while
the pilots are randomly interleaved with the signals at the even uses of the channel.

In a small fading interval $\delta\tau$, we have $p_{\delta}m_{\text{P}}$
pilots to be used for the estimation of $\tau$. From these pilots, the parties
derive $p_{\delta}m_{\text{P}}\nu_{\text{det}}$ pairs $\{x_{i},y_{i}\}$ of
sampling variables $x_{i}=\sqrt{2\bar{n}_{\text{P}}}$ and $y_{i}=\sqrt{\tau
}x_{i}+z_{i}$. They then build the estimator%
\begin{equation}
\hat{T}_{\text{P}}:=\frac{1}{p_{\delta}m_{\text{P}}\nu_{\text{det}}}\sum
_{i}\frac{y_{i}}{x_{i}},
\end{equation}
with mean $\sqrt{\tau}$ and variance $\sigma_{z}^{2}/(2\bar{n}_{\text{P}%
}p_{\delta}m_{\text{P}}\nu_{\text{det}})$. The latter variance goes to zero
for suitably large $\bar{n}_{\text{P}}$, so that the parties achieve a
practically-perfect estimate of $\tau$ already for $m_{\text{P}}%
\simeq\mathcal{O}(1)$. In other words, we may consider $\hat{T}_{\text{P}%
}=\sqrt{\tau}$, meaning that the parties can perform real-time tracking of the
transmissivity $\tau$ with negligible error.

\subsection{Post-selection interval and lattice allocation}

While monitoring the transmissivity $\tau$ with the pilots, the parties only
keep the data points exchanged within an agreed post-selection\ interval
$\Delta:=[\tau_{\text{min}},\tau_{\text{max}}]$, with associated probability
$p_{\Delta}=p(\tau_{\text{min}},\tau_{\text{max}})$ as computed from
Eq.~(\ref{pt1t2}). Thus, from a total of $N$ exchanged pulses, only a portion
$S_{\Delta}:=(N-m_{\text{P}})p_{\Delta}$ of signals is selected for further
processing. The interval is chosen so that $S_{\Delta}\gg1$, leading to
sufficient statistics for parameter estimation. The parties may choose
$\tau_{\text{max}}=\eta:=\eta_{\text{st}}\eta_{\text{atm}}\eta_{\text{eff}}$,
which is the maximum value achievable by a perfectly-aligned beam, and then
take $\tau_{\text{min}}=f_{\text{th}}\eta$ for a threshold value
$f_{\text{th}}\in(0,1)$.

Within the post-selection interval $\Delta$, Alice and Bob introduce a regular
lattice with step $\delta\tau$, so that there are a number of transmissivity
slots/bins $\Delta_{k}:=[\tau_{k},\tau_{k+1}]$ with $\tau_{k}:=\tau
_{\text{min}}+(k-1)\delta\tau$, for $k=1,\ldots,M$ and $M=(\tau_{\text{max}%
}-\tau_{\text{min}})/\delta\tau$. In this coarse graining of the
transmissivity, each slot $\Delta_{k}$ is populated with probability
$p_{k}=p(\tau_{k},\tau_{k+1})$ according to the fading distribution in
Eq.~(\ref{pt1t2}). This means that slot $\Delta_{k}$ has $S_{k}%
:=(N-m_{\text{P}})p_{k}$ signals to be used for parameter estimation and key
generation. For a sufficiently narrow slot, these signals provide
$\nu_{\text{det}}S_{k}$ pairs of points $\{x_{i},y_{i}\}$ that satisfy the
input-output relation
\begin{equation}
y^{k}\simeq\sqrt{\tau_{k}}x+z^{k}, \label{kslotmap}%
\end{equation}
where $z^{k}=z(\tau_{k})$ is a noise variable [cf. Eq.~(\ref{IOnew})] with
variance%
\begin{equation}
\sigma_{z}^{2}(\tau_{k}):=\mathrm{var}(z^{k})=2\bar{n}(\tau_{k})+\nu
_{\text{det}}.
\end{equation}

A potential strategy consists of processing each slot $\Delta_{k}$
independently from the others, by performing parameter estimation over a
corresponding set of sacrificed signals, and then going through the next steps
of data processing. This approach is based on the fact that we can consider
the transmissivity $\tau_{k}$ and the noise-variance $\sigma_{z}^{2}(\tau
_{k})$ to be approximately constant for all data points in the same slot (so
that there is a well-defined thermal-loss channel associated with it). In turn
this means that we can directly apply the procedures of
Sec.~\ref{Sec_stableCVQKD} valid for a stable quantum channel. As a result,
each slot $\Delta_{k}$ will provide a slot-rate $R_{k}$ with corresponding
epsilon security $\varepsilon_{k}$. The total finite-size key rate of the link
is the average of $R_{k}$ over the slots, i.e.,%
\begin{equation}
\bar{R}=\sum_{k=1}^{M}p_{k}\max\{0,R_{k}\}, \label{FiniteSizeAVVV}%
\end{equation}
with total security $\varepsilon=\sum_{k=1}^{M}p_{k}\varepsilon_{k}$. Because
this solution may suffer from insufficient statistics in the various slots, we
adopt the procedure of the following subsection.

\subsection{De-fading\label{defadingSEC}}

The parties can process their data in order to eliminate/reduce the fading and
create an overall stable channel at the cost of using the minimum
transmissivity within the post-selection interval. This procedure of de-fading
is one of the possible strategies and is used to provide an achievable lower
bound for the secret key rate.

\begin{figure}[t]
\vspace{-1.7cm}
\par
\begin{center}
\includegraphics[width=1\columnwidth] {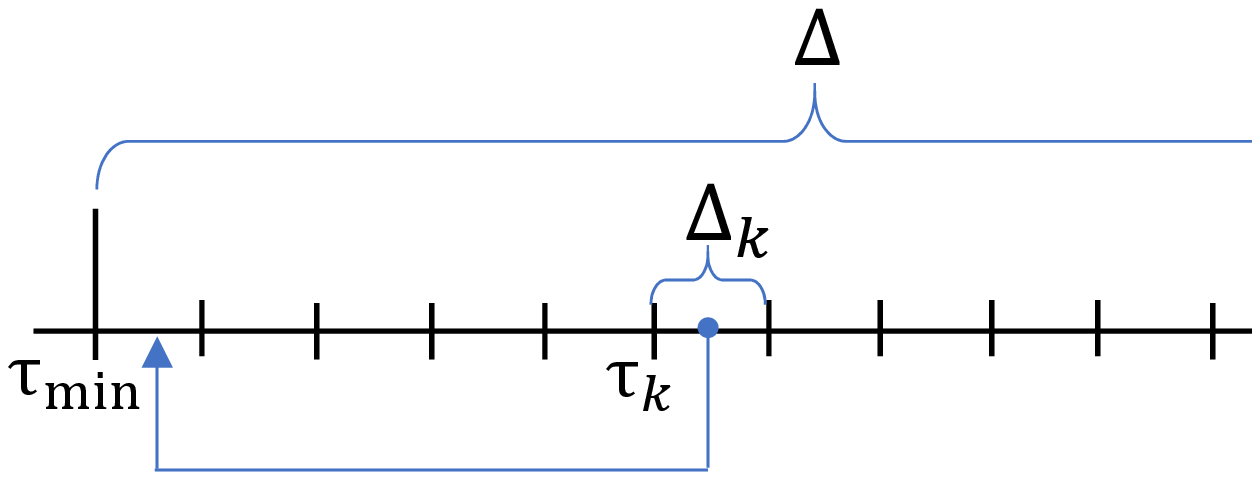}
\end{center}
\par
\vspace{-2.2cm}\caption{De-fading of data. See text for details}%
\label{defaaa}%
\end{figure}

In this procedure, Bob maps all his $\nu_{\text{det}}S_{k}$ data points
$y^{k}$ from the generic $k^{\text{th}}$ slot $\Delta_{k}$ to the first slot
$\Delta_{1}$ in the post-selection interval, by using the following
\textquotedblleft downlift\textquotedblright\ transformation
\begin{equation}
y^{k}\rightarrow y^{\prime k}:=\sqrt{\frac{\tau_{\text{min}}}{\tau_{k}}}%
y^{k}=\sqrt{\tau_{\text{min}}}x+z^{\prime k}, \label{defadingIO}%
\end{equation}
where $z^{\prime k}$ is a Gaussian noise variable with variance \textrm{var}%
$(z^{\prime k})=\sigma_{z}^{2}(\tau_{k})\tau_{\text{min}}/\tau_{k}$. See also
Fig.~\ref{defaaa}.

While Eq.~(\ref{defadingIO}) is certainly a valid post-processing of data, it
is not guaranteed that the entire input-output transformation $x\rightarrow
y^{\prime k}$\ can be made equivalent to the action of a quantum channel
(which is a useful condition for our theoretical treatment). This is due to
the noise reduction induced by the re-scaling $\tau_{\text{min}}/\tau_{k}%
\leq1$, so that \textrm{var}$(z^{\prime k})$ might become $<\nu_{\text{det}}$,
which is the minimum noise associated with the final quantum measurement.

This problem is solved if Bob applies a classical Gaussian channel $y^{\prime
k}\rightarrow y^{\prime\prime k}:=y^{\prime k}+\xi_{\text{add}}^{k}$ with
additive noise $\mathrm{var}(\xi_{\text{add}}^{k})=(1-\tau_{\text{min}}%
/\tau_{k})\nu_{\text{det}}$. In this way, Bob generates%
\begin{equation}
y^{\prime\prime k}=\sqrt{\tau_{\text{min}}}x+z^{\prime\prime k},
\end{equation}
where $z^{\prime\prime k}$ is a Gaussian variable with variance
\begin{equation}
\sigma_{k}^{2}:=\mathrm{var}(z^{\prime\prime k})=2\bar{n}(\tau_{k}%
)\tau_{\text{min}}/\tau_{k}+\nu_{\text{det}}\geq\nu_{\text{det}}.
\end{equation}

We see that the transformation $y^{k}\rightarrow y^{\prime\prime k}$ is a
slot-dependent beam-splitter channel $\mathcal{C}_{k}$ performed over the
data, with transmissivity $\iota_{k}:=\tau_{\text{min}}/\tau_{k}$ and
environmental noise-variance equal to $\nu_{\text{det}}$. Equivalently, this
can be represented by a virtual beam splitter directly applied to the pulses
allocated to slot $\Delta_{k}$\ followed by the measurement. In other words,
Alice and Bob's input-output relation $x\rightarrow y^{\prime\prime k}$ is
equivalent to the action of a composite Gaussian channel $\mathcal{F}%
_{k}:=\mathcal{C}_{k}\circ\mathcal{E}_{k}$, where $\mathcal{E}_{k}$ is a
thermal-loss channel with transmissivity $\tau_{k}$ and thermal number
$\bar{n}(\tau_{k})$, followed by Bob's measurement.

Assuming that the transformation $y^{k}\rightarrow y^{\prime\prime k}$ is
performed for all the $M$ slots of the interval, Bob creates a new variable
$y^{\prime\prime}$ which satisfies%
\begin{equation}
y^{\prime\prime}=\sqrt{\tau_{\text{min}}}x+z^{\prime\prime}, \label{zprimePB2}%
\end{equation}
where $z^{\prime\prime}$ is non-Gaussian. Since $x\rightarrow y\rightarrow
y^{\prime\prime}$ is a Markov chain, Bob's post-processing can only decrease
the mutual information $I(x:y^{\prime\prime})\leq I(x:y)$. The noise variable
$z^{\prime\prime}$ can be written as an ensemble $\{\pi_{k},z^{\prime\prime
k}\}$, with the independent element $z^{\prime\prime k}$ being selected with
probability $\pi_{k}=p_{k}/p_{\Delta}$. Thus, it has zero mean and variance
\begin{equation}
\sigma_{z^{\prime\prime}}^{2}=\sum\limits_{k=1}^{M}\pi_{k}\sigma_{k}^{2}%
=\nu_{\text{det}}+2\tau_{\text{min}}\sum\limits_{k=1}^{M}\frac{\pi_{k}}%
{\tau_{k}}\bar{n}(\tau_{k}). \label{sigmazprimeN}%
\end{equation}
Overall, the transformation of Eq.~(\ref{zprimePB2}) is equivalently obtained
by measuring the output of a non-Gaussian channel $\mathcal{F}$, which is
described by the ensemble $\{\pi_{k},\mathcal{F}_{k}\}$ and assumed to be
completely controlled by Eve.

Due to the optimality of collective Gaussian attacks for Gaussian-modulated
coherent-state protocols, the parties may assume the worst-case scenario where
the non-Gaussian channel $\mathcal{F}$ is replaced by a thermal-loss Gaussian
channel $\mathcal{E}_{\tau_{\text{min}},\bar{n}_{\text{G}}}$ with the same
transmissivity $\tau_{\text{min}}$ and thermal number
\begin{equation}
\bar{n}_{\text{G}}=\tau_{\text{min}}\sum\limits_{k=1}^{M}\frac{\pi_{k}}%
{\tau_{k}}\bar{n}(\tau_{k}),
\end{equation}
so that it has noise variance $\sigma_{\text{G}}^{2}=2\bar{n}_{\text{G}}%
+\nu_{\text{det}}$ equal to $\sigma_{z^{\prime\prime}}^{2}$ of
Eq.~(\ref{sigmazprimeN}). This means that the noise variable $z^{\prime\prime
}$ in Eq.~(\ref{zprimePB2}) can be replaced by a Gaussian variable
$z_{\text{G}}$, and the total input-output relation is assumed to be%
\begin{equation}
y^{\prime\prime}=\sqrt{\tau_{\text{min}}}x+z_{\text{G}}. \label{Gaussdata}%
\end{equation}

Thus, we lower-bound Alice and Bob's performance by considering the
post-processed variables $\{x,y^{\prime\prime}\}$ connected by the
input-output relation of Eq.~(\ref{Gaussdata}), after de-fading and assuming a
Gaussian attack (`Gaussianification'). This leads to the asymptotic key rate%
\begin{equation}
R_{\text{asy}}=\beta I(x:y^{\prime\prime})_{\tau_{\text{min}},\bar
{n}_{\text{G}}}-\chi(\mathbf{E}:y^{\prime\prime})_{\tau_{\text{min}},\bar
{n}_{\text{G}}}, \label{asyFSP}%
\end{equation}
which can be computed from Eq.~(\ref{rateASYm}). The explicit expressions for
the mutual information $I$ and the Holevo bound $\chi$ are given in
Secs.~\ref{descriptionPROTsec} and~\ref{subsectionAKR}, for the homodyne
($R_{\text{asy}}^{\text{hom}}$) and heterodyne protocol ($R_{\text{asy}%
}^{\text{het}}$)~\cite{NoteBETA}.

Because we have reduced the fading process to a stable thermal-loss channel
$\mathcal{E}_{\tau_{\text{min}},\bar{n}_{\text{G}}}$, we can exploit the
methodology of Sec.~\ref{Sec_stableCVQKD}. In particular, we can apply the
tools of Sec.~\ref{PEmainSEC} to compute the estimators/worst-case estimators
for $\tau_{\text{min}}$ and $\bar{n}_{\text{G}}$, to be employed in the key rate.

\subsection{Estimating the channel parameters}

In the parameter estimation step, Alice and Bob sacrifice some of their
signals in order to estimate the actual values of the minimum transmissivity
$\tau_{\min}=T_{\text{min}}^{2}$ and the Gaussian noise $\sigma_{\text{G}}%
^{2}$ (or $\bar{n}_{\text{G}}$) up to an acceptable error probability. Note
that, in general, the actual value of $\tau_{\min}$ might be different from
what determined via the pilots, so that its estimation via the signals is
needed. In fact, Eve might try to use a QND measurement to distinguish between
pilots and signals. After such QND measurement (with loss $\tau_{k}$), Eve may
apply an additional measurement (with loss $\tilde{\tau}$) only to the
signals. This means that, after de-fading, the input-output relation of
Eq.~(\ref{Gaussdata}) would become $\tilde{y}^{\prime\prime}=\sqrt{\tilde
{\tau}_{\text{min}}}x+\tilde{z}_{\text{G}}$, with lower transmissivity
$\tilde{\tau}_{\text{min}}:=\tilde{\tau}\tau_{\text{min}}$ and generally
higher noise $\tilde{z}_{\text{G}}$.

Because parameter estimation is performed over a subset of the signals, the
parties will detect these discrepancies with respect to the pilots. Most
importantly, they will derive the corresponding estimators for the lower
transmissivity $\tilde{\tau}_{\text{min}}$ and the different noise level, to
be used in the calculation of their secret key. Of course, Eve might be more
disruptive over the signals so that their transmissivity might be sensibly
different from that of the corresponding pilots, but the point is that any
such a perturbation will be anyway detected/estimated by the parties. If the
discrepancy between pilots and signals is too strong, the noise level detected
by the parties becomes too high for secure communication (denial of service).
In the following, we make the realistic assumption that Eve acts universally
over pilots and signals, so that $\tilde{\tau}_{\text{min}}=\tau_{\text{min}}%
$. However, we point out that this is only a simplification, not a limitation
of the approach whose application to $\tilde{\tau}_{\text{min}}\neq
\tau_{\text{min}}$ is immediate.

In order to create their estimators, the parties sacrifice $mp_{\Delta}$
signals from those they have post-selected. This corresponds to $m_{\Delta
}:=\nu_{\text{det}}mp_{\Delta}$ pairs of data points $\{x,y^{\prime\prime}\}$,
and we can also write $m_{\Delta}=\sum\nolimits_{k=1}^{M}m_{k}$, where
$m_{k}:=\nu_{\text{det}}mp_{k}$ is the contribution coming from the generic
slot $\Delta_{k}$. In writing $mp_{\Delta}$, we implicitly assume that $m$ is
the equivalent number of signals that would have been sacrificed by the
parties in the absence of post-selection. This notation is theoretical useful
to describe scenarios where the same protocol (with fixed $m$) is implemented
over different distances over which the value of $p_{\Delta}$ can be optimized.

For the square-root transmissivity, Alice and Bob build the estimator%
\begin{equation}
\hat{T}_{\text{min}}:=\left.  \sum\limits_{i=1}^{m_{\Delta}}x_{i}y_{i}%
^{\prime\prime}\right/  \sum\limits_{i=1}^{m_{\Delta}}x_{i}^{2}\simeq\frac
{1}{m_{\Delta}\sigma_{x}^{2}}\sum\limits_{i=1}^{m_{\Delta}}x_{i}y_{i}%
^{\prime\prime}.
\end{equation}
It is easy to check that this is unbiased (i.e., its mean is $\simeq
T_{\text{min}}$) and its variance is given by%
\begin{align}
\mathrm{var}(\hat{T}_{\text{min}})  &  \simeq\frac{1}{m_{\Delta}^{2}\sigma
_{x}^{4}}\sum\limits_{i=1}^{m_{\Delta}}\mathrm{var}(x_{i}y_{i}^{\prime\prime
})\\
&  =\frac{1}{m_{\Delta}^{2}\sigma_{x}^{4}}\sum\limits_{k=1}^{M}\sum
\limits_{i_{k}=1}^{m_{k}}\mathrm{var}(x_{i_{k}}y_{i_{k}}^{\prime\prime k})\\
&  \simeq\frac{1}{m_{\Delta}^{2}\sigma_{x}^{4}}\sum\limits_{k=1}^{M}%
m_{k}\mathrm{var}(xy^{\prime\prime k})\\
&  =\frac{1}{m_{\Delta}^{2}\sigma_{x}^{4}}\sum\limits_{k=1}^{M}m_{k}%
(2\tau_{\text{min}}\sigma_{x}^{4}+\sigma_{x}^{2}\sigma_{k}^{2})\\
&  =\frac{1}{m_{\Delta}}\left(  2\tau_{\text{min}}+\frac{1}{m_{\Delta}%
\sigma_{x}^{2}}\sum\limits_{k=1}^{M}m_{k}\sigma_{k}^{2}\right) \\
&  =\frac{2\tau_{\text{min}}+\sigma_{\text{G}}^{2}/\sigma_{x}^{2}}{m_{\Delta}%
}. \label{finaljj}%
\end{align}

Let us build an estimator for the variance $\sigma_{\text{G}}^{2}$ of the
thermal noise $z_{\text{G}}$. This is given by%
\begin{align}
\widehat{\sigma_{\text{G}}^{2}}  &  :=\frac{1}{m_{\Delta}}\sum\limits_{i=1}%
^{m_{\Delta}}\left(  y_{i}^{\prime\prime}-\hat{T}_{\text{min}}x_{i}\right)
^{2}\\
&  \simeq\frac{1}{m_{\Delta}}\sum\limits_{k=1}^{M}\sum\limits_{i_{k}=1}%
^{m_{k}}\left(  y_{i_{k}}^{\prime\prime k}-T_{\text{min}}x_{i_{k}}\right)
^{2}\\
&  =\frac{1}{m_{\Delta}}\sum\limits_{k=1}^{M}\sigma_{k}^{2}Y_{k},~~Y_{k}%
:=\sum\limits_{i_{k}=1}^{m_{k}}\frac{(z_{i_{k}}^{\prime\prime k})^{2}}%
{\sigma_{k}^{2}},
\end{align}
where $Y_{k}$ is distributed according to a $\chi^{2}$ distribution with
$m_{k}$ degrees of freedom. It is easy to check that the estimator is
unbiased, i.e., we have
\begin{equation}
\langle\widehat{\sigma_{\text{G}}^{2}}\rangle=\frac{1}{m_{\Delta}}%
\sum\limits_{k=1}^{M}\sigma_{k}^{2}\langle Y_{k}\rangle\simeq\sigma_{\text{G}%
}^{2}. \label{nnbb1}%
\end{equation}
Then, for the variance we compute%
\begin{equation}
\mathrm{var}(\widehat{\sigma_{\text{G}}^{2}})=\frac{1}{m_{\Delta}^{2}}%
\sum\limits_{k=1}^{M}\sigma_{k}^{4}\mathrm{var}(Y_{k})\simeq\frac{2}%
{m_{\Delta}^{2}}\sum\limits_{k=1}^{M}m_{k}\sigma_{k}^{4}. \label{nnbb2}%
\end{equation}
Equivalently, in terms of number of thermal photons $\bar{n}_{\text{G}%
}:=(\sigma_{\text{G}}^{2}-\nu_{\text{det}})/2$, we write the estimator%
\begin{equation}
\widehat{\bar{n}_{\text{G}}}:=(\widehat{\sigma_{\text{G}}^{2}}-\nu
_{\text{det}})/2,
\end{equation}
which is unbiased $\langle\widehat{\bar{n}_{\text{G}}}\rangle\simeq\bar
{n}_{\text{G}}$ with $\mathrm{var}(\widehat{\bar{n}_{\text{G}}})=\mathrm{var}%
(\widehat{\sigma_{\text{G}}^{2}})/4$.

It is important to note that all the mean values and variances above are
computable by the parties by replacing estimators in the right-hand sides of
the formulas. In fact, once $\hat{T}_{\text{min}}$ and $\widehat
{\sigma_{\text{G}}^{2}}$ have been computed, these can be replaced in
Eq.~(\ref{finaljj}) to provide $\mathrm{var}(\hat{T}_{\text{min}})$. To
compute $\mathrm{var}(\widehat{\sigma_{\text{G}}^{2}})$, the parties need to
derive estimators of $\sigma_{k}^{2}$, i.e.,
\begin{equation}
\widehat{\sigma_{k}^{2}}:=\frac{1}{m_{k}}\sum\limits_{i_{k}=1}^{m_{k}}\left(
y_{i_{k}}^{\prime\prime k}-\hat{T}_{\text{min}}x_{i_{k}}\right)  ^{2},
\end{equation}
whose squares go in Eq.~(\ref{nnbb2}).

\subsection{Worst-case estimators and bounds}

According to Eq.~(\ref{Gaussdata}), Alice and Bob's post-processed data is
generated by a thermal-loss channel $\mathcal{E}_{\tau_{\text{min}},\bar
{n}_{\text{G}}}$ with transmissivity $\tau_{\text{min}}$ and thermal number
$\bar{n}_{\text{G}}$. For the transmissivity and the thermal number, we write
the worst-case estimators
\begin{align}
\tau_{\text{min}}^{\prime}  &  :=\left[  \hat{T}_{\text{min}}-w~\sqrt
{\mathrm{var}(\hat{T}_{\text{min}})}\right]  ^{2}\nonumber\\
&  \simeq\hat{T}_{\text{min}}^{2}-2w\hat{T}_{\text{min}}\sqrt{\mathrm{var}%
(\hat{T}_{\text{min}})}+\mathcal{O}(m_{\Delta}^{-1}),\label{LBpara}\\
\bar{n}_{\text{G}}^{\prime}  &  :=\widehat{\bar{n}_{\text{G}}}+w~\sqrt
{\mathrm{var}(\widehat{\bar{n}_{\text{G}}})}, \label{nGestimator}%
\end{align}
where the confidence parameter $w$ is connected to the error $\varepsilon
_{\text{pe}}$ according to Eq.~(\ref{wSTVALUE}) or Eq.~(\ref{wTAIL}).

For the sake of the theoretical analysis, it is useful to introduce bounds for
$\tau_{\text{min}}^{\prime}$\ and $\bar{n}_{\text{G}}^{\prime}$. Consider the
worst-case noise variance $\sigma_{\text{wc}}^{2}=2\bar{n}_{\text{wc}}%
+\nu_{\text{det}}$ such that $\sigma_{\text{wc}}^{2}\geq\sigma_{k}^{2}$ for
any slot $k$. Then, we may write%
\begin{align}
\mathrm{var}(\hat{T}_{\text{min}})  &  \lesssim\frac{2\tau_{\text{min}}%
+\sigma_{\text{wc}}^{2}/\sigma_{x}^{2}}{m_{\Delta}},\label{nnbb0}\\
\langle\widehat{\bar{n}_{\text{G}}}\rangle &  \lesssim\bar{n}_{\text{wc}%
},~\mathrm{var}(\widehat{\bar{n}_{\text{G}}})\lesssim\frac{(2\bar
{n}_{\text{wc}}+\nu_{\text{det}})^{2}}{2m_{\Delta}}.
\end{align}
As a result, we have the bounds%
\begin{align}
\tau_{\text{min}}^{\prime}  &  \gtrsim\tau_{\text{LB}}:=\tau_{\text{min}%
}-2w\sqrt{\frac{2\tau_{\text{min}}^{2}+\tau_{\text{min}}\sigma_{\text{wc}}%
^{2}/\sigma_{x}^{2}}{m_{\Delta}}},\label{bbb1}\\
\bar{n}_{\text{G}}^{\prime}  &  \lesssim\bar{n}_{\text{UB}}:=\bar
{n}_{\text{wc}}+w\frac{2\bar{n}_{\text{wc}}+\nu_{\text{det}}}{\sqrt
{2m_{\Delta}}}. \label{bbb2}%
\end{align}

Let us now evaluate the worst-case thermal number $\bar{n}_{\text{wc}}$ to be
used in the bounds above. We write
\begin{equation}
\bar{n}_{\text{wc}}=\eta_{\text{eff}}\bar{n}_{B}+\bar{n}_{\text{ex,wc}%
},\label{UB1e}%
\end{equation}
where $\bar{n}_{\text{ex,wc}}:=\bar{n}_{\text{ex}}(\tau_{\text{wc}})\geq
\bar{n}_{\text{ex}}(\tau)$ is computed over the worst-case value$\ \tau
_{\text{wc}}$. The latter may be chosen to be$\ \tau_{\text{wc}}%
=\tau_{\text{min}}$ for the TLO and $\tau_{\text{wc}}=\tau_{\text{max}}$ for
the LLO [due to the fact that $\bar{n}_{\text{ex}}(\tau)$ has different
monotonicity in $\tau$, as discussed in Sec.~\ref{subSEC_options}]. In other
words, for $\bar{n}_{\text{ex,wc}}$, we may consider the two estimates%
\begin{align}
\bar{n}_{\text{ex,wc}}^{\text{TLO}} &  \simeq\Theta_{\text{el}}/\tau
_{\text{min}},\label{exwc}\\
\bar{n}_{\text{ex,wc}}^{\text{LLO}} &  \simeq\Theta_{\text{el}}+\pi
\tau_{\text{max}}\sigma_{x}^{2}C^{-1}l_{\text{W}},\label{exwc2}%
\end{align}
where $\Theta_{\text{el}}$ is the electronic noise term in
Eq.~(\ref{eleCONTRIBUT}).

In our numerical investigations, we assume the bounds $\tau_{\text{LB}}$ and
$\bar{n}_{\text{UB}}$ in Eqs.~(\ref{bbb1}) and~(\ref{bbb2}), which take
different expressions for TLO\ and LLO depending on Eqs.~(\ref{exwc})
and~(\ref{exwc2}). Since each of these worst-case estimators is correct up to
an error $\varepsilon_{\text{pe}}$, the total error affecting the procedure of
parameter estimation is $\simeq2\varepsilon_{\text{pe}}$.

\subsection{Composable key rate for free-space CV-QKD}

Let us summarize the scenario. Alice and Bob perform a Gaussian-modulated
coherent-state (homodyne or heterodyne) protocol with variance $\sigma_{x}%
^{2}=\mu-1$ over a free-space channel with instantaneous transmissivity
$\eta_{\text{ch}}$ and background thermal noise $\bar{n}_{B}$. The receiver
has setup efficiency $\eta_{\text{eff}}$\ and setup noise $\bar{n}_{\text{ex}%
}$, so that the total thermal noise is $\bar{n}=\eta_{\text{eff}}\bar{n}%
_{B}+\bar{n}_{\text{ex}}$. The overall instantaneous transmissivity from Alice
to Bob is given by $\tau=\eta_{\text{ch}}\eta_{\text{eff}}$ and it fluctuates
following a fading distribution $P_{0}(\tau)$ as in Eq.~(\ref{P0tau}). Because
$\bar{n}_{\text{ex}}=\bar{n}_{\text{ex}}(\tau)$ [see Sec.~\ref{subSEC_options}%
], we also have thermal-noise fluctuations $\bar{n}=\bar{n}(\tau)$. The
physical scenario is depicted in Fig.~\ref{detectorPIC},\ and also modelled in
Fig.~\ref{setupandWC} for each fixed value of the transmissivity.

Alice sends to Bob a total of $N$ pulses which are multiplexed with an LO in
polarization (TLO) or in time (LLO). Note that in terms of throughput
(bits/sec), given by the rate (bits/use) times the clock $C$ (uses/sec), one
should account for the additional uses of the link associated with the LO.
Thus, there is a factor of $1/2\ $for the LLO, unless this is compensated by
using two polarizations for the quantum signals (see Sec.~\ref{LLOTLO}).

Within the total set of $N$ pulses, there are $m_{\text{P}}$ pilots that are
prepared in a bright coherent state and are randomly interleaved with the
$N-m_{\text{P}}$ signal pulses. Thanks to these pilots, the parties monitor
the instantaneous transmissivity $\tau$ and they create a post-selection
interval $\Delta:=[\tau_{\min},\tau_{\max}]$, where $\tau_{\text{max}}%
=\eta:=\eta_{\text{st}}\eta_{\text{atm}}\eta_{\text{eff}}$ is the maximum
value achievable and $\tau_{\text{min}}=f_{\text{th}}\eta$ for some threshold
value $f_{\text{th}}\in(0,1)$. The interval $\Delta$ post-selects a portion
$S_{\Delta}=(N-m_{\text{P}})p_{\Delta}$ of the signals, where the probability
$p_{\Delta}=p(\tau_{\text{min}},\tau_{\text{max}})$ is given in
Eq.~(\ref{pt1t2}). Then, the interval is further divided into a lattice of $M$
slots with small step $\delta\tau$, so that each slot $\Delta_{k}:=[\tau
_{k},\tau_{k+1}]$ collects signals with almost-equal transmissivity
$\tau\simeq\tau_{k}:=\tau_{\text{min}}+(k-1)\delta\tau$.

The post-selected $S_{\Delta}$ signals provide $\nu_{\text{det}}S_{\Delta}%
$\ pairs of data points $\{x_{i},y_{i}\}$ where $x$ is Alice's generic
quadrature encoding and $y$ is Bob's corresponding decoding. The outcomes
$\{y_{i}\}$ are all mapped into the first slot $\Delta_{1}$ with minimum
transmissivity $\tau_{\text{min}}$, by means of the de-fading channel
$y\rightarrow$ $y^{\prime\prime}$ described in Sec.~\ref{defadingSEC}. As a
result, Alice and Bob's data points satisfy the input-output relation
$y^{\prime\prime}=\sqrt{\tau_{\text{min}}}x+z_{\text{G}}$ of
Eq.~(\ref{Gaussdata}), which is equivalent to a thermal-loss channel
$\mathcal{E}_{\tau_{\text{min}},\bar{n}_{\text{G}}}$ with transmissivity
$\tau_{\text{min}}$ and thermal number $\bar{n}_{\text{G}}$, so that
$\sigma_{\text{G}}^{2}=2\bar{n}_{\text{G}}+\nu_{\text{det}}$.

Alice and Bob sacrifice $mp_{\Delta}$ signals to derive worst-case estimators
$\tau_{\text{min}}^{\prime}$ and $\bar{n}_{\text{G}}^{\prime}$ according to
Eqs.~(\ref{LBpara}) and~(\ref{nGestimator}), where the confidence parameter
$w$ is determined by the error $\varepsilon_{\text{pe}}$ according to
Eq.~(\ref{wSTVALUE}) or Eq.~(\ref{wTAIL}). These estimators are used to
compute the asymptotic key rate affected by parameter estimation
\begin{equation}
R_{\text{pe}}=R_{\text{asy}}(\tau_{\text{min}}^{\prime},\bar{n}_{\text{G}%
}^{\prime}),
\end{equation}
where $R_{\text{asy}}$ is given in Eq.~(\ref{asyFSP}). For the theoretical
analysis, we consider the further lower bound%
\begin{equation}
R_{\text{pe}}\geq R_{\text{LB}}:=R_{\text{asy}}(\tau_{\text{LB}},\bar
{n}_{\text{UB}}),
\end{equation}
which is based on $\tau_{\text{LB}}$ and $\bar{n}_{\text{UB}}$ from
Eqs.~(\ref{bbb1}) and~(\ref{bbb2}).

The signals remaining for key generation are $np_{\Delta}$, where
$n=N-(m+m_{\text{P}})$. Thus, after parameter estimation, the parties process
their $np_{\Delta}\nu_{\text{det}}$ key generation points $\{x_{i}%
,y_{i}^{\prime\prime}\}$ via the procedures of error correction and privacy
amplification. Depending on the reconciliation parameter $\beta$ (related to
the rate of the error-correcting code) and the correctness $\varepsilon
_{\text{cor}}$ (related to the probability of residual errors in Alice's and
Bob's corrected strings), the step of error correction has an associated
success probability $p_{\text{ec}}$ to promote the block of points to the next
step of privacy amplification. The latter procedure is ideal (i.e., decouples
Eve) up to an error quantified by the secrecy parameter $\varepsilon
_{\text{sec}}=\varepsilon_{\text{s}}+\varepsilon_{\text{h}}$, in turn
decomposed into a smoothing ($\varepsilon_{\text{s}}$) and a hashing parameter
($\varepsilon_{\text{h}}$). After privacy amplification, an average number of
$np_{\Delta}p_{\text{ec}}$ signals contribute to the final key, leading to an
overall factor $np_{\Delta}p_{\text{ec}}/N$ in front of the rate.

The composable finite-size key rate associated with the post-selection
interval $\Delta$\ is bounded by
\begin{equation}
R\geq\frac{np_{\Delta}p_{\text{ec}}}{N}\left(  R_{\text{LB}}-\frac
{\Delta_{\text{aep}}}{\sqrt{np_{\Delta}}}+\frac{\Theta}{np_{\Delta}}\right)  ,
\label{erreDELTA}%
\end{equation}
where the two terms $\Delta_{\text{aep}}$ and $\Theta$ are given in
Eqs.~(\ref{deltaAEPPP}) and~(\ref{bigOMEGA}) for some value $\log_{2}d$ of
digitalization. This rate is $\varepsilon$-secure against collective Gaussian
attacks, where $\varepsilon=2p_{\text{ec}}\varepsilon_{\text{pe}}%
+\varepsilon_{\text{cor}}+\varepsilon_{\text{sec}}$. The expression of the key
rate in Eq.~(\ref{erreDELTA}) can be specified for the homodyne/heterodyne
protocol and for the two types of LO (TLO/LLO).

For the heterodyne protocol, we can extend the key rate to composable
finite-size security against general coherent attacks (see
Sec.~\ref{SecKEYgeneral}). This is done by adopting a suitable symmetrization
and including energy tests, both operations to be performed on the data points
$\{x_{i},y_{i}^{\prime\prime}\}$. The number of energy tests is set to be
$p_{\Delta}m_{\mathrm{et}}$, where $m_{\mathrm{et}}=f_{\mathrm{et}}n$ for some
factor $f_{\mathrm{et}}<1$. Thus, the final key generation signals will be
$np_{\Delta}p_{\text{ec}}$ with%
\begin{equation}
n=N-(m+m_{\text{P}}+m_{\text{et}})=\frac{N-(m+m_{\text{P}})}{1+f_{\mathrm{et}%
}}.
\end{equation}

The composable key rate is bounded as follows%
\begin{equation}
R_{\text{gen}}^{\text{het}}\geq\frac{np_{\Delta}p_{\text{ec}}}{N}\left(
R_{\text{LB}}^{\text{het}}-\frac{\Delta_{\text{aep}}}{\sqrt{np_{\Delta}}%
}+\frac{\Theta-\Phi_{np_{\Delta}}}{np_{\Delta}}\right)  , \label{hetDELTAgen}%
\end{equation}
where the extra term $\Phi_{n}$ is defined as in Eq.~(\ref{FInGEN}) and is
expressed in terms of $K_{n}$ of Eq.~(\ref{KnGEN}), for which we choose the
dimensions $d_{R}=d_{T}\simeq\bar{n}_{T}=\sigma_{x}^{2}/2$ (so that the energy
test succeeds with probability $p_{\text{et}}\simeq1$).

Note that the key rate $R_{\text{gen}}^{\text{het}}$ is secure up to an
epsilon security $\varepsilon^{\prime}=K_{np_{\Delta}}^{4}\varepsilon/50$.
This means that, in order to get $\varepsilon^{\prime}\simeq10^{-10}$ against
general attacks, we need to start from a security of $\varepsilon
\simeq10^{-43}$ against collective Gaussian attacks. In turn, this also
implies $\varepsilon_{\text{pe}}\simeq10^{-43}$, so we need to use
Eq.~(\ref{wTAIL}) for the worst-case estimators.

\subsection{Numerical simulations}

In our numerical investigations we consider the heterodyne protocol, for which
we study the free-space composable key rate under collective and coherent
attacks, assuming the two types of LO. The free-space model is the same as in
Sec.~\ref{Sec_PLOBfree} and depicted in Fig.~\ref{detectorPIC}. We consider
the $z$-propagation of a collimated Gaussian beam which is subject to
diffraction, atmospheric extinction $\eta_{\text{atm}}$ [as quantified by the
Beer-Lambert equation of Eq.~(\ref{BLmain})], pointing error $\sigma
_{\text{P}}^{2}\simeq(10^{-6}z)^{2}$ (for an error of $1~\mu$rad at the
transmitter), and Rytov-Yura weak turbulence ($\sigma_{\text{Rytov}}^{2}<1$)
under the Hufnagel-Valley model of atmosphere (see Appendix~\ref{TurbSECTION}%
). Turbulence leads to beam broadening, with short-term transmissivity
$\eta_{\text{st}}$, and centroid wandering, with variance $\sigma_{\text{TB}%
}^{2}$. Including the setup efficiency $\eta_{\text{eff}}$, we have a maximum
transmissivity $\eta:=\eta_{\text{st}}\eta_{\text{atm}}\eta_{\text{eff}}$ when
the beam is perfectly-aligned. The overall wandering, with variance
$\sigma^{2}=\sigma_{\text{P}}^{2}+\sigma_{\text{TB}}^{2}$, leads to the
distribution $P_{0}(\tau)$ of Eq.~(\ref{P0tau}) for the instantaneous
transmissivity $\tau$ of the link. Thermal background follows the description
of Sec.~\ref{SEC_thermal_noise} for cloudy day-time conditions (but suppressed
by the homodyne filter). In particular, we assume the physical parameters
listed in Table~\ref{TablePhysical}.

\begin{table}[h]
\vspace{0.2cm}
\begin{tabular}
[c]{|l|l|l|}\hline
Physical parameter & Symbol & Value\\\hline\hline
Beam curvature & $R_{0}$ & $\infty$\\\hline
Wavelength & $\lambda$ & $800~$nm\\\hline
Beam spot size & $w_{0}$ & $5~$cm\\\hline
Receiver aperture & $a_{R}$ & $5~$cm\\\hline
Receiver field of view & $\Omega_{\text{fov}}$ & $10^{-10}~$sr\\\hline
Homodyne filter & $\Delta\lambda$ & $0.1~\text{pm}$\\\hline
Detector efficiency & $\eta_{\text{eff}}$ & $0.5$\\\hline
Detector bandwidth & $W$ & $100~$MHz\\\hline
Noise equivalent power & NEP & $6~$pW/$\sqrt{\text{Hz}}$\\\hline
Linewidth & $l_{\text{W}}$ & $1.6$~KHz\\\hline
LO power & $P_{\text{LO}}$ & $100~$mW\\\hline
Clock & $C$ & $5~$MHz\\\hline
Pulse duration & $\Delta t,\Delta t_{\text{LO}}$ & $10~$ns\\\hline
Altitude & $h$ & $30~$m\\\hline
Structure constant (day) & $C_{n}^{2}$ & $2.06\times10^{-14}~$m$^{-2/3}%
$\\\hline
$%
\begin{array}
[c]{l}%
\text{Background noise}\\
\text{(day, }\Delta\lambda=0.1~\text{pm)}%
\end{array}
$ & $\bar{n}_{B}$ & $4.75\times10^{-7}$\\\hline
\end{tabular}
\caption{Physical parameters.}%
\label{TablePhysical}%
\end{table}

The steps of the protocol are those explained in the previous subsection,
where Alice and Bob assume a post-selection interval $\Delta:=[\tau_{\min
},\tau_{\max}]$ with $\tau_{\text{max}}=\eta$ and $\tau_{\text{min}%
}=f_{\text{th}}\eta$ for some threshold value $f_{\text{th}}\in(0,1)$. In
particular, we choose the parameters listed in Table~\ref{TableProtocol}.

\begin{table}[t]%
\begin{tabular}
[c]{|l|l|l|l|}\hline
$%
\begin{array}
[c]{l}%
\text{Protocol}\\
\text{parameter}%
\end{array}
$ & Symbol & $%
\begin{array}
[c]{l}%
\text{Collective}\\
\text{attacks}%
\end{array}
$ & $%
\begin{array}
[c]{l}%
\text{General}\\
\text{attacks}%
\end{array}
$\\\hline\hline
Total pulses & $N$ & $5\times10^{7}$ & $5\times10^{7}$\\\hline
Pilot pulses & $m_{\text{P}}$ & $0.1\times N$ & $0.1\times N$\\\hline
PE signals & $m$ & $0.1\times N$ & $0.1\times N$\\\hline
Energy tests & $f_{\text{et}}$ & $-$ & $0.2$\\\hline
KG signals & $n$ & $0.8\times N$ & $\simeq3.33\times10^{7}$\\\hline
Digitalization & $d$ & $2^{5}$ & $2^{5}$\\\hline
Rec. efficiency & $\beta$ & $0.98$ & $0.98$\\\hline
EC success prob & $p_{\text{ec}}$ & $0.9$ & $0.5$\\\hline
Epsilons & $\varepsilon_{\text{h,s,\ldots}}$ & $2^{-33}\simeq10^{-10}$ &
$10^{-43}$\\\hline
Confidence & $w$ & $\simeq6.34$ & $\simeq14.07$\\\hline
Security & $\varepsilon,\varepsilon^{\prime}$ & $\simeq5.6\times10^{-10}$ &
$\lesssim1.3\times10^{-9}$\\\hline
Modulation & $\mu$ & variable & $%
\begin{array}
[c]{l}%
20~\text{(TLO)}\\
8.4~\text{(LLO)}%
\end{array}
$\\\hline
Threshold & $f_{\text{th}}~$ & variable & $0.84$\\\hline
\end{tabular}
\caption{Protocol parameters.}%
\label{TableProtocol}%
\end{table}

\begin{figure}[t]
\vspace{0.2cm}
\par
\begin{center}
\includegraphics[width=0.85\columnwidth] {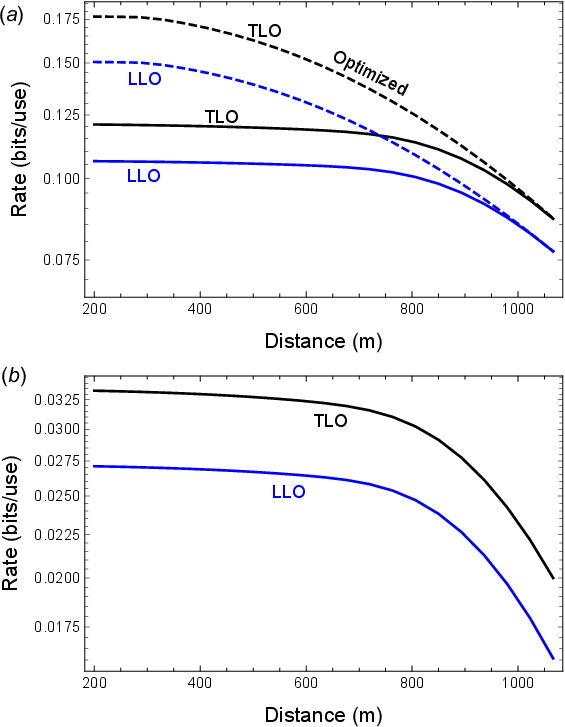}
\end{center}
\par
\vspace{-0.3cm}\caption{Composable secret-key rates (bits/use) versus distance
(m) for free-space QKD in the regime of weak turbulence and for cloudy
day-time operation. We consider a coherent-state protocol with heterodyne
detection, pilot-guided and operated in post-selection as described in the
main text. Physical and protocol parameters are listed in
Tables~\ref{TablePhysical} and~\ref{TableProtocol}. (a)~We plot the secret key
rate of Eq.~(\ref{erreDELTA}) assuming a TLO (black curves) and an LLO (blue
curves). In particular, we plot the performances at fixed post-selection
threshold $f_{\text{th}}=0.84$\ and fixed input modulation, $\mu=20$ for TLO
and $\mu=8.4$ for LLO (solid curves). These are chosen to optimize the rates
at the maximum distance ($z=1066~$m). We compare these performances with those
achievable by optimizing the rates over $\mu$ and $f_{\text{th}}$ at each
distance (dashed curves). (b)~We plot the rate of Eq.~(\ref{hetDELTAgen})
against general attacks for TLO (black line) and LLO (blue line). These
performances are not optimized and refer to fixed threshold $f_{\text{th}%
}=0.84$\ and input modulation ($\mu=20$ for TLO and $\mu=8.4$ for LLO).}%
\label{RatesDayPic}%
\end{figure}

As we can see from Fig.~\ref{RatesDayPic}(a), the composable key rates against
collective attacks are sufficiently high, even though these values actually
represent lower bounds to what achievable by Alice and Bob. As a matter of
fact, in most of the weak-turbulence range, these rates are within one order
of magnitude of the ultimate loss-based upper bound of Eq.~(\ref{PLOBdelta})
which is plotted as red dashed line in Fig.~\ref{FSpics}(a), computed for
day-time and the same physical parameters considered here. In
Fig.~\ref{RatesDayPic}(a), we study the rates that are achievable with the TLO
and the LLO. In one setting (solid curves), we fix the value of the threshold
parameter for post-selection $f_{\text{th}}$ to $84\%$ and we also fix the
value of the input Gaussian modulation ($\mu=20$ for TLO and $\mu=8.4$ for
LLO). These values are chosen to maximize the rates at the maximum distance
$z=1066~$m, but they are not the optimal choices for the other distances. In
another approach, we maximize the rates over $f_{\text{th}}$ and $\mu$ at each
distance, finding substantially improved performances (dashed lines).

In Fig.~\ref{RatesDayPic}(b), we plot the composable key rates achievable
against general attacks assuming no optimization in $f_{\text{th}}$ and $\mu$.
On the one hand, these rates are not far from the corresponding results
against collective attacks. On the other hand, the choice of parameters in
Table~\ref{TableProtocol} may be far more challenging for this general case
(e.g., in terms of $\beta$ and $p_{\text{ec}}$ for such a low value of
$\varepsilon_{\text{cor}}$). Also note that the final epsilon security
$\varepsilon^{\prime}$ depends on the distance. For the parameters chosen,
this ranges from $\simeq1.38\times10^{-11}$ for the LLO at $z=1066~$m and
$\simeq1.32\times10^{-9}$ for the TLO at $z=200~$m.

A final important observation (already mentioned before but here relevant to
stress) is that the rates shown in Fig.~\ref{RatesDayPic} refer to bits per
use of the quantum communication channel, without accounting for the
transmission of the LO-reference pulses. If we include the clock of the system
(uses/second) and compute the throughput of the communication (bits/second),
then we need to include the uses of the link dedicated to the LO. Thus, for
the LLO, we should halve the final rate (with respect to the TLO) due to the
time multiplexing of the LO. However, it is also true that, with the LLO, one
could use both polarizations in the transmission of the signals, so that the
factor $1/2$ in the final rate can be fully compensated.

\section{Conclusions\label{SEC_conclusios}}

In conclusion, we have established the ultimate bounds for free-space quantum
communications under general conditions of diffraction, atmospheric
extinction, pointing errors, turbulence, and background thermal noise. We have
first developed the theory for the regime of weak turbulence, crucial for free
space ground-communications in a relatively short range, and then extended the results
to the case of stronger turbulence. In the short range, we have then derived
achievable and composable key rates for free-space CV-QKD, proving that these
rates are sufficiently close to the ultimate limits. This shows the robustness
and suitability of free-space channels for implementing high-rate
quantum-secured communications.

The achievable rates are derived by first formulating a general theory of
composable finite-size security for Gaussian-modulated coherent-state
protocols under conditions of channel stability, and then extending this
theory to considering fading (non Gaussian) channels, which can be dealt via
the introduction of pilot modes and suitable post-processing techniques. In
this way we have been able to handle the difficult step of parameter
estimation and to reduce the problem to the easier framework of a stable
Gaussian channel. Fully assessing the practical security of CV-QKD in strong
turbulent channels is an interesting future direction of investigation.

In conclusion our work not only established the ultimate limits and benchmarks
for free-space quantum communications but also provided a comprehensive
machinery for studying the composable finite-size security of
CV-QKD\ protocols both in stable conditions (e.g., in standard fiber-based
connections) and unstable conditions (i.e., in free-space links subject to
fading effects).

\textit{Acknowledgements}.--~The author acknowledges funding from the European
Union's Horizon 2020 research and innovation programme under grant agreement
No 820466 (Quantum-Flagship Project CiViQ: \textquotedblleft Continuous
Variable Quantum Communications\textquotedblright).

\appendix

\section{Propagation of Gaussian beams\label{APP_free_basics}}

Most of the contents of this Appendix are basic notions of quantum optics.
They are given here to set the general notation of the work and for the sake
of completeness.

\subsection{Free-space diffraction}

Consider an optical bosonic mode with wavelength $\lambda$, angular frequency
$\omega=2\pi c/\lambda$, and wavenumber $k=\omega/c=2\pi/\lambda$. Under the
scalar approximation (single and uniform polarization) and the paraxial wave
approximation, the electric field takes the form
\begin{equation}
E(x,y,z,t)=u(x,y,z)\exp[i(kz-\omega t)],
\end{equation}
where the field amplitude $u(x,y,z)$ is a slowly varying function in the
longitudinal propagation direction $z$, with $x$ and $y$ being the transverse
coordinates and $t$ the time coordinate. The possible expressions for the
field amplitude $u$ must satisfy the Fresnel-Kirchoff integral in the Fresnel
approximation~\cite[Eq.~(4.6.9)]{svelto}. A solution of this integral which
maintains its functional form, i.e., an eigensolution, is the Gaussian beam.

In particular, assume free-space propagation along the $z$ direction with no
limiting apertures in the transverse plane, for which we introduce the radial
coordinate $r=\sqrt{x^{2}+y^{2}}$. Then, the lowest order ($\mathrm{TEM}_{00}%
$) single-mode Gaussian beam takes a simple analytical expression. At the
initial position $z=0$, its field amplitude has the form%
\begin{equation}
u(0,r)=\exp(-r^{2}/w_{0}^{2})\exp{[-ikr^{2}/(2R_{0})]}, \label{iniGauss}%
\end{equation}
where $w_{0}$ is the beam spot size and $R_{0}$ is the phase-front radius of
curvature. For beam spot size we precisely mean the `field' spot size,
corresponding to the radial distance at which the amplitude of the field
decays to $1/e$ of its maximum value. Note that the intensity of the beam is
given by $I(0,r)=\exp\left(  -2r^{2}/w_{0}^{2}\right)  $, so that one can
define an `intensity' spot size $w_{0}^{\text{I}}=w_{0}/\sqrt{2}$, that is
also widely used in the literature (e.g., in Refs.~\cite{Yura73,Fante75}).

Let us introduce the `Rayleigh range'
\begin{equation}
z_{R}:=\frac{\pi w_{0}^{2}}{\lambda},
\end{equation}
and the Fresnel number of the beam
\begin{equation}
f:=\frac{\pi w_{0}^{2}}{\lambda z}=\frac{z_{R}}{z},
\end{equation}
so that the far-field regime ($z\gg z_{R}$) corresponds to $f\ll1$. Following
the notation of Ref.~\cite{Andrews93}, we also introduce the Fresnel ratio
$\Omega:=f^{-1}$ and the curvature parameter $\Omega_{0}:=1-z/R_{0}$. Note
that a collimated beam ($R_{0}=+\infty$) corresponds to $\Omega_{0}=1$, while
a convergent beam ($R_{0}>0$) to $\Omega_{0}<1$, and a divergent beam
($R_{0}<0$) to $\Omega_{0}>1$.

In terms of the previous parameters, we can write the field at any distance
$z$ as~\cite{Andrews93,Andrews94}%
\begin{equation}
u(z,r)=\frac{w_{0}}{w_{z}}\exp(-r^{2}/w_{z}^{2})\exp{[-ikr^{2}/(2R_{z}%
)-i\phi_{z}]}, \label{gbeam}%
\end{equation}
where $\omega_{z}$ is the spot size at position $z$, $R_{z}$ is the
corresponding curvature at $z$, and $\phi_{z}$ is its longitudinal phase at
$z$, also known as Guoy phase shift~\cite[Sec.~17.4]{Siegman}. These
quantities take the following expressions
\begin{align}
w_{z}^{2}  &  =w_{0}^{2}\left(  \Omega_{0}^{2}+\Omega^{2}\right)
,\label{wzGEN}\\
{R_{z}}  &  {=}\frac{z(\Omega_{0}^{2}+\Omega^{2})}{\Omega_{0}(1-\Omega
_{0})-\Omega^{2}},\\
\phi_{z}  &  =\tan^{-1}(\Omega/\Omega_{0}). \label{GuoyGEN}%
\end{align}
More explicitly, we may write
\begin{align}
w_{z}^{2}  &  =w_{0}^{2}\left[  \left(  1-\frac{z}{R_{0}}\right)  ^{2}+\left(
\frac{z}{z_{R}}\right)  ^{2}\right] \label{spotFUNC}\\
&  =w_{0}^{2}\left(  1-\frac{z}{R_{0}}\right)  ^{2}+\frac{\lambda^{2}z^{2}%
}{\pi^{2}w_{0}^{2}}.
\end{align}

A typical assumption is to adopt the planar approximation of a collimated beam
at the transmitter ($\Omega_{0}=1$). In such a case, it is immediate to check
that
\begin{align}
w_{z}^{2}  &  =w_{0}^{2}[1+(z/z_{R})^{2}],\label{bwaist}\\
R_{z}  &  =-z[1+(z_{R}/z)^{2}],\\
\phi_{z}  &  =\tan^{-1}(z/z_{R}). \label{Guoy}%
\end{align}
Note that $w_{z}^{2}$ is the sum of the initial (minimum) condition $w_{0}%
^{2}$ and a term $w_{0}^{2}(z/z_{R})^{2}$ which is due to diffraction. In the
far-field, the latter term is dominant and we have
\begin{equation}
w_{z}\simeq w_{0}(z/z_{R})=\frac{\lambda z}{\pi w_{0}}, \label{outputwz}%
\end{equation}
which increases linearly with the distance $z$. Defining beam divergence as
$\theta:=w_{z}/z$, we write $\theta\simeq\lambda/(\pi w_{0})$, which increases
with the wavelength (as expected). For a collimated beam, the curvature is
minimal at $z=z_{R}$ and then goes as $\simeq z$ at large distance, so that
the beam asymptotically becomes a spherical wave.

From Eq.~(\ref{gbeam}), we see that the beam intensity at longitudinal
distance $z$ is given by
\begin{equation}
I(z,r)=I_{\max}^{z}\exp\left(  -2r^{2}/w_{z}^{2}\right)  ,~I_{\max}^{z}%
:=w_{0}^{2}/w_{z}^{2}. \label{Irz}%
\end{equation}
Assume that the beam is orthogonally intercepted by a receiver, which is
described as a sharped-edged circular aperture with radial size $a_{R}$,
therefore with total detection area $\pi a_{R}^{2}$. Let us compute the total
power impinging on the finite-size detector by integrating over the radial
coordinates $0\leq r\leq a_{R}$ and $0\leq\varphi\leq2\pi$. We easily find%
\begin{align}
P(z,a_{R})  &  :=%
{\displaystyle\int\limits_{0}^{2\pi}}
d\varphi%
{\displaystyle\int\limits_{0}^{a_{R}}}
rdrI(z,r)\nonumber\\
&  =P_{z}\left(  1-e^{-2a_{R}^{2}/w_{z}^{2}}\right)  , \label{ILexp}%
\end{align}
where $P_{z}:=(\pi w_{z}^{2}I_{\max}^{z})/2$ represents the total power in the
optical beam at distance $z$ (corresponding to a receiver of infinite radius
$a_{R}\rightarrow\infty$). Note that we may also rewrite Eq.~(\ref{Irz}) as
\begin{equation}
I(z,r)=(2P_{z}/\pi w_{z}^{2})\exp(-2r^{2}/w_{z}^{2}).
\end{equation}

The diffraction-limited transmissivity $\eta_{\text{d}}$ associated with the
finite size of the receiver is given by%
\begin{equation}
\eta_{\text{d}}:=P(z,a_{R})/P_{z}=1-e^{-2a_{R}^{2}/w_{z}^{2}},
\label{etaSatellite}%
\end{equation}
where we may explicitly express $w_{z}^{2}$ as in Eq.~(\ref{spotFUNC}).\ In
the far field, we have $\Omega\gg1$ in Eq.~(\ref{wzGEN}), so that$\ w_{z}\gg
w_{0}$. Assuming that the receiver's aperture radius $a_{R}$ is comparable to
the spot size $w_{0}$, then we have $w_{z}\gg a_{R}$ and we can expand
Eq.~(\ref{etaSatellite}) into%
\begin{equation}
\eta_{\text{d}}\simeq\eta_{\text{d}}^{\text{far}}:=\frac{2a_{R}^{2}}{w_{z}%
^{2}}\ll1. \label{etadAPPROX}%
\end{equation}
In particular, for a collimated beam we can use the approximation in
Eq.~(\ref{outputwz}) and write the far-field expression%
\begin{equation}
\eta_{\text{d}}\simeq\eta_{\text{d}}^{\text{far,coll}}:=2\left(  \frac{\pi
w_{0}a_{R}}{\lambda z}\right)  ^{2}. \label{longDIST}%
\end{equation}
Recognizing that $A_{0}=\pi w_{0}^{2}$ and $A_{R}=\pi a_{R}^{2}$\ as the
effective transversal areas of the beam and the receiver's aperture, we note
that we may write Eq.~(\ref{longDIST}) as $\eta_{\text{d}}\simeq2f_{0R}$
where
\begin{equation}
f_{0R}:=A_{0}A_{R}/(\lambda z)^{2} \label{FresnelProduct}%
\end{equation}
is the Fresnel number product associated with the beam and the receiver.


\subsection{Diffraction at the transmitter\label{APP_diffraction_subsection}}

Any realistic transmitter involves an aperture with finite radius $a_{T}$.
This means that the Gaussian profile of the beam could be truncated outside
that radius causing diffraction. However, if the aperture $a_{T}$ is
sufficiently larger than $w_{0}$, diffraction becomes negligible.

Assume that the transmitter has a plane exit pupil $\mathcal{A}_{0}$ of area
$A_{0}$ while the receiver has an entrance pupil $\mathcal{A}_{z}$ of area
$A_{z}$. We consider the quasi-monochromatic approximation where the
transmitter excites planar modes within a narrow band of frequencies, centered
around the carrier (angular) frequency $\omega$, and the receiver only detects
planar modes within this bandwidth. We then consider the usual scalar
approximation (i.e., a single and uniform polarization) and the paraxial wave
approximation (so that the transverse components of the wave-vector are
negligible at the receiver).

Let us write $\mathbf{x}:=(x,y)\in\mathcal{A}_{0}$ to be the transverse
coordinates at the transmitter and $\mathbf{x}^{\prime}:=(x^{\prime}%
,y^{\prime})\in\mathcal{A}_{z}$\ those at the receiver. The electric field at
the transmitter can then be expressed as~\cite{Jeff}
\begin{equation}
E_{0}(\mathbf{x},t)=%
{\displaystyle\sum\limits_{k,l}}
\hat{a}_{k,l}\Phi_{k}(\mathbf{x})\Psi_{l}(t),
\end{equation}
where $\Phi_{k}(\mathbf{x})\Psi_{l}(t)$ are orthonormal spatiotemporal modes
defined over $\mathcal{A}_{0}$ and $0\leq t\leq t_{\text{max}}$, with
$t_{\text{max}}$ being the time duration of the transmitter's signal. These
modes have corresponding annihilation operators $\hat{a}_{k,l}$. Thanks to
this normal-mode decomposition, one can express the electric field at the
receiver, which is given by~\cite{Jeff}%
\begin{align}
E_{z}(\mathbf{x}^{\prime},t)  &  =%
{\displaystyle\sum\limits_{k,l}}
\left(  \sqrt{\eta_{k}}\hat{a}_{k,l}+\sqrt{1-\eta_{k}}\hat{e}_{k,l}\right)
\nonumber\\
&  \times\Phi_{k}(\mathbf{x}^{\prime})\Psi_{l}(t-c^{-1}z),
\end{align}
for modes defined over $\mathcal{A}_{z}$ and $0\leq t-c^{-1}z\leq
t_{\text{max}}$. Above, $\hat{e}_{k,l}$ are the annihilation operators
associated with environmental modes impinging on the pupil of the receiver,
which are generally described by thermal states.

Free-space diffraction-limited quantum communication can therefore be
completed described by the input-output relations
\begin{equation}
\hat{a}_{k,l}\rightarrow\hat{b}_{k,l}=\sqrt{\eta_{k}}\hat{a}_{k,l}%
+\sqrt{1-\eta_{k}}\hat{e}_{k,l},
\end{equation}
which correspond to a collection of thermal-loss channels (beam-splitter
transformations with thermal environment). It is important to note that
\begin{equation}%
{\displaystyle\sum\limits_{k}}
\eta_{k}=\frac{A_{0}A_{z}}{(\lambda z)^{2}}:=n_{\text{f}},
\end{equation}
which is equal the Fresnel number product $n_{\text{f}}$\ of the two
pupils~\cite[Eq.~(37)]{Jeff}. In the far-field regime ($n_{\text{f}}\ll1$),
only one mode is effectively transmitted from transmitter to receiver, with
transmissivity $\eta_{\text{far}}\simeq n_{\text{f}}$.

For circular apertures $A_{0}=\pi a_{T}^{2}$ and $A_{z}=\pi a_{R}^{2}$, we
therefore have
\begin{equation}
\eta_{\text{far}}=\left(  \frac{\pi a_{T}a_{R}}{\lambda z}\right)  ^{2}.
\label{etaJJJ}%
\end{equation}
From Eq.~(\ref{etaJJJ}), we see that we obtain the far-field collimated-beam
transmissivity in Eq.~(\ref{longDIST}) by setting $a_{T}=\sqrt{2}w_{0}%
\simeq1.41w_{0}$. In other words, by choosing such a\ value for the
transmitter's aperture, we may neglect its far-field contribution to
diffraction from the point of view of the transmissivity (otherwise
$a_{T}=w_{0}$ would cause a $3$dB loss). That being said, the choice
$a_{T}=\sqrt{2}w_{0}$ may still be too generous because the profile of the
Gaussian beam could be affected in the far field by non-negligible intensity
ripples and peak intensity reductions.

In order to preserve the Gaussian profile with excellent approximation, a more
conservative choice is $a_{T}\geq2w_{0}$, e.g., $a_{T}\simeq2.3w_{0}%
$~\cite[Sec.~17.1]{Siegman}. Let us write Eq.~(\ref{ILexp}) at $z=0$ for the
transmitter's aperture $a_{T}$. Then, we see that the total power passing
through the transmitter is given by $P_{0}(1-e^{-2a_{T}^{2}/w_{0}^{2}})$. If
we choose $a_{T}\geq2w_{0}$ then $\geq99.97\%$ of $P_{0}$ is transmitted. This
estimate provides an idea of the extremely small perturbation that such a
large aperture ($a_{T}\geq2w_{0}$) causes to the Gaussian beam.

\section{Diffraction-limited free-space bounds\label{APP_diffraction_limited}}

Quantum mechanically, the propagation of the Gaussian beam from transmitter to
receiver can be represented by a single mode whose annihilation operator
$\hat{a}$ at the transmitter undergoes the following input-output Bogoliubov
transformation
\begin{equation}
\hat{a}\rightarrow\hat{b}=\sqrt{\eta_{\text{d}}}\hat{a}+\sqrt{1-\eta
_{\text{d}}}\hat{e}, \label{IOsat}%
\end{equation}
where $\hat{b}$ is the annihilation operator of the signal mode at the
receiver, and $\hat{e}$ is the annihilation operator of an environmental mode
impinging on the receiver and coupling with the output signal mode. Mode
$\hat{e}$ is generally described by a thermal state whose mean number of
photons $\bar{n}_{e}$ depends on various factors. Its typical values largely
vary between night-time and day-time operation, weather conditions etc. The
basic process described in Eq.~(\ref{IOsat}) is also known as single-mode
thermal-loss channel~\cite{RMP}, here denoted by $\mathcal{E}_{\eta_{\text{d}%
}}^{\bar{n}_{e}}$. (Note that, in the main text and other parts of these
appendices, we use the different notation $\mathcal{E}_{\eta_{\text{d}},\bar{n}}$ to indicate a thermal-loss channel with transmissivity
$\eta_{\text{d}}$ and $\bar{n}_{e}=\bar{n}/(1-\eta_{\text{d}})$, so that
$\bar{n}$ thermal photons are added to its output).

In order to give a universal upper bound which is valid in every condition, we
neglect thermal noise, so that Eq.~(\ref{IOsat}) describes a pure-loss channel
$\mathcal{E}_{\eta_{\text{d}}}:\hat{a}\rightarrow\sqrt{\eta_{\text{d}}}\hat
{a}+\sqrt{1-\eta_{\text{d}}}\hat{v}$, where the environmental mode $\hat{v}$
is associated with a vacuum state. Thermal noise can be neglected from an
information-theoretical point of view, because an upper bound on a pure-loss
channel would automatically be an upper bound on a thermal-loss channel. In
fact, a thermal-loss channel $\mathcal{E}_{\eta_{\text{d}}}^{\bar{n}_{e}}$ as
in Eq.~(\ref{IOsat}) is equivalent to a composition of a pure-loss channel
$\mathcal{E}_{\eta_{\text{d}}}$\ followed by an additive-noise Gaussian
channel $\mathcal{A}_{\eta_{\text{d}}}^{\xi}:\hat{a}\rightarrow\hat{a}%
+\sqrt{1-\eta_{\text{d}}}\xi$, where the variable $\xi$ is taken with noise
variance $\left\langle \xi^{2}\right\rangle =\bar{n}_{e}$, so that
\begin{align}
\hat{a}  &  \overset{\mathcal{E}_{\eta_{\text{d}}}}{\rightarrow}\sqrt
{\eta_{\text{d}}}\hat{a}+\sqrt{1-\eta_{\text{d}}}\hat{v}\nonumber\\
&  \overset{\mathcal{A}_{\eta_{\text{d}}}^{\xi}}{\rightarrow}\sqrt
{\eta_{\text{d}}}\hat{a}+\sqrt{1-\eta_{\text{d}}}(\hat{v}+\xi)\nonumber\\
&  =\sqrt{\eta_{\text{d}}}\hat{a}+\sqrt{1-\eta_{\text{d}}}\hat{e}.
\end{align}
Because we have $\mathcal{E}_{\eta_{\text{d}}}^{\bar{n}_{e}}=\mathcal{A}%
_{\eta_{\text{d}}}^{\xi}\circ\mathcal{E}_{\eta_{\text{d}}}$, we may apply data
processing for any functional that is decreasing under completely positive
trace-preserving (CPTP) maps. This is a property which can be exploited for
the relative entropy of entanglement (REE).

Given two states $\rho$ and $\sigma$, their relative entropy is defined by
$S(\rho||\sigma):=\mathrm{Tr}[\rho(\log_{2}\rho-\log_{2}\sigma)]$. Then, the
REE of a bipartite state $\rho_{AB}$\ is defined by
\begin{equation}
E_{\text{R}}(\rho_{AB}):=\inf_{\sigma\in\text{SEP}}S(\rho_{AB}||\sigma_{AB}),
\end{equation}
where SEP is the set of separable states. Now we observe that the relative
entropy is monotonic under the same CPTP map $\mathcal{N}$ applied to both its
arguments, i.e., $S[\mathcal{N}(\rho)||\mathcal{N}(\sigma)]\leq S(\rho
||\sigma)$. This allows one to show that, for any bipartite state $\rho_{AB}$,
we may also write
\begin{equation}
E_{\text{R}}[\mathcal{I}\otimes\mathcal{N}(\rho_{AB})]\leq E_{\text{R}}%
(\rho_{AB}). \label{DPI}%
\end{equation}
In fact, it is quite easy to check that
\begin{align}
E_{\text{R}}[\mathcal{I}\otimes\mathcal{N}(\rho_{AB})]  &  =\inf_{\sigma
\in\text{SEP}}S[\mathcal{I}\otimes\mathcal{N}(\rho_{AB})||\sigma
_{AB}]\nonumber\\
&  \overset{(1)}{\leq}\inf_{\sigma\in\text{SEP}}S[\mathcal{I}\otimes
\mathcal{N}(\rho_{AB})||\mathcal{I}\otimes\mathcal{N}(\sigma_{AB})]\nonumber\\
&  \overset{(2)}{\leq}\inf_{\sigma\in\text{SEP}}S(\rho_{AB}||\sigma
_{AB})\nonumber\\
&  :=E_{\text{R}}(\rho_{AB}),
\end{align}
where (1) exploits the fact that $\mathcal{I}\otimes\mathcal{N}(\sigma_{AB})$
represent a subset of all possible separable states, and (2) exploits the
monotonicity of the relative entropy under the CPTP map $\mathcal{I}%
\otimes\mathcal{N}$.

The ultimate rates at which two remote parties can generate a key (secret key
capacity $K$), or distribute entanglement (two-way assisted entanglement
distribution capacity $E$, also denoted by $D_{2}$), or teleport/transfer
quantum states (two-way assisted quantum capacity $Q_{2}$) at the two ends of
a bosonic single-mode Gaussian channel $\mathcal{G}$\ are all limited by the
following REE bound~\cite{QKDpaper}%
\begin{equation}
Q_{2}=E\leq K\leq\Phi(\mathcal{G}):=\underset{\mu\rightarrow\infty}{\lim\inf
}E_{\text{R}}[\mathcal{I}\otimes\mathcal{G}(\Phi_{AB}^{\mu})], \label{PLOB}%
\end{equation}
where $\Phi_{AB}^{\mu}$ is a TMSV state with variance
$\mu$, i.e., $(\mu-1)/2$ mean number of photons in each mode.

For any composition of Gaussian channels, we can combine Eq.~(\ref{PLOB}) with
the data processing inequality in Eq.~(\ref{DPI}). In particular, for the
secret key capacity (SKC) of a thermal-loss channel $\mathcal{E}%
_{\eta_{\text{d}}}^{\bar{n}_{e}}$ we may write%
\begin{equation}
K\leq\Phi(\mathcal{E}_{\eta_{\text{d}}}^{\bar{n}_{e}})\leq\Phi(\mathcal{E}%
_{\eta_{\text{d}}}),
\end{equation}
where
\begin{equation}
\Phi(\mathcal{E}_{\eta_{\text{d}}})=\Phi(\eta_{\text{d}}):=-\log_{2}%
(1-\eta_{\text{d}}) \label{PLOBbound}%
\end{equation}
is the PLOB bound~\cite{QKDpaper}. For $\eta_{\text{d}}\simeq0$, we have the
approximation%
\begin{equation}
\Phi(\eta_{\text{d}})\simeq\eta_{\text{d}}/\ln2=1.44\eta_{\text{d}}\text{
(bits per channel use).}%
\end{equation}

Consider now free-space line-of-sight quantum communication at wavelength
$\lambda$, between a transmitter, generating a Gaussian beam with spot size
$w_{0}$ and curvature radius $R_{0}$, and a remote receiver, with aperture
radius $a_{R}$ at slant distance $z$. The corresponding expression for the
diffraction-induced transmissivity $\eta_{\text{d}}$ is explicitly given in
Eq.~(\ref{etaSatellite}). By replacing it in the PLOB bound $\Phi
(\eta_{\text{d}})$, we see that the maximum rate for QKD and, therefore, any other form of quantum communication, is bounded by%
\begin{equation}
K\leq\mathcal{U}(z):=\frac{2}{\ln2}\left(  \frac{a_{R}}{w_{z}}\right)  ^{2},
\label{FPbound}%
\end{equation}
where $w_{z}$ is the spot-size function of Eq.~(\ref{spotFUNC}). More
explicitly, we may write%
\begin{equation}
\mathcal{U}(z)=\frac{2}{\ln2}\frac{a_{R}^{2}}{w_{0}^{2}}\left[  \left(
1-\frac{z}{R_{0}}\right)  ^{2}+\frac{z^{2}}{z_{R}^{2}}\right]  ^{-1}.
\label{FPboundEXP}%
\end{equation}

From Eq.~(\ref{FPboundEXP}), we see that the bound is maximized by a focused
beam ($z=R_{0}$). In such a case, we derive
\begin{equation}
\mathcal{U}_{\text{foc}}(z)=\frac{2}{\ln2}\frac{a_{R}^{2}}{w_{0}^{2}}%
\frac{z_{R}^{2}}{z^{2}}=\frac{2}{\ln2}\left(  \frac{\pi w_{0}a_{R}}{\lambda
z}\right)  ^{2}=\frac{2f_{0R}}{\ln2},
\end{equation}
where $f_{0R}$ is the Fresnel number product associated to the beam and the
receiver, as in Eq.~(\ref{FresnelProduct}). Instead, for a collimated beam
($R_{0}=+\infty$), the upper bound simplifies to the following expression%
\begin{align}
\mathcal{U}_{\text{coll}}(z)  &  =\frac{2}{\ln2}\frac{a_{R}^{2}}{w_{0}%
^{2}[1+z^{2}/z_{R}^{2}]},\\
&  \simeq\mathcal{U}_{\text{foc}}(z)\text{,~in the far field.}%
\end{align}

\section{Atmospheric turbulence\label{TurbSECTION}}

A crucial parameter in the study of atmospheric turbulence is the refraction
index structure constant $C_{n}^{2}$~\cite{AndrewsBook,Hemani}. This measures
the strength of the fluctuations in the refraction index, due to spatial
variations of temperature and pressure. There are several models which provide
$C_{n}^{2}$ with a functional expression in terms of the altitude $h$ in
meters above sea-level. The most known is the Hufnagel-Valley (H-V)
model~\cite{Stanley,Valley}%
\begin{align}
C_{n}^{2}(h)  &  =5.94\times10^{-53}\left(  \frac{v}{27}\right)  ^{2}%
h^{10}e^{-h/1000}\nonumber\\
&  +2.7\times10^{-16}e^{-h/1500}+Ae^{-h/100}, \label{HFmodel}%
\end{align}
where $v$ is the windspeed (m/s) and $A\simeq C_{n}^{2}(0)$. Assuming
high-altitude low-wind $v=21~$m/s and the ground-level night-time value
$A=1.7\times10^{-14}~$m$^{-2/3}$, one has the H-V$_{5/7}$
model~\cite[Sec.~12.2.1]{AndrewsBook}. However, during the day, we may have
$A\simeq2.75\times10^{-14}~$m$^{-2/3}$~\cite{BrussSAT}. In our work, we assume
$v=21~$m/s, the day-value $A\simeq2.75\times10^{-14}~$m$^{-2/3}$, and an
altitude of $h=30$~m, so that $C_{n}^{2}\simeq2.06\times10^{-14}~$m$^{-2/3}$.

The structure constant is at the basis of other important parameters such as
the scintillation index~\cite{AndrewsBook} and the Rytov
variance~\cite{RytovAPPROX}, which is given by
\begin{equation}
\sigma_{\text{Rytov}}^{2}=1.23C_{n}^{2}k^{7/6}z^{11/6}. \label{Rytov}%
\end{equation}
The condition $\sigma_{\text{Rytov}}^{2}<1$ corresponds to the regime of weak
turbulence, where scintillation (i.e., random fluctuations of the intensity)
can be considered to be negligible, and the mean intensity of the beam can
still be approximated by a Gaussian spatial profile. An alternative condition
was considered by Yura~\cite{Yura73} and Fante~\cite{Fante75} in terms of the
spherical-wave coherence length $\rho_{0}$, which is closely related to the
Fried's parameter~\cite{Fried,BelandBook}. For a fixed (or mean) value of the
structure constant $C_{n}^{2}$, this length is expressed by
\begin{equation}
\rho_{0}=(0.548k^{2}C_{n}^{2}z)^{-3/5}. \label{fixCLeq}%
\end{equation}
Then, weak turbulence corresponds to the condition
\begin{equation}
z\lesssim k\left[  \min\{2a_{R},\rho_{0}\}\right]  ^{2}. \label{YuraWR}%
\end{equation}
We note that, in our numerical investigations, Eq.~(\ref{Rytov}) turns out to
be more stringent than the condition in Eq.~(\ref{YuraWR}). In fact, for the
regime of day-time parameters considered in Fig.~\ref{FSpics} of the main
text, $\sigma_{\text{Rytov}}^{2}<1$ leads to $z\lesssim1066$~m, while
Eq.~(\ref{YuraWR}) implies $z\lesssim1657$~m.

In the regime of weak turbulence, we may distinguish the actions of small and
large turbulent eddies: Those smaller than the beam waist act on a fast
time-scale and broaden the waist; those larger than the beam waist act on a
slow time-scale ($10-100$ms) and randomly deflect the beam~\cite{Fante75}. The
overall action can be decomposed in the sum of two contributions, the
broadening of the diffraction-limited beam waist $w_{z}$ into the short-term
spot size $w_{\text{st}}$, and the random wandering of the beam centroid with
variance $\sigma_{\text{TB}}^{2}$. Averaging over all the dynamics, one has
the long-term spot size~\cite[Eq.~(32)]{Fante75}
\begin{equation}
w_{\text{lt}}^{2}=w_{\text{st}}^{2}+\sigma_{\text{TB}}^{2}. \label{turbmodel}%
\end{equation}

If we assume the validity of Yura's condition~\cite{Yura73,Fante75}
\begin{equation}
\phi:=0.33\left(  \frac{\rho_{0}}{w_{0}}\right)  ^{1/3}\ll1, \label{condsYURA}%
\end{equation}
then we can write decomposition in Eq.~(\ref{turbmodel}) where the long- and
short-term spot sizes take the following forms~\cite{Yura73,Fante75} (see also
Refs.~\cite{Poirier72,Bunkin70,Dios04,Belmonte})
\begin{align}
w_{\text{lt}}^{2}  &  \simeq w_{z}^{2}+2\left(  \frac{\lambda z}{\pi\rho_{0}%
}\right)  ^{2},\label{LTexpression}\\
w_{\text{st}}^{2}  &  \simeq w_{z}^{2}+2\left(  \frac{\lambda z}{\pi\rho_{0}%
}\right)  ^{2}(1-\phi)^{2}, \label{STexpression}%
\end{align}
and we may also expand%
\begin{equation}
(1-\phi)^{2}\simeq1-0.66\left(  \frac{\rho_{0}}{w_{0}}\right)  ^{1/3}.
\label{GammaFF2}%
\end{equation}
As a result, for the variance of centroid wandering, we derive the following
expression~\cite{Yura73}%
\begin{equation}
\sigma_{\text{TB}}^{2}=w_{\text{lt}}^{2}-w_{\text{st}}^{2}\simeq
\frac{0.1337\lambda^{2}z^{2}}{w_{0}^{1/3}\rho_{0}^{5/3}}. \label{SigmaCwander}%
\end{equation}

Note that, while the expression in Eq.~(\ref{LTexpression}) of the long-term
spot size $w_{\text{lt}}^{2}$ is valid under general
conditions~\cite[Eq.~(37)]{Fante75}, Yura's short-term expressions in
Eqs.~(\ref{STexpression}) and~(\ref{SigmaCwander}) are rigorous in the limit
$\phi\ll1$. These short-term expressions can also be considered good
approximations for $\rho_{0}/w_{0}<1$, i.e., for $\phi<0.33$. In the regime of
day-time parameters considered for Fig.~\ref{FSpics} of the main text, we have
that $\phi<0.33$ implies a minimum distance $z\gtrsim200~$m. (In other words,
our numerical investigation in that figure meets the `sweet spot' provided by
the range $200\leq z\leq1066$, where turbulence is weak and Yura's analytical
expansions are approximately correct).

When $\phi$ passes its threshold (i.e., $\rho_{0}/w_{0}\gtrsim1$), the
expansions in Eqs.~(\ref{STexpression}) and~(\ref{SigmaCwander}) become
imprecise and the correct value of $w_{\text{st}}^{2}$ needs to be numerically
derived from the $1/e$ point of the spherical-wave short-term mutual coherence
function (see Ref.~\cite{Yura73}). Alternatively, one can exploit
Eqs.~(41a),(41b) and Fig.~3 of Ref.~\cite{Fante75}. Once $w_{\text{st}}^{2}$
is known, then Eq.~(\ref{turbmodel}) can be used to derive $\sigma_{\text{TB}%
}^{2}$. When $\rho_{0}/w_{0}\gg1$, $\sigma_{\text{TB}}^{2}$ is negligible and
$w_{\text{st}}^{2}$ is equal to the long-term value $w_{\text{lt}}^{2}$ in
Eq.~(\ref{LTexpression}). The long-term spot-size $w_{\text{lt}}^{2}$ also
applies in the regime of strong turbulence $z\gg k\left[  \min\{2a_{R}%
,\rho_{0}\}\right]  ^{2}$, where the beam is broken up into multiple patches;
in this case, $w_{\text{lt}}^{2}$ describes the radius of the mean region
where the multiple patches are observed.

\begin{remark}
\label{AppNOTATION}The expressions in Eqs.~(\ref{LTexpression})
and~(\ref{STexpression}) are derived from Ref.~\cite[Eqs.~(16-18)]{Yura73} and
Ref.~\cite[Eq.~(37)]{Fante75}, changing their notation from intensity spot
size ($w^{\text{I}}$) to field spot size ($w=\sqrt{2}w^{\text{I}}$). In
Ref.~\cite{Fante75}, instead of $(1-\phi)^{2}$, we find%
\begin{equation}
\Psi=\left[  1-0.5523\left(  \frac{\rho_{0}}{w_{0}}\right)  ^{1/3}\right]
^{6/5}.
\end{equation}
Despite slightly different, its expansion for $\rho_{0}/w_{0}\ll1$ is the same
as in Eq.~(\ref{GammaFF2}). As a result the centroid wandering is
characterized by the same variance $\sigma_{\text{TB}}^{2}$ as in
Eq.~(\ref{SigmaCwander}), which is equivalent to Eq.~(40) of
Ref.~\cite{Fante75}. Also note that Yura's expressions take different forms in
terms of the Fried's parameter $\rho_{\mathrm{F}}=2.088\rho_{0}$%
~\cite{Fante80}. In fact, one may also write~\cite{Dios04,Belmonte}
\begin{equation}
w_{\mathrm{st}}^{2}\simeq w_{z}^{2}+2\left(  \frac{2.088\lambda z}{\pi
\rho_{\mathrm{F}}}\right)  ^{2}\left[  1-0.26\left(  \frac{\rho_{\mathrm{F}}%
}{w_{0}}\right)  ^{1/3}\right]  ^{2}.
\end{equation}

\end{remark}

\section{Random walk of the beam centroid\label{APP_centroid}}

Consider a random walk of the beam centroid $\vec{x}_{C}$ around an average
point $\vec{x}_{P}$ at distance $d$ from the center of the receiver $\vec
{x}_{R}$, following a Gaussian distribution with variance $\sigma^{2}$. The
distribution for the instantaneous deflection distance $r=\left\Vert \vec
{x}_{C}-\vec{x}_{R}\right\Vert \geq0$ will be Rician with parameters $d$ and
$\sigma$, i.e.,%
\begin{equation}
p(r|d,\sigma)=\frac{r}{\sigma^{2}}\exp\left(  -\frac{r^{2}+d^{2}}{2\sigma^{2}%
}\right)  I_{0}\left(  \frac{rd}{\sigma^{2}}\right)  . \label{RicianCentroid}%
\end{equation}
When the mean deflection is zero ($d=0$), Eq.~(\ref{RicianCentroid}) can be
simplified to the Weibull distribution $P_{\text{WB}}(r):=p(r|0,\sigma)$ of
Eq.~(\ref{WeibullDISTRIBUTION}) of the main text.

By combining the Rice distribution of the centroid $r$ given in
Eq.~(\ref{RicianCentroid}) with%
\begin{equation}
r=r_{0}\left(  \ln\frac{\eta}{\tau}\right)  ^{\frac{1}{\gamma}}:=r_{0}\Sigma,
\end{equation}
which is the inverse of Eq.~(\ref{Eqnst}) of the main text, one can easily
compute the probability distribution for the deflected transmissivity%
\begin{equation}
P(\tau)=\left[  p(r|d,\sigma)\right]  _{r=r(\tau)}\left\vert \frac{dr}{d\tau
}\right\vert .
\end{equation}
Explicitly, this takes the following form%
\begin{align}
&  P(\tau)=\frac{r_{0}^{2}\Sigma^{2-\gamma}}{\gamma\sigma^{2}\tau}I_{0}\left(
\frac{r_{0}d}{\sigma^{2}}\Sigma\right) \nonumber\\
&  \times\exp\left(  -\frac{r_{0}^{2}\Sigma^{2}+d^{2}}{2\sigma^{2}}\right)
,~\text{for }0<\tau\leq\eta, \label{PDTE}%
\end{align}
and zero otherwise.

The latter equation can also be derived by combining Ref.~\cite[Eq.~(8)]%
{Vasy12}, there written for the transmittance coefficient $\sqrt{\tau}$, with
the probability density of the squared variable $P(\tau)=(2\sqrt{\tau}%
)^{-1}P(\sqrt{\tau})$. Also note that the distribution in Eq.~(\ref{PDTE}) can
be bounded exploiting the inequality $I_{0}(x)\leq\cosh(x)\leq\exp(x)$ valid
for any $x\geq0$. For $x=0$ the equality holds, while for $x>0$ the upper
bound comes from the fact that we may write $I_{n}(x)<\frac{x^{n}}{2^{n}%
n!}\cosh(x)$ for $n=0,1,...$which can be easily proven starting from
Ref.~\cite[Eq.~(6.25)]{Luke72}. After simple algebra, we therefore find
\begin{align}
P(\tau)  &  \leq\frac{r_{0}^{2}\Sigma^{2-\gamma}}{\gamma\sigma^{2}\tau}%
\exp\left[  -\frac{\left(  r_{0}\Sigma-d\right)  ^{2}}{2\sigma^{2}}\right] \\
&  \leq\frac{r_{0}^{2}}{\gamma\sigma^{2}\tau}\left(  \ln\frac{\eta}{\tau
}\right)  ^{\frac{2}{\gamma}-1}.
\end{align}

Assuming zero mean deflection ($d=0$), Eq.~(\ref{PDTE}) simplifies to
$P_{0}(\tau)$ in Eq.~(\ref{P0tau}) of the main text. The probability
distribution $P_{0}(\tau)$ describes the statistics of the fading channel by
providing the instantaneous value of the deflected transmissivity $\tau$ for
the case where the average position of the beam centroid is aligned with the
center of the receiver's aperture.

\section{Achievability of the loss-based bounds\label{App_achievability}}

As long as the instantaneous (short-term) quantum channels can be approximated
to pure-loss channels $\mathcal{E}_{\tau}$, the upper bound in
Eq.~(\ref{PLOBdelta}) of the main text is an achievable rate for secret key
generation and entanglement distribution. In fact, the PLOB upper-bound
$\Phi(\tau)=-\log_{2}(1-\tau)$ of each $\mathcal{E}_{\tau}$ is achievable,
i.e., there are optimal protocols whose rates saturate this ultimate limit for
all the relevant capacities, so that we have $Q_{2}(\mathcal{E}_{\tau}%
)=D_{2}(\mathcal{E}_{\tau})=K(\mathcal{E}_{\tau})=\Phi(\tau)$. In fact, a
pure-loss channel is known to be distillable~\cite{QKDpaper}, which means that
the upper bound $\Phi(\tau)$, based on the REE, is achievable by a protocol of
entanglement distribution, leading to $D_{2}(\mathcal{E}_{\tau})=\Phi(\tau)$.

In particular, it is sufficient to consider a protocol where the entanglement is
distributed and then distilled with the help of a single round of feedback classical communication~\cite{CInfo}. This protocol may achieve a rate that is at least the reverse
coherent information of the channel $I_{\text{RCI}}(\mathcal{E}_{\tau}%
)=-\log_{2}(1-\tau)$~\cite{RCI}. Once this entanglement has been distilled, it
can also be used to transmit qubits via teleportation or to generate secret keys.


If we are interested in QKD\ only, then there are different asymptotic ways to
reach the PLOB upper bound, i.e., the secret key capacity $K$ of the pure-loss
channel. This is certainly possible by using a QKD protocol equipped with a
quantum memory as discussed in Ref.~\cite{QKDpaper}. An alternative method is
to use a strongly-biased QKD protocol with squeezed states~\cite{Cerf}.
Suppose that, with probability $p$, the transmitter prepares a
position-squeezed state with CM $\mathrm{diag}(\mu
^{-1},\mu)$. With probability $1-p$, it instead prepares a momentum-squeezed
state with CM $\mathrm{diag}(\mu,\mu^{-1})$. In each case, the mean value of the squeezed quadrature is Gaussianly modulated with variance $\mu-\mu^{-1}$, so
that the average output state is an isotropic thermal state with variance
$\mu=2\bar{n}_{T}+1$, where $\bar{n}_{T}$ is the mean number of photons. These
states are sent through the link and measured at the receiver by an homodyne
detector switching between position and momentum with the same probability
distribution of the transmitter. Finally, the parties perform a sifting
process where they only select their matching choices of the quadrature, which
happens with frequency $p^{2}+(1-p)^{2}$.

Assume that the communication is long enough (asymptotic limit of infinite
signals exchanged), so that the parties access many times the
instantaneous\ pure-loss channel $\mathcal{E}_{\tau}$ for some $\tau$ (within
some small resolution $\delta\tau$). For large $\mu$, we can compute the
following mutual information between transmitter and receiver%
\begin{equation}
I_{TR|p,\tau}\simeq\frac{p^{2}+(1-p)^{2}}{2}\log_{2}\left(  \frac{\tau\mu
}{1-\tau}\right)  .
\end{equation}
Assuming reverse reconciliation, where the variable to be inferred is the
outcome of the receiver, we have that the eavesdropper's information cannot
exceed the Holevo bound%
\begin{equation}
\chi_{ER|\tau}\simeq\frac{1}{2}\log_{2}[(1-\tau)\tau\mu].
\end{equation}
The asymptotic (conditional) rate is equal to
\begin{equation}
R_{\text{sq}}(p,\tau):=I_{TR|p,\tau}-\chi_{ER|\tau}.
\end{equation}
For an unbiased protocol ($p=1/2$), we have $R_{\text{sq}}(1/2,\tau)=\Phi
(\tau)/2$. In the limit of a completely biased protocol ($p\rightarrow1$), we
instead find $R_{\text{sq}}(1^{-},\tau)\rightarrow\Phi(\tau)$.

It is clear that this is the same performance that could be achieved by an
equivalent entanglement-based protocol where the transmitter sends the
$B$-modes of TMSV states (with large variance $\mu$), keeps their $A$-modes in
a quantum memory, and finally homodynes the $A$-modes once the receiver
classically communicates which detection was in the position quadrature and
which was in the momentum one~\cite{RCI,QKDpaper}.

Let us now account for the fading process, according to which the
instantaneous transmissivity $\tau$ occurs with probability density
$P_{0}(\tau)$. In a coarse-graining description of the process, one has a
large number of instantaneous channels with transmissivities contained in
slots $[0,\delta\tau],[\delta\tau,2\delta\tau],\ldots\lbrack(k-1)\delta
\tau,k\delta\tau],\ldots$ up to a maximum value $\eta$, given by
$\eta_{\text{st}}\eta_{\text{eff}}\eta_{\text{atm}}$. Each slot is used a
large (virtually infinite) number of times. Therefore, we can take a suitable
joint limit for small $\delta\tau$, and approximate the weighted sum of rates
with an integral. For the case of the squeezed-state protocol, we write the
average rate
\begin{equation}
R_{\text{sq}}(p)=\int_{0}^{\eta}d\tau~P_{0}(\tau)R_{\text{sq}}(p,\tau).
\end{equation}
In the biased limit $p\rightarrow1^{-}$, we have that the achievable rate of
the fading channel $\{P_{0}(\tau),\mathcal{E}_{\tau}\}$ is%
\begin{equation}
R_{\text{sq}}(1^{-})\rightarrow\int_{0}^{\eta}d\tau~P_{0}(\tau)\Phi(\tau),
\end{equation}
which coincides with the upper bound of Eq.~(\ref{PLOBdelta}) of the main
text. In other words, this bound is asymptotically achievable by this ideal
QKD protocol.

It is clear that the squeezed-state protocol just represents a theoretical
tool to demonstrate the achievability of the bound, but it is not realizable
with current technology. Consider now the protocol of Ref.~\cite{GG02}, where
the transmitter Gaussianly modulates coherent states and the receiver performs
homodyne detection switching between the two quadratures. In the large
modulation limit, one computes the instantaneous rate $R_{\text{coh}}%
(\tau)=\Phi(\tau)/2$, so that we have the average value
\begin{equation}
R_{\text{coh}}=\frac{1}{2}\int_{0}^{\eta}d\tau~P_{0}(\tau)\Phi(\tau),
\end{equation}
achieving half of the bound.

\section{Free-space bounds with thermal noise\label{APP_ThermalBounds}}

\subsection{Thermal-noise model\label{APP_subsection_Thermal}}

During day-time operation, background thermal noise may become non-trivial.
For this reason, we need to suitably modify the description of the free-space
channel and derive more appropriate bounds. In the presence of non-negligible
noise, an instantaneous (short-term) quantum channel can be approximated by an
overall thermal-loss channel $\mathcal{E}_{\tau,\bar{n}}$ between transmitter
and receiver. More precisely, assume that $\bar{n}_{T}$ is the mean number of
photons in the mode generated by the transmitter. Then, the mean number of
photons $\bar{n}_{R}$ reaching the receiver's detector is given by the
input-output relation%
\begin{equation}
\bar{n}_{T}\rightarrow\bar{n}_{R}=\tau\bar{n}_{T}+\bar{n}, \label{IOenergy}%
\end{equation}
where $\tau$ is the instantaneous transmissivity, and $\bar{n}=\eta
_{\text{eff}}\bar{n}_{B}+\bar{n}_{\text{ex}}$ is the channel's thermal number,
given by the detected environmental photons $\eta_{\text{eff}}\bar{n}_{B}$
plus extra photons $\bar{n}_{\text{ex}}$ added by the receiver's setup. To
understand Eq.~(\ref{IOenergy}), see also Fig.~\ref{detectorPIC} of the main text.

The instantaneous channel $\mathcal{E}_{\tau,\bar{n}}$ can equivalently be
described by a beam splitter with transmissivity $\tau$ mixing an input mode
with an environmental mode with mean number of photons $\bar{n}_{e}=\bar
{n}/(1-\tau)$. Channel's transmissivity $\tau$ varies between $0$ and a
maximum value $\eta$ according to the probability density $P_{0}(\tau)$
determined by turbulence and pointing error. The mean number of thermal
photons $\bar{n}$ can be assumed to be constant by assuming a
suitably-stabilized receiver setup (with negligible fluctuations in $\bar
{n}_{\text{ex}}$) and stable conditions for the external background (so that
the photons collected within the field of view are approximately constant). If
this assumption is not met, then we can always make $\bar{n}$ constant by
maximizing it over $\tau$ (worst-case scenario, suitable for the lower bound)
or minimizing it over $\tau$ (best-case scenario, suitable for the upper
bound). For this reason, we can always model the free-space fading channel
$\mathcal{E}$ as an ensemble $\{P_{0}(\tau),\mathcal{E}_{\tau,\bar{n}}\}$,
whose elements have variable $\tau$ but constant $\bar{n}$.

\subsection{Upper and lower bounds}

Given the asymptotic rate $R(\mathcal{E}_{\tau,\bar{n}})$ associated with a
generic instantaneous channel $\mathcal{E}_{\tau,\bar{n}}$, the asymptotic
rate of the free-space link $\mathcal{E}$ is given by the average
\begin{equation}
R=\int_{0}^{\eta}d\tau~P_{0}(\tau)R(\mathcal{E}_{\tau,\bar{n}}).
\label{toreplaceinf}%
\end{equation}
This rate is asymptotically achievable if the fading dynamics is perfectly
resolved by detectors and a large (virtually infinite) number of signals are
allocated to each infinitesimal slot $[\tau,\tau+d\tau]$. It also assumes that
the adaptive optics completely eliminates any average offset $d$ of the beam's
centroid [otherwise $P_{0}$ is replaced by the more general distribution in
Eq.~(\ref{PDTE})].

Because the instantaneous channel is a thermal-loss channel $\mathcal{E}%
_{\tau,\bar{n}}$, we do not know its two-way assisted capacities
$D_{2}(\mathcal{E}_{\tau,\bar{n}})=Q_{2}(\mathcal{E}_{\tau,\bar{n}})\leq
K(\mathcal{E}_{\tau,\bar{n}})$ and we are limited to consider upper and lower
bounds. The secret key capacity\ is upperbounded by the thermal-loss version
of the PLOB bound $K(\mathcal{E}_{\tau,\bar{n}})\leq\Phi(\tau,\bar{n})$, given
by%
\begin{equation}
\Phi(\tau,\bar{n})=-\log_{2}\left[  (1-\tau)\tau^{\frac{\bar{n}}{1-\tau}%
}\right]  -h\left(  \frac{\bar{n}}{1-\tau}\right)  ,
\end{equation}
for $\bar{n}\leq\tau$, while $\Phi(\tau,\bar{n})=0$ for $\bar{n}\geq\tau$. In
the previous formula, the entropic quantity $h$\ is defined as in
Eq.~(\ref{hFUNCTIONmain}), i.e., we have
\begin{equation}
h\left(  x\right)  :=(x+1)\log_{2}(x+1)-x\log_{2}x. \label{hENTfunction}%
\end{equation}
As a result, any key rate associated with the fading channel $\mathcal{E}%
=\{P_{0}(\tau),\mathcal{E}_{\tau,\bar{n}}\}$ cannot exceed the thermal bound
\begin{equation}
R\leq\int_{\bar{n}}^{\eta}d\tau~P_{0}(\tau)\Phi(\tau,\bar{n}),
\label{ThBoundEEE}%
\end{equation}
which is different from zero when $\bar{n}\leq\eta=\eta_{\text{st}}%
\eta_{\text{eff}}\eta_{\text{atm}}$.

Let us define the normalization factor%
\begin{align}
&  \mathcal{N}(\bar{n},\eta,\sigma):=\int_{\bar{n}}^{\eta}d\tau~P_{0}(\tau)\\
&  =1-\exp\left\{  -\frac{r_{0}^{2}}{2\sigma^{2}}\left[  \ln\left(  \frac
{\eta}{\bar{n}}\right)  \right]  ^{\frac{2}{\gamma}}\right\}  ,
\end{align}
and the following entropic quantity%
\begin{align}
g\left(  \bar{n}\right)   &  :=\frac{\bar{n}\log_{2}\bar{n}}{1-\bar{n}%
}+h\left(  \bar{n}\right) \\
&  =(\bar{n}+1)\log_{2}(\bar{n}+1)+\frac{\bar{n}^{2}\log_{2}\bar{n}}{1-\bar
{n}}.
\end{align}

For $\bar{n}\leq\eta$, we may therefore write%
\begin{align}
R  &  \leq-\int_{\bar{n}}^{\eta}d\tau~P_{0}(\tau)\left[  \log_{2}%
(1-\tau)\right. \nonumber\\
&  \left.  +\frac{\bar{n}}{1-\tau}\log_{2}\tau+h\left(  \frac{\bar{n}}{1-\tau
}\right)  \right] \\
&  \leq-\int_{\bar{n}}^{\eta}d\tau~P_{0}(\tau)\log_{2}(1-\tau)\nonumber\\
&  -\left[  \frac{\bar{n}\log_{2}\bar{n}}{1-\bar{n}}+h\left(  \bar{n}\right)
\right]  \int_{\bar{n}}^{\eta}d\tau~P_{0}(\tau)\\
&  \leq\mathcal{B}(\eta,\sigma)-\mathcal{T}(\bar{n},\eta,\sigma),
\label{Th-UB}%
\end{align}
where $\mathcal{B}(\eta,\sigma)=-\Delta(\eta,\sigma)\log_{2}(1-\eta)$ is the
pure-loss upper bound [cf. Eqs.~(\ref{PLOBdelta}) and~(\ref{deltaEXPRESSION})
of the main text], and $\mathcal{T}(\bar{n},\eta,\sigma)$ is a thermal
correction given by
\begin{equation}
\mathcal{T}(\bar{n},\eta,\sigma)=g\left(  \bar{n}\right)  \mathcal{N}(\bar
{n},\eta,\sigma)-\Delta(\bar{n},\sigma)\log_{2}(1-\bar{n}).
\end{equation}

Let us now discuss lower bounds. For each short-term instantaneous channel, an
asymptotically achievable rate $R(\mathcal{E}_{\tau,\bar{n}})$ is given by the
reverse coherent information~\cite{RCI}, here taking the following form
\begin{equation}
I_{\text{RCI}}(\mathcal{E}_{\tau,\bar{n}})=-\log_{2}(1-\tau)-h\left(
\frac{\bar{n}}{1-\tau}\right)  .
\end{equation}
Replacing this expression in Eq.~(\ref{toreplaceinf}) provides an achievable
rate for entanglement distribution and secret key generation via the
free-space link. Explicitly, we write%
\begin{align}
R  &  \geq\mathcal{B}(\eta,\sigma)-\int_{0}^{\eta}d\tau~P_{0}(\tau)h\left(
\frac{\bar{n}}{1-\tau}\right) \label{RCI-LB}\\
&  \geq\mathcal{B}(\eta,\sigma)-h\left(  \frac{\bar{n}}{1-\eta}\right)  .
\label{Th-LB}%
\end{align}

If we look at QKD, we can consider two specific protocols. For an
asymptotically-biased squeezed-state protocol ($p\rightarrow1^{-}$), we can
write the short-term rate $R_{\text{sq}}(1^{-},\tau,\bar{n})\rightarrow
I_{\text{RCI}}(\mathcal{E}_{\tau,\bar{n}})$. For the coherent-state protocol,
we can instead write%
\begin{align}
R_{\text{coh}}(\tau,\bar{n})  &  =\Phi(\tau)-h\left(  \frac{\bar{n}}{1-\tau
}\right) \nonumber\\
&  +\frac{1}{2}\log_{2}\left(  1-\frac{\tau}{2\bar{n}+1}\right)  .
\label{homprotRATE}%
\end{align}
Replacing these expressions in Eq.~(\ref{toreplaceinf}) provides
asymptotically-achievable QKD rates for the free-space link. In particular
note that, for small $\bar{n}$, we can expand
\begin{equation}
R_{\text{coh}}(\tau,\bar{n})\simeq\frac{\Phi(\tau)}{2}-h\left(  \frac{\bar{n}%
}{1-\tau}\right)  ,
\end{equation}
and write the following rate for the link%
\begin{equation}
R_{\text{coh}}\geq\frac{\mathcal{B}(\eta,\sigma)}{2}-h\left(  \frac{\bar{n}%
}{1-\eta}\right)  . \label{coh-untrusted}%
\end{equation}

In conclusion, according to our derivations, the optimal rates for
entanglement distribution and key generation in the presence of background
thermal noise can be bounded by the following sandwich relation
\begin{equation}
\mathcal{B}(\eta,\sigma)-h\left(  \frac{\bar{n}}{1-\eta}\right)  \leq
R\leq\mathcal{B}(\eta,\sigma)-\mathcal{T}(\bar{n},\eta,\sigma).
\label{sandRATE}%
\end{equation}
One can check that these inequalities collapse to single loss-based bound
$R\simeq\mathcal{B}(\eta,\sigma)$ for small thermal numbers $\bar{n}$ (e.g.
compatible with night-time operation).

\section{More details on the composable security of
CV-QKD\label{MoreDetailsAPP}}

\subsection{Composable key rate under collective attacks}

Consider a CV-QKD protocol where $N$ modes are transmitted from Alice $A$
(transmitter) to Bob $B$ (receiver). A portion $n$ of these modes will be used
for key generation, while the remaining part is used for parameter estimation
(and other potential operations). Here we start by assuming perfect knowledge
of the channel parameters; afterwards we will include the effect of imperfect
knowledge as coming from parameter estimation.

Let us call $x$ Alice's variable and $y$ Bob's variable. In the homodyne
protocol, the relevant quadrature is selected by Bob's randomly-switched
measurement of $\hat{q}$ and $\hat{p}$. Therefore, $x$ and $y$ represent
Alice's quadrature encoding and the corresponding Bob's outcome after the
random selection imposed by the measurement. In the heterodyne protocol, these
variables are instead bi-dimensional real vectors associated to both
quadratures, so that we have $x=(q_{A},p_{A})$ and $y=(q_{B},p_{B})$. The
continuous variables are subject to analog-to-digital conversion (ADC), so
that $x\overset{\text{ADC}}{\rightarrow}k$ and $y\overset{\text{ADC}%
}{\rightarrow}l$, where $k$ and $l$ are $d$-bit strings. Note that, for the
heterodyne protocol, ADC\ may occur independently for each quadrature
$(q_{A},p_{A})\overset{\text{ADC}}{\rightarrow}(l_{q},l_{p})$ after which one
may concatenate $l=l_{q}l_{p}$. In such a case, we assume that each quadrature
component is digitalized with $d/2$ bits (for even $d$).

Under the action of a collective attack, the output classical-quantum (CQ) state of
Alice ($A$), Bob ($B$) and Eve ($E$) has the tensor-structure form
$\rho^{\otimes n}$, where%
\begin{equation}
\rho=%
{\displaystyle\sum\limits_{k,l}}
p(k,l)\left\vert k\right\rangle _{A}\left\langle k\right\vert \otimes
\left\vert l\right\rangle _{B}\left\langle l\right\vert \otimes\rho_{E}(k,l),
\label{sharedSSS}%
\end{equation}
and $p(k,l)$ is a joint probability distribution. For $n$ uses, there will be
two sequences, $k^{n}$ and $l^{n}$, with binary length $n\log_{2}d$ and
associated probability $p(k^{n},l^{n})$. Alice and Bob will then perform
procedures of error correction and privacy amplification over the state
$\rho^{\otimes n}$ in order to approximate the $s_{n}$-bit ideal CQ state%
\begin{equation}
\rho_{\text{id}}:=2^{-s_{n}}%
{\displaystyle\sum\limits_{z=0}^{2^{s_{n}}-1}}
\left\vert z\right\rangle _{A^{n}}\left\langle z\right\vert \otimes\left\vert
z\right\rangle _{B^{n}}\left\langle z\right\vert \otimes\rho_{E^{n}},
\end{equation}
where Alice's and Bob's classical systems contain the same random sequence
$z$\ of binary length $s_{n}$ from which Eve is completely decoupled.

In reverse reconciliation, it is Alice attempting to reconstruct Bob's
sequence $l^{n}$. During the step of error correction, Bob reveals
$\mathrm{leak}_{\text{ec}}$ bits of information to help Alice to compute her
guess $\tilde{l}^{n}$ of $l^{n}$ starting from her local data $k^{n}$. In a
practical scheme, these $\mathrm{leak}_{\text{ec}}$ bits of information
correspond to a syndrome that Bob computes over his sequence $l^{n}$,
interpreted as noisy codeword of a linear error-correcting code agreed with Alice.

Then, as a verification, Alice and Bob publicly compare hashes computed over
$l^{n}$ and $\tilde{l}^{n}$. If these hashes coincide, the two parties go
ahead with probability $p_{\text{ec}}$, otherwise they abort the protocol. The
hash comparison requires Bob sending $\left\lceil -\log_{2}\varepsilon
_{\text{cor}}\right\rceil $ bits to Alice for some\ suitable $\varepsilon
_{\text{cor}}$ (the number of these bits is negligible in comparison to
$\mathrm{leak}_{\text{ec}}$). Parameter $\varepsilon_{\text{cor}}$ is called
$\varepsilon$-correctness~\cite[Sec.~4.3]{Portmann} and it bounds the
probability that the sequences are different even if their hashes coincide.
The probability of such an error is bounded by~\cite{TomaEpsCorr}
\begin{equation}
p_{\text{ec}}\mathrm{Prob}(\tilde{l}^{n}\neq l^{n})\leq p_{\text{ec}%
}2^{-\left\lceil -\log_{2}\varepsilon_{\text{cor}}\right\rceil }%
\leq\varepsilon_{\text{cor}}.
\end{equation}
Note that $p_{\text{ec}}$ and $\varepsilon_{\text{cor}}$ are implicitly
related. In fact, the lower is the value of $\varepsilon_{\text{cor}}$, the
stronger is the hash-verification test made over the sequences $l^{n}$ and
$\tilde{l}^{n}$, which results into a lower probability of success
$p_{\text{ec}}$.

Error correction can be simulated by a projection $\Pi_{\mathcal{S}}$ of
Alice's and Bob's classical systems $A^{n}$ and $B^{n}$\ onto a
\textquotedblleft good\textquotedblright\ set $\mathcal{S}$ of sequences. With
success probability
\begin{equation}
p_{\text{ec}}=\mathrm{Tr}(\Pi_{\mathcal{S}}\rho^{\otimes n}), \label{pecPPPP}%
\end{equation}
this operation generates a CQ state
\begin{equation}
\tilde{\rho}^{n}:=p_{\text{ec}}^{-1}~\Pi_{\mathcal{S}}\rho^{\otimes n}%
\Pi_{\mathcal{S}}, \label{CQProjected}%
\end{equation}
which is restricted to those good sequences $\{k^{n},l^{n}\}$ that can be
transformed into a successful pair $\{\tilde{l}^{n},l^{n}\}$ by Alice's
transformation $k^{n}\rightarrow\tilde{l}^{n}$. We implicitly assume that the
latter transformation is performed on the state $\tilde{\rho}^{n}$ so that it
provides the pair $\{\tilde{l}^{n},l^{n}\}$ for next manipulations.

With probability $p_{\text{ec}}$ the protocol proceeds to privacy
amplification, where the parties apply a two-way hash function over
$\tilde{\rho}^{n}$ which outputs the privacy amplified state $\bar{\rho}^{n}$,
i.e., $\rho^{\otimes n}\overset{\text{ec}}{\longrightarrow}\tilde{\rho}%
^{n}\overset{\text{pa}}{\longrightarrow}\bar{\rho}^{n}$. The latter state
approximates the ideal private state $\rho_{\text{id}}$, so that we may write
$p_{\text{ec}}D(\bar{\rho}^{n},\rho_{\text{id}})\leq\varepsilon_{\text{sec}}$
where $\varepsilon_{\text{sec}}$ is the $\varepsilon$-secrecy of the
protocol~\cite[Sec.~4.3]{Portmann}. Via the triangle inequality, this
condition implies~\cite[Th.~4.1]{Portmann}
\begin{equation}
p_{\text{ec}}D(\tilde{\rho}^{n},\rho_{\text{id}})\leq\varepsilon
:=\varepsilon_{\text{cor}}+\varepsilon_{\text{sec}}, \label{traceIII}%
\end{equation}
and the protocol is said to be $\varepsilon$-secure.

Thanks to the procedure of two-universal hashing applied to $\tilde{\rho}^{n}%
$, Alice and Bob's state $\bar{\rho}^{n}$ will contain $s_{n}$ bits of shared
uniform randomness. According to Ref.~\cite{TomaRenner} (see also Ref.~\cite[Eq.~(8.7)]{TomaThesis}), we have that $s_{n}$ satisfies the direct
leftover hash bound%
\begin{equation}
s_{n}\geq H_{\text{min}}^{\varepsilon_{\text{s}}}(l^{n}|E^{n})_{\tilde{\rho
}^{n}}+2\log_{2}\sqrt{2}\varepsilon_{\text{h}}-\mathrm{leak}_{\text{ec}}.
\label{eeq2}%
\end{equation}
Here $H_{\text{min}}^{\varepsilon_{\text{s}}}(l^{n}|E^{n})_{\tilde{\rho}^{n}}$
is the smooth min-entropy of Bob's sequence $l^{n}$ conditioned on Eve's
system $E^{n}$, and the smoothing $\varepsilon_{\text{s}}$ and hashing
$\varepsilon_{\text{h}}$ parameters\ satisfy
\begin{equation}
\varepsilon_{\text{s}}+\varepsilon_{\text{h}}=\varepsilon_{\text{sec}}.
\end{equation}
In Eq.~(\ref{eeq2}) we explicitly account for the bits leaked to Eve during
error correction. In fact, one may write $s_{n}\geq H_{\text{min}%
}^{\varepsilon_{\text{s}}}(l^{n}|E^{n}R)_{\tilde{\rho}^{n}}+2\log_{2}\sqrt
{2}\varepsilon_{\text{h}}$ where $R$ is a register of dimension $d_{R}%
=2^{\mathrm{leak}_{\text{ec}}}$, while $E^{n}$ are the systems used by Eve
during the quantum communication. Then, the chain rule for the smooth-min
entropy leads to $H_{\text{min}}^{\varepsilon_{\text{s}}}(l^{n}|E^{n}%
R)_{\tilde{\rho}^{n}}\geq H_{\text{min}}^{\varepsilon_{\text{s}}}(l^{n}%
|E^{n})_{\tilde{\rho}^{n}}-\log_{2}d_{R}$.

As next step, we revise and improve a previous result which connects the
smooth-min entropies of $\tilde{\rho}^{n}$ and $\rho^{\otimes n}$. In fact, we
may show that%
\begin{align}
H_{\text{min}}^{\varepsilon_{\text{s}}}(l^{n}|E^{n})_{\tilde{\rho}^{n}}  &
\geq H_{\text{min}}^{p_{\text{ec}}\varepsilon_{\text{s}}^{2}/3}(l^{n}%
|E^{n})_{\rho^{\otimes n}}\nonumber\\
&  +\log_{2}[p_{\text{ec}}(1-\varepsilon_{\text{s}}^{2}/3)]. \label{eeq3}%
\end{align}
Because $H_{\text{min}}^{\varepsilon_{\text{s}}}$\ only depends on Bob and
Eve's parts of the state $\tilde{\rho}^{n}$, one could trace Alice's system
$\tilde{\rho}^{n}\rightarrow$\textrm{tr}$_{A}\tilde{\rho}^{n}$ and write the
bound above directly for the reduced state. See Appendix~\ref{proofPC}\ for a
proof of Eq.~(\ref{eeq3}) which exploits tools from
Refs.~\cite{TomaThesis,Cosmo_comp}.

Next, we simplify the smooth-min entropy term via the asymptotic equipartition
property~\cite[Cor.~6.5]{TomaThesis}%
\begin{align}
H_{\text{min}}^{p_{\text{ec}}\varepsilon_{\text{s}}^{2}/3}(l^{n}|E^{n}%
)_{\rho^{\otimes n}}  &  \geq nH(l|E)_{\rho}\nonumber\\
&  -\sqrt{n}\Delta_{\text{aep}}\left(  p_{\text{ec}}\varepsilon_{\text{s}}%
^{2}/3,d\right)  , \label{eeq4}%
\end{align}
where $H(l|E)_{\rho}$ is the conditional von Neumann entropy computed over the
single-copy state $\rho$, and~\cite[Th.~6.4]{TomaThesis}
\begin{align}
\Delta_{\text{aep}}(\varepsilon_{\text{s}},d)  &  :=4\log_{2}\left(  2\sqrt
{d}+1\right)  \sqrt{-\log_{2}\left(  1-\sqrt{1-\varepsilon_{\text{s}}^{2}%
}\right)  }\nonumber\\
&  \simeq4\log_{2}\left(  2\sqrt{d}+1\right)  \sqrt{\log_{2}(2/\varepsilon
_{\text{s}}^{2})}, \label{AEPd}%
\end{align}
with $d$ being the cardinality of the discretized variable $l$.

The combination of Eqs.~(\ref{eeq2}), (\ref{eeq3}) and~(\ref{eeq4}) allows us
to write the following lower bound%
\begin{gather}
s_{n}\geq nH(l|E)_{\rho}-\sqrt{n}\Delta_{\text{aep}}\left(  p_{\text{ec}%
}\varepsilon_{\text{s}}^{2}/3,d\right) \nonumber\\
+\log_{2}[p_{\text{ec}}(1-\varepsilon_{\text{s}}^{2}/3)]+2\log_{2}\sqrt
{2}\varepsilon_{\text{h}}-\mathrm{leak}_{\text{ec}}. \label{form}%
\end{gather}
Note that, for the conditional entropy, we have
\begin{equation}
H(l|E)_{\rho}=H(l)-\chi(l:E)_{\rho}, \label{ff1}%
\end{equation}
where $H(l)$ is the Shannon entropy of $l$, and $\chi(l:E)_{\rho}$ is Eve's
Holevo bound with respect to $l$. Because of the data processing inequality,
we have $\chi(l:E)_{\rho}\leq\chi(y:E)_{\rho}$\ under digitalization
$y\overset{\text{ADC}}{\rightarrow}l$, so that we may write
\begin{equation}
H(l|E)_{\rho}\geq H(l)-\chi(y:E)_{\rho}. \label{ff1b}%
\end{equation}

Moreover, we may define the reconciliation parameter $\beta\in\lbrack0,1]$ by
setting
\begin{equation}
H(l)-n^{-1}\mathrm{leak}_{\text{ec}}=\beta I(x:y), \label{ff2}%
\end{equation}
where $I(x:y)\geq I(k:l)$ is Alice and Bob's mutual information computed over
their continuous variables. By replacing Eqs.~(\ref{ff1b}) and~(\ref{ff2}) in
Eq.~(\ref{form}), we derive%
\begin{align}
s_{n}  &  \geq nR_{\infty}-\sqrt{n}\Delta_{\text{aep}}\left(  p_{\text{ec}%
}\varepsilon_{\text{s}}^{2}/3,d\right) \nonumber\\
&  +\log_{2}[p_{\text{ec}}(1-\varepsilon_{\text{s}}^{2}/3)]+2\log_{2}\sqrt
{2}\varepsilon_{\text{h}}, \label{snprima}%
\end{align}
where we have introduced the asymptotic rate%
\begin{equation}
R_{\infty}=\beta I(x:y)-\chi(y:E)_{\rho}.
\end{equation}
The lower bound in Eq.~(\ref{snprima}) refers to a protocol with security
$\varepsilon=\varepsilon_{\text{cor}}+\varepsilon_{\text{s}}+\varepsilon
_{\text{h}}$ and success probability $p_{\text{ec}}$.

Let us account for the effect of parameter estimation. The asymptotic key rate
$R_{\infty}$ depends on a number $n_{\text{pm}}$ of parameters $\mathbf{p}$
(e.g., transmissivity and thermal noise of the channel). By sacrificing $m$
modes, Alice and Bob compute maximum likelihood estimators $\mathbf{\hat{p}}$
with associated mean values $\mathbf{\bar{p}}$\ and error-variances
$\sigma_{\mathbf{p}}^{2}$. Then, they compute worst-case estimators
$\mathbf{p}_{\text{wc}}$\ which are $w$ standard-deviations away from the mean
values of the estimators or they are computed by employing suitable tail
bounds for the variables involved. Each worst-case estimator bounds the
corresponding actual parameter up to an error probability $\varepsilon
_{\text{pe}}=\varepsilon_{\text{pe}}(w)$, so that all together the $n_{\text{pm}}$
worst-case estimators $\mathbf{p}_{\text{wc}}$\ bound the parameters
$\mathbf{p}$\ up to a total error probability $\simeq n_{\text{pm}}%
\varepsilon_{\text{pe}}$. Correspondingly, the key rate $R_{\infty}%
(\mathbf{p})$ is replaced by $R_{\text{pe}}:=R_{\infty}(\mathbf{p}_{\text{wc}%
})$.

Note that assuming $\mathbf{p}_{\text{wc}}$ for the quantum channel is
equivalent to change the global output $\tilde{\rho}^{n}$ of Alice, Bob and
Eve with a worst-case state $\tilde{\rho}_{\text{wc}}^{n}$ (described by
parameters that are at least as good as the worst-case estimators). However,
with probability $n_{\text{pm}}\varepsilon_{\text{pe}}$, one could have a
different state $\tilde{\rho}_{\text{bad}}^{n}$ with a lower rate (where one
or more parameters violate the worst-case estimators). On average, the state
could be modelled as $\rho_{\text{pe}}:=(1-n_{\text{pm}}\varepsilon
_{\text{pe}})\tilde{\rho}_{\text{wc}}^{n}+n_{\text{pm}}\varepsilon_{\text{pe}%
}\tilde{\rho}_{\text{bad}}^{n}$ with trace distance $D(\rho_{\text{pe}}%
,\tilde{\rho}_{\text{wc}}^{n})\leq n_{\text{pm}}\varepsilon_{\text{pe}}$. From
$D(\tilde{\rho}_{\text{wc}}^{n},\rho_{\text{id}})\leq\varepsilon/p_{\text{ec}%
}$ [cf. Eq.~(\ref{traceIII})]\ and the triangle inequality, we compute%
\begin{equation}
D(\rho_{\text{pe}},\rho_{\text{id}})\leq\varepsilon/p_{\text{ec}}%
+n_{\text{pm}}\varepsilon_{\text{pe}}. \label{fromEQQ}%
\end{equation}
Thus, the average state $\rho_{\text{pe}}$ is $(\varepsilon/p_{\text{ec}%
}+n_{\text{pm}}\varepsilon_{\text{pe}})$-close to an ideal private state
$\rho_{\text{id}}$ whose number of secret bits $s_{n}$ is lower-bounded by
Eq.~(\ref{snprima}) up to replacing $R_{\infty}\rightarrow R_{\text{pe}}$. It
is clear that parameter estimation adds an overall error $p_{\text{ec}%
}n_{\text{pm}}\varepsilon_{\text{pe}}$ to the $\varepsilon$-security of the
protocol, so that we have $\varepsilon\rightarrow\varepsilon+p_{\text{ec}%
}n_{\text{pm}}\varepsilon_{\text{pe}}$, as is clear from Eq.~(\ref{fromEQQ}).

Replacing $R_{\infty}\rightarrow R_{\text{pe}}$ in Eq.~(\ref{snprima}),
dividing by $N=n+m$ and including $p_{\text{ec}}$, we derive the following
bound for the composable secret key rate (bits per use) of a generic
CV-QKD\ protocol under collective attacks%
\begin{align}
R_{n}  &  :=\frac{p_{\text{ec}}s_{n}}{N}\geq\nonumber\\
&  \frac{p_{\text{ec}}}{N}\left\{  nR_{\text{pe}}-\sqrt{n}\Delta_{\text{aep}%
}\left(  p_{\text{ec}}\varepsilon_{\text{s}}^{2}/3,d\right)  \right.
\nonumber\\
&  \left.  +\log_{2}[p_{\text{ec}}(1-\varepsilon_{\text{s}}^{2}/3)]+2\log
_{2}\sqrt{2}\varepsilon_{\text{h}}\right\}  , \label{sckeee2}%
\end{align}
which is valid for a protocol with success probability $p_{\text{ec}}$ (or
frame error rate $1-p_{\text{ec}}$)\ and overall security
\begin{equation}
\varepsilon=\varepsilon_{\text{cor}}+\varepsilon_{\text{s}}+\varepsilon
_{\text{h}}+p_{\text{ec}}n_{\text{pm}}\varepsilon_{\text{pe}}.
\end{equation}
The expression in Eq.~(\ref{sckeee2}) corresponds to Eq.~(\ref{sckeee}) in the
main text.

\subsection{Proof of Eq.~(\ref{eeq3})\label{proofPC}}

Consider an arbitrary Hilbert space $\mathcal{H}$ and\ two generally
sub-normalized states $\rho,\rho_{\ast}\in S_{\leq}(\mathcal{H})$ with
$\mathrm{Tr}\rho,\mathrm{Tr}\rho_{\ast}\leq1$. We may consider the purified
distance~\cite{purified} $P(\rho,\rho_{\ast})=\sqrt{1-F_{G}(\rho,\rho_{\ast
})^{2}}$, where $F_{G}$ is the generalized quantum fidelity~\cite[Def. 3.3,
Lemma~3.1]{TomaThesis}%
\begin{align}
F_{G}(\rho,\rho_{\ast})  &  :=F(\rho,\rho_{\ast})+\sqrt{(1-\mathrm{Tr}%
\rho)(1-\mathrm{Tr}\rho_{\ast})},\\
F(\rho,\rho_{\ast})  &  :=||\sqrt{\rho}\sqrt{\rho_{\ast}}||_{1}~.
\end{align}
Using the Fuchs-van de Graaf inequalities~\cite{Fuchs}, one may check that
$D_{G}\leq P\leq\sqrt{2D_{G}-D_{G}^{2}}\leq\sqrt{2D_{G}}$, where $D_{G}$ is
the generalized trace distance~\cite[Def. 3.1]{TomaThesis}
\begin{align}
D_{G}(\rho,\rho_{\ast})  &  :=D(\rho,\rho_{\ast})+\frac{1}{2}|\mathrm{Tr}%
\rho-\mathrm{Tr}\rho_{\ast}|,\\
D(\rho,\rho_{\ast})  &  :=\frac{1}{2}||\rho-\rho_{\ast}||_{1}=\frac{1}%
{2}\mathrm{Tr}\left\vert \rho-\rho_{\ast}\right\vert ~.
\end{align}

In particular, consider the CQ states
\begin{align}
\rho &  =\sum_{x\in\aleph}P(x)\left\vert x\right\rangle _{C}\left\langle
x\right\vert \otimes\omega(x),\\
\rho_{\ast}  &  =\sum_{x\in\aleph}P_{\ast}(x)\left\vert x\right\rangle
_{C}\left\langle x\right\vert \otimes\omega_{\ast}(x),
\end{align}
where the classical system $C$ is equivalent to an alphabet $\aleph$ of
dimension $d$, and the quantum system $Q$\ has dimension $d_{Q}\geq d$. Here
$P(x)$ and $P_{\ast}(x)$ are probability distributions, while $\omega(x)$ and
$\omega_{\ast}(x)$ are generally sub-normalized states defined over system
$Q$. In the following, we assume that the state $\rho$ is normalized to $1$,
also denoted by $\rho\in S_{=}(\mathcal{H})$.

For any normalized\ state $\rho$ of two quantum systems $A$ and $B$, we may
write~\cite[Def. 5.2]{TomaThesis}%
\begin{equation}
H_{\text{min}}^{\varepsilon}(A|B)_{\rho}=\max_{\rho_{\ast}\in\mathcal{B}%
^{\varepsilon}(\rho)}H_{\text{min}}(A|B)_{\rho_{\ast}}, \label{torrrrGG}%
\end{equation}
where
\begin{equation}
\mathcal{B}^{\varepsilon}(\rho):=\{\rho^{\prime}:\mathrm{Tr}\rho^{\prime}%
\leq1,P(\rho^{\prime},\rho)\leq\varepsilon<1\}
\end{equation}
is a ball of generally sub-normalized states around $\rho$. In particular, for
any normalized CQ state $\rho$, we can find a (generally sub-normalized) CQ
state $\rho_{\ast}\in\mathcal{B}^{\varepsilon}(\rho)$ such that~\cite[Prop.
5.8]{TomaThesis}%
\begin{equation}
H_{\text{min}}^{\varepsilon}(C|Q)_{\rho}=H_{\text{min}}(C|Q)_{\rho_{\ast}}.
\label{torrr}%
\end{equation}

Consider a projector $\Pi:=\sum_{x\in\beth}\left\vert x\right\rangle
_{C}\left\langle x\right\vert $ defined over a reduced alphabet $\beth
\subseteq\aleph$ for the classical system $C$. Also consider two CQ states,
$\rho\in S_{=}(\mathcal{H}_{CQ})$ and $\rho_{\ast}\in S_{\leq}(\mathcal{H}%
_{CQ})$, the latter with normalization
\begin{equation}
\mathcal{N}:=\mathrm{Tr}\rho_{\ast}=\sum_{x\in\aleph}P_{\ast}(x)\mathrm{Tr}%
[\omega_{\ast}(x)]\leq1.
\end{equation}
We may write the two projected states
\begin{align}
\sigma &  =p^{-1}\Pi\rho\Pi=p^{-1}\sum_{x\in\beth}P(x)\left\vert
x\right\rangle _{C}\left\langle x\right\vert \otimes\omega(x),\\
\sigma_{\ast}  &  =p_{\ast}^{-1}\Pi\rho_{\ast}\Pi=p_{\ast}^{-1}\sum_{x\in
\beth}P_{\ast}(x)\left\vert x\right\rangle _{C}\left\langle x\right\vert
\otimes\omega_{\ast}(x),
\end{align}
with associated probabilities
\begin{align}
p  &  =\mathrm{Tr}\left(  \Pi\rho\right)  =\sum_{x\in\beth}P(x),\\
p_{\ast}  &  =\mathcal{N}^{-1}\mathrm{Tr}\left(  \Pi\rho_{\ast}\right)
=\mathcal{N}^{-1}\sum_{x\in\beth}P_{\ast}(x)\mathrm{Tr}[\omega_{\ast}(x)].
\end{align}
For $\rho_{\ast},\sigma_{\ast}\in S_{\leq}(\mathcal{H}_{CQ})$, we may then
write
\begin{equation}
H_{\text{min}}(C|Q)_{\sigma_{\ast}}\geq H_{\text{min}}(C|Q)_{\rho_{\ast}}%
+\log_{2}p_{\ast}. \label{toPPP}%
\end{equation}

In order to prove Eq.~(\ref{toPPP}) we adopt the approach of
Ref.~\cite[Lemma~1]{Cosmo_comp} (for normalized states) but starting from a
different result that is valid for sub-normalized states. For any
$\sigma_{\ast}\in S_{\leq}(\mathcal{H}_{CQ})$, we may write~\cite[Eq.~(4.6)]%
{TomaThesis}
\begin{equation}
2^{-H_{\text{min}}(C|Q)_{\sigma_{\ast}}}=\max_{\mathcal{E}_{Q\rightarrow
Q^{\prime}}}\left\langle \Gamma_{CQ^{\prime}}\right\vert \mathcal{I}%
\otimes\mathcal{E}(\sigma_{\ast})\left\vert \Gamma_{CQ^{\prime}}\right\rangle
,
\end{equation}
where $\mathcal{E}$ is a CPTP map (quantum channel) acting on system $Q$, and
\begin{equation}
\left\vert \Gamma_{CQ^{\prime}}\right\rangle :=\sum_{x\in\aleph}\left\vert
x\right\rangle _{C}\left\vert x\right\rangle _{Q^{\prime}}%
\end{equation}
is a non-normalized entangled state defined over the orthonormal set of states
$\{\left\vert x\right\rangle \}$. The latter is a basis for $C$ and a set for
$Q^{\prime}$, which is assumed to have $d_{Q^{\prime}}\geq d$. It is easy to
see that
\begin{align}
&  \left\langle \Gamma\right\vert \mathcal{I}\otimes\mathcal{E}(\sigma_{\ast
})\left\vert \Gamma\right\rangle \nonumber\\
&  =p_{\ast}^{-1}\sum_{x\in\beth}P_{\ast}(x)\left\langle \Gamma\right\vert
\left\{  \left\vert x\right\rangle _{C}\left\langle x\right\vert
\otimes\mathcal{E}[\omega_{\ast}(x)]\right\}  \left\vert \Gamma\right\rangle
\nonumber\\
&  \leq p_{\ast}^{-1}\sum_{x\in\aleph}P_{\ast}(x)\left\langle \Gamma
\right\vert \left\{  \left\vert x\right\rangle _{C}\left\langle x\right\vert
\otimes\mathcal{E}[\omega_{\ast}(x)]\right\}  \left\vert \Gamma\right\rangle
\nonumber\\
&  =p_{\ast}^{-1}\left\langle \Gamma\right\vert \mathcal{I}\otimes
\mathcal{E}(\rho_{\ast})\left\vert \Gamma\right\rangle .
\end{align}
This leads to
\begin{align}
2^{-H_{\text{min}}(C|Q)_{\sigma_{\ast}}}  &  \leq p_{\ast}^{-1}\max
_{\mathcal{E}_{Q\rightarrow Q^{\prime}}}\left\langle \Gamma_{CQ^{\prime}%
}\right\vert \mathcal{I}\otimes\mathcal{E}(\rho_{\ast})\left\vert
\Gamma_{CQ^{\prime}}\right\rangle \nonumber\\
&  =p_{\ast}^{-1}2^{-H_{\text{min}}(C|Q)_{\rho_{\ast}}}.
\end{align}
Taking the log we obtain Eq.~(\ref{toPPP}).

For the projected states, $\sigma$ and $\sigma_{\ast}$, and their
probabilities, $p$ and $p_{\ast}$, we may write the following inequalities
(proven below)
\begin{align}
\left\vert p-p_{\ast}\right\vert  &  \leq D_{G}(\rho,\rho_{\ast}%
),\label{mmm1}\\
D_{G}(\sigma,\sigma_{\ast})  &  \leq\frac{3}{2p}D_{G}(\rho,\rho_{\ast}).
\label{mmm2}%
\end{align}
In fact, consider the normalized state $\rho_{\ast N}:=\mathcal{N}^{-1}%
\rho_{\ast}$ so that $p_{\ast}=\mathrm{Tr}(\Pi\rho_{\ast N})$. Recall that the
trace distance between two normalized states $\rho$ and $\rho_{\ast N}$ is
equal to the maximum Kolmogorov distance between the probability distributions
generated by the application of a POVM. Considering the (generally
non-optimal) POVM\ $\{\Pi_{k}\}=\{\Pi,I-\Pi\}$, we may write
\begin{equation}
\left\Vert \rho-\rho_{\ast N}\right\Vert _{1}\geq\sum_{k}\left\vert
\mathrm{Tr}(\Pi_{k}\rho)-\mathrm{Tr}(\Pi_{k}\rho_{\ast N})\right\vert
=2\left\vert p-p_{\ast}\right\vert .
\end{equation}
Using the result above and the triangle inequality, we get
\begin{equation}
\left\vert p-p_{\ast}\right\vert \leq D(\rho,\rho_{\ast N})\leq D(\rho
,\rho_{\ast})+D(\rho_{\ast},\rho_{\ast N}).
\end{equation}
It is easy to check that
\begin{align}
D(\rho_{\ast},\rho_{\ast N})  &  =D(\rho_{\ast},\mathcal{N}^{-1}\rho_{\ast
})=\frac{1}{2}\mathrm{Tr}\left\vert (1-\mathcal{N}^{-1})\rho_{\ast}\right\vert
\nonumber\\
&  =\frac{\mathcal{N}^{-1}-1}{2}\mathrm{Tr}\rho_{\ast}=\frac{1-\mathrm{Tr}%
\rho_{\ast}}{2},
\end{align}
leading to
\begin{equation}
\left\vert p-p_{\ast}\right\vert \leq D(\rho,\rho_{\ast})+\frac{1-\mathrm{Tr}%
\rho_{\ast}}{2}=D_{G}(\rho,\rho_{\ast}).
\end{equation}

In order to prove Eq.~(\ref{mmm2}), we suitably extend the approach of
Ref.~\cite[Lemma~2]{Cosmo_comp} to include sub-normalized states. First
observe that
\begin{align}
D(\rho,\rho_{\ast})  &  =\sum_{\substack{x\in\aleph}}D\left[  P(x)\omega
(x),P_{\ast}(x)\omega_{\ast}(x)\right]  ,\\
D(\sigma,\sigma_{\ast})  &  =\sum_{x\in\beth}D\left[  p^{-1}P(x)\omega
(x),p_{\ast}^{-1}P_{\ast}(x)\omega_{\ast}(x)\right] \\
&  \leq p^{-1}\sum_{x\in\beth}D\left[  P(x)\omega(x),P_{\ast}(x)\omega_{\ast
}(x)\right] \label{gg3}\\
&  +\sum_{x\in\beth}D\left[  p^{-1}P_{\ast}(x)\omega_{\ast}(x),p_{\ast}%
^{-1}P_{\ast}(x)\omega_{\ast}(x)\right]  , \label{gg4}%
\end{align}
where we have used the triangle inequality for the trace distance (here
applied to Hermitian operators). It is easy to show that the term in
Eq.~(\ref{gg3}) can be bounded as follows%
\begin{align}
&  p^{-1}\sum_{x\in\beth}D(...)\leq p^{-1}\sum_{x\in\aleph}D(...)\\
&  =p^{-1}D(\rho,\rho_{\ast}).
\end{align}
For the second term in Eq.~(\ref{gg4}), we write%
\begin{align}
&  \sum_{x\in\beth}D[p^{-1}P_{\ast}(x)\omega_{\ast}(x),p_{\ast}^{-1}P_{\ast
}(x)\omega_{\ast}(x)]\\
&  =\sum_{x\in\beth}\frac{1}{2}\mathrm{Tr}|(p^{-1}-p_{\ast}^{-1})P_{\ast
}(x)\omega_{\ast}(x)|\\
&  =\frac{1}{2}|p^{-1}-p_{\ast}^{-1}|\sum_{x\in\beth}P_{\ast}(x)\mathrm{Tr}%
[\omega_{\ast}(x)]\\
&  =\frac{1}{2}p^{-1}p_{\ast}^{-1}|p-p_{\ast}|\mathcal{N}p_{\ast}\leq
\frac{|p-p_{\ast}|}{2p}\\
&  \leq\frac{1}{2p}D_{G}(\rho,\rho_{\ast}). \label{2repl}%
\end{align}
By combining the two terms, we find
\begin{equation}
D(\sigma,\sigma_{\ast})\leq\frac{1}{p}D(\rho,\rho_{\ast})+\frac{1}{2p}%
D_{G}(\rho,\rho_{\ast}).
\end{equation}
From the inequality above and the fact that $\mathrm{Tr}\sigma_{\ast
}=\mathrm{Tr}\rho_{\ast}$, we may derive the following%
\begin{align}
D_{G}(\sigma,\sigma_{\ast})  &  \leq\frac{1}{p}D(\rho,\rho_{\ast}%
)+\frac{1-\mathrm{Tr}\rho_{\ast}}{2}+\frac{1}{2p}D_{G}(\rho,\rho_{\ast
})\nonumber\\
&  =\frac{3}{2p}D_{G}(\rho,\rho_{\ast})-(1-p)\frac{1-\mathrm{Tr}\rho_{\ast}%
}{2p},
\end{align}
which leads to Eq.~(\ref{mmm2}).

We now have all the ingredients to conclude the proof. Given a normalized CQ
state $\rho$, take a generally sub-normalized CQ state $\rho_{\ast}%
\in\mathcal{B}^{\varepsilon}(\rho)$ which realizes Eq.~(\ref{torrr}), i.e.,%
\begin{equation}
H_{\text{min}}(C|Q)_{\rho_{\ast}}=H_{\text{min}}^{\varepsilon}(C|Q)_{\rho}.
\label{DGGG0}%
\end{equation}
For the projected states $\sigma$ and $\sigma_{\ast}$, we may replace
$D_{G}(\rho_{\ast},\rho)\leq P(\rho_{\ast},\rho)\leq\varepsilon$ in
Eqs.~(\ref{mmm1}) and~(\ref{mmm2}), and write%
\begin{align}
p_{\ast}  &  \geq p-\varepsilon,\label{DGGG1}\\
D_{G}(\sigma,\sigma_{\ast})  &  \leq\frac{3\varepsilon}{2p}. \label{DGGG2}%
\end{align}

From Eq.~(\ref{DGGG2}) we see that $P(\sigma,\sigma_{\ast})\leq\sqrt
{3\varepsilon/p}:=\varepsilon^{\prime}$, so that $\sigma_{\ast}\in
\mathcal{B}^{\varepsilon^{\prime}}(\sigma)$. Assume that $p>0$ and
$\varepsilon<p/3$ so that $\varepsilon^{\prime}<1$ and the $\varepsilon
^{\prime}$-ball is well defined (this is typically the case because
$p=\mathcal{O}(1)$ and $\varepsilon\simeq10^{-10}$). Therefore, from
Eq.~(\ref{torrrrGG}) we derive%
\begin{equation}
H_{\text{min}}^{\varepsilon^{\prime}}(C|Q)_{\sigma}\geq H_{\text{min}%
}(C|Q)_{\sigma_{\ast}}.
\end{equation}
We can combine the inequality above with Eq.~(\ref{toPPP}) which leads to%
\begin{equation}
H_{\text{min}}^{\varepsilon^{\prime}}(C|Q)_{\sigma}\geq H_{\text{min}%
}(C|Q)_{\rho_{\ast}}+\log_{2}p_{\ast}.
\end{equation}
Now using Eqs.~(\ref{DGGG0}) and~(\ref{DGGG1}), we get%
\begin{equation}
H_{\text{min}}^{\varepsilon^{\prime}}(C|Q)_{\sigma}\geq H_{\text{min}%
}^{\varepsilon}(C|Q)_{\rho}+\log_{2}(p-\varepsilon).
\end{equation}
Finally, by replacing $\varepsilon\rightarrow p\varepsilon^{2}/3$ so that
$\varepsilon^{\prime}\rightarrow\varepsilon$, we write%
\begin{equation}
H_{\text{min}}^{\varepsilon}(C|Q)_{\sigma}\geq H_{\text{min}}^{p\varepsilon
^{2}/3}(C|Q)_{\rho}+\log_{2}[p(1-\varepsilon^{2}/3)].
\end{equation}
The latter inequality provides Eq.~(\ref{eeq3}) up to performing the correct
replacements ($\sigma\rightarrow\tilde{\rho}_{n}$, $\rho\rightarrow
\rho^{\otimes n}$, $C\rightarrow l^{n}$, $Q\rightarrow E^{n}$ etc.)


\begin{thebibliography}{99}
\bibitem {NoteIntegral}Via direct numerical integration, one can also check
that $\int_{0}^{\eta}d\tau~\tau P_{0}(\tau)\lesssim\eta_{\text{tot}}$.
\bibitem {NoteCLOCK}Note that the repetition rate/clock $C$ needs to be less
than $1/3$ of the detector bandwidth, and the pulse duration is typically
chosen to be $0.1C^{-1}$ or a bit less than that.
\bibitem {NoteBETA}In an experimental implementation, the lower-bound $\beta
I(x:y^{\prime\prime})_{\tau_{\text{min}},\bar{n}_{\text{G}}}$ is replaced by
an empirical quantity $I_{\text{emp}}$ based on the entropy of the final key,
the rate of the code used for error correction, and the number of bits used
for digitalizing the continuous variables~\cite{Alex}. The value of
$I_{\text{emp}}$ may be higher than $I(x:y^{\prime\prime})_{\tau_{\text{min}%
},\bar{n}_{\text{G}}}$. One certainly retrieves $\beta\leq1$, when
$I_{\text{emp}}$ is compared with the optimal mutual information between Alice
and Bob.
\end{thebibliography}
\end{document}